\documentclass[12pt,preprint]{aastex}

\newcommand{\Msun}{M_{\sun}}

\newcommand{\Ha}{\mathrm{H\alpha}}
\newcommand{\Hb}{\mathrm{H\beta}}
\newcommand{\HI}{H {\sc{i}} }
\newcommand{\HII}{H {\sc{ii}} }
\newcommand{\CIV}{C {\sc{iv}} }
\newcommand{\NIII}{[N {\sc{iii}}] }
\newcommand{\NV}{N {\sc{v}} }
\newcommand{\OIII}{[O {\sc{iii}}] }
\newcommand{\SII}{[S {\sc{ii}}] }
\newcommand{\NaI}{Na {\sc{i}} }

\shorttitle{Massive stars and energy balance of ISM}
\shortauthors{Freyer, Hensler, \& Yorke}

\begin{document}


\title{Massive stars and the energy balance of the interstellar medium. II.
       The 35 $\Msun$ star and a solution to the ``missing wind problem''}


\author{Tim Freyer}
\affil{Institut f\"ur Theoretische Physik und Astrophysik
       der Universit\"at, D-24098 Kiel, Germany}
\email{freyer@astrophysik.uni-kiel.de}

\author{Gerhard Hensler}
\affil{Institut f\"ur Astronomie, University Observatory of Vienna,
       T\"urkenschanzstr. 17, A-1180 Vienna, Austria}
\email{hensler@astro.univie.ac.at}

\and

\author{Harold W. Yorke}
\affil{Jet Propulsion Laboratory, California Institute of Technology,
       MS 169-506, 4800 Oak Grove Drive, Pasadena, CA 91109, USA}
\email{Harold.Yorke@jpl.nasa.gov}




\begin{abstract}
  We continue our numerical analysis of the morphological and energetic
  influence of massive stars on their ambient interstellar medium for a
  35 $\Msun$ star that evolves from the main sequence through red supergiant
  and Wolf-Rayet phases, until it ultimately explodes as a supernova. We find
  that structure formation in the circumstellar gas during the early
  main-sequence evolution occurs as in the 60 $\Msun$ case but is much less
  pronounced because of the lower mechanical wind luminosity of the star.
  Since on the other hand the shell-like structure of the \HII region is
  largely preserved, effects that rely on this symmetry become more important.
  At the end of the stellar lifetime 1\% of the energy released as Lyman
  continuum radiation and stellar wind has been transferred to the
  circumstellar gas. From this fraction 10\% is kinetic energy of bulk motion,
  36\% is thermal energy, and the remaining 54\% is ionization energy of
  hydrogen. The sweeping up of the slow red supergiant wind by the fast
  Wolf-Rayet wind produces remarkable morphological structures and emission
  signatures, which are compared with existing observations of the Wolf-Rayet
  bubble S308, whose central star has probably evolved in a manner very
  similar to our model star. Our model reproduces the correct order of
  magnitude of observed X-ray luminosity, the temperature of the emitting
  plasma as well as the limb brightening of the intensity profile. This is
  remarkable, because current analytical and numerical models of Wolf-Rayet
  bubbles fail to consistently explain these features. A key result is that 
  almost the entire X-ray emission in this stage comes from the shell of
  red supergiant wind swept up by the shocked Wolf-Rayet wind rather than
  from the shocked Wolf-Rayet wind itself as hitherto assumed and modeled.
  This offers a possible solution to what is called the
  ``missing wind problem'' of Wolf-Rayet bubbles.  
\end{abstract}


\section{Introduction}
\label{sec_intro}

  In the first paper of this series
  \citep*[subsequently referred to as Paper I]{freyer03} we studied the
  evolution of the circumstellar gas around an isolated 60 $\Msun$ star
  by means of numerical two-dimensional radiation hydrodynamic simulations.
  We found that the interaction of the photoionized \HII region with the
  stellar wind bubble (SWB) strongly influences the morphological evolution
  during the early main-sequence (MS) phase of the star. On the one hand,
  the results show that the dynamical interaction processes contribute to
  the formation of complex structures which can be found in \HII regions.
  On the other hand, these processes also impact on how and to what
  extent the stellar energy input (wind and H-ionizing radiation) is
  supplied to the interstellar medium (ISM), distributed among different
  forms of energy, and ultimately radiated from the system. While the
  consideration of the stellar wind strongly enhances the kinetic energy of
  bulk motion present in the system, ionization energy and the associated
  thermal energy of warm gas are generally lowered because the stellar wind
  intensifies the formation of high-density structures (clumps) with shorter
  hydrogen recombination times and stronger cooling.

  With this paper we continue our numerical analysis of the morphological and
  energetic influence of massive stars on their ambient ISM for a 35 $\Msun$
  star that evolves from the MS through the red supergiant (RSG) and the
  Wolf-Rayet (W-R) phases until it explodes as a supernova (SN). The goals of
  this paper are to examine the combined influence of wind and ionizing
  radiation on the dynamical evolution of circumstellar matter for this second
  set of stellar parameters, to compare and contrast the two numerical models
  we have obtained so far and to compare the model with observations of
  bubbles around W-R stars that have undergone an RSG phase. We will complete
  our little sequence of circumstellar gas models for a 85 $\Msun$ star
  (D. Kr\"oger et al. 2005, in preparation) and a 15 $\Msun$ star,
  representing the upper and lower end, respectively, of the stellar
  mass range that we are investigating.
  
  The remainder of this paper has the following structure: In section
  \ref{sec_observations} we review some recent observations of
  W-R stars with circumstellar nebulae that are conjectured to be ejected
  during the RSG phase of the star. Section \ref{sec_numerics} briefly
  reflects on the numerical methods and initial conditions used to produce
  the results presented in this paper and describes the set of stellar
  parameters used as time-dependent boundary conditions that drive the
  evolution. The results of the model calculations along with a comparison
  to observations and analytical models are presented in section
  \ref{sec_results}. We summarize our main results and conclusions in
  section \ref{sec_conclusions}.

\section{Recent Observations}
\label{sec_observations}

  From the theoretical point of view, each star that has a supersonic wind
  with a sufficiently high mass-loss rate should be capable of blowing a SWB
  into the ISM. The crucial indicator for the existence of such SWBs is the
  hot gas produced at the reverse shock. The X-ray emission of the hot
  ($10^6-10^8~\mathrm{K}$) gas should be detectable if the surface brightness
  is high enough for the respective telescope in use. Although theory predicts
  that even massive stars on the MS are supposed to produce SWBs, they are
  expected to be large and diffuse with a low surface brightness in X-rays.
  Thus, the vicinities of W-R stars seem to be promising candidates
  for the observation of SWBs (also called W-R bubbles if the central star
  is in its W-R stage). However, only 1/4 to 1/3 of the known galactic
  W-R stars seem to have associated ring nebulae and only 10 are wind-driven
  bubbles \citep{wrigge99}, according to the optically derived kinematics of
  the shell \citep{chu81}. Up to now only 2 of these SWB candidates have
  actually been detected in X-rays: NGC\,6888 \citep{bochkarev88, wrigge94}
  and S308 \citep{wrigge99, chu03}. Since both central W-R stars are thought
  to have undergone the MS $\rightarrow$ RSG $\rightarrow$ W-R evolution that
  we are investigating in this paper, we will briefly review the recent
  observational work for a careful comparison with our numerical results.

  NGC 6888 was among the first galactic ring nebulae whose formation have been
  attributed to mass ejection and radiation from a W-R star \citep{johnson65}.
  Its relative proximity and thus large angular size ($18' \times 12'$) has
  made it to one of the best studied examples of this class of objects. At a
  distance of 1.45 kpc \citep{wendker75} the physical radius of the nebula is
  3.8 pc (major axis) $\times$ 2.5 pc (minor axis). The WN6 star HD 192163
  \citep[= WR 136 according to the list of][]{vanderHucht01} is close to the
  center of the nebula. The ellipsoidal shell appears to be geometrically very
  thin and has a highly filamentary structure. With a mean electron density in
  the filaments of the nebula of 400 $\mathrm{cm^{-3}}$ \citep{parker64},
  \citet{wendker75} derived a mass of 5 $\Msun$ for the ionized shell and a
  mean shell thickness of 0.01 pc. The shell expansion velocity varies among
  different authors in the range $75 - 93~\mathrm{km~s^{-1}}$
  \citep{treffers82, marston88, moore00}.

  It is important to distinguish two morphological features that are
  theoretically expected to evolve during the RSG and W-R stage of the star:
  A thin and dense ``RSG shell'' forms around the outflowing RSG wind when
  the reverse shock becomes radiative due to the high density of the wind in
  this stage. When the fast W-R wind turns on, the non-radiative reverse shock
  reestablishes itself and the shocked W-R wind sweeps up the RSG wind in the
  so-called ``W-R shell''. The dynamical age of the W-R shell
  \begin{equation}
    t_{\mathrm{dyn}} = \frac{\eta R_{\mathrm{shell}}}{v_{\mathrm{shell}}}
    \label{eq_t_dyn} 
  \end{equation}
  depends on the expansion velocity and on the assumption of the density
  profile into which the shell expands. Typically, two cases are considered:
  a constant ambient density \citep[$\eta = 0.6$ according to][]{weaver77}
  and a $\rho \propto 1/r^2$ density profile
  \citep[$\eta = 1.0$ according to][]{garcia95a}.
  With the expansion velocities quoted above, a mean shell radius of
  3.2 pc, and the assumption that the optically visible nebula can be
  associated with the W-R shell, one obtains a minimum value for the
  dynamical age of NGC 6888 of $2.0 \times 10^4~\mathrm{yr}$
  ($\eta = 0.6$ and $v_{\mathrm{shell}}$ = $93~\mathrm{km~s^{-1}}$)
  and a maximum value of $4.2 \times 10^4~\mathrm{yr}$
  ($\eta = 1.0$ and $v_{\mathrm{shell}}$ = $75~\mathrm{km~s^{-1}}$).
 
  \citet{moore00} used the WFPC2 on the {\it Hubble Space Telescope (HST)}
  to examine the filaments of the bright northeast rim of NGC 6888 in the
  light of the $\Ha$ $\lambda 6563$, \OIII $\lambda 5007$, and
  \SII $\lambda\lambda 6717,~6731$ lines. They found filament densities of
  $1000-1600~\mathrm{cm^{-3}}$. The dense shell is enveloped by a skin of
  emission most evident in \OIII $\lambda 5007$ and it is proposed that this
  skin arises in a cooling regime behind a radiative shock driven into the
  medium around the shell. For the low density that is expected in the
  MS bubble without heat conduction a forward shock ahead of the nebula shell
  would not be visible. Thus, the authors propose that a considerable fraction
  of the approximately 18 $\Msun$ RSG wind is possibly present in the
  MS bubble as a low-density ($2~\mathrm{cm^{-3}}$) exterior layer with low
  $\Ha$ surface brightness, visible only in \OIII $\lambda 5007$ postshock
  emission as the skin engulfing the $\Ha$ filaments when it becomes shocked
  by some combination of nebular shell expansion and the pressure of the
  postshock W-R wind overtaking the RSG shell. The large discrepancy
  between the density in the MS bubble of the model and the observed density
  of the skin could be explained by thermal evaporation of RSG wind material
  into the MS bubble. Spectroscopy of the nebular shell shows that the ionized
  gas is enriched with nitrogen and helium and underabundant in oxygen.
  Possible explanations might be the transport of chemically enriched material
  from the core of the star to the outer layers that are ejected during the
  RSG stage \citep{esteban92} or the mixing of W-R and RSG wind material
  \citep{kwitter81}. The high abundance of nitrogen in the nebula is
  consistent with the classification of HD 192163 as a WN6 star.   

  There is some debate on how much neutral gas is present in
  NGC 6888. The answer to this question has implications on the
  interpretation of the dynamics of the bubble as being either energy or
  momentum conserving. \citet{marston88} found from {\it IRAS} observations
  (assuming a gas to dust mass ratio of 100) a shell of 40 $\Msun$ neutral gas
  directly outside the ionized shell. The difficulty in the
  interpretation of the observed data is the correct estimate of the 
  forbidden line contribution in the far infrared. \citet{vanBuren88}
  concluded that the emission in the {\it IRAS} 60 $\mathrm{\mu m}$ and
  100 $\mathrm{\mu m}$ bands is predominantly from forbidden \OIII lines
  indicating that the continuum emission from the dust and thus the neutral
  mass would be negligible. \citet{marston91} claims that the flux
  contribution from forbidden \OIII and \NIII lines is only $8-20$\% and that
  the major reason for these deviating results is the higher flux from the
  nebula which he derived because of a differing background removal procedure.
  However, \citet{moore00} conclude from their {\it HST} WFPC2 observations
  of the nebula that there cannot be a significant amount of neutral material
  close to the optical nebula. Their result for the hydrogen ionizing flux of
  $10^{49.3} \mathrm{s^{-1}}$ is in reasonable agreement with the value of
  $10^{49} \mathrm{s^{-1}}$ from the W-R models of \citet{crowther96} and
  does perfectly match the value from \citet{garcia96b} that we use for
  our calculations presented in this paper, but it is roughly a factor of
  50 higher than the Lyman continuum flux of $10^{47.6} \mathrm{s^{-1}}$
  which \citet{marston88} found as necessary to maintain the observed
  $\Ha$ brightness of the nebula. In other words: Only 2\% of the
  ionizing photons from HD 192163 are used to ionize the observable
  nebular shell of NGC 6888 - it is ``density-bounded'' not
  ``ionization-bounded''. 98\% would still be available to ionize
  the neutral parts of the shell. Since the shock front ahead of the
  shell is partially shadowed from the stellar Lyman continuum flux by dense
  clumps within the shell, it is obvious that there is indeed some neutral
  material in the shell, but on the other hand the shell is very leaky to
  Lyman continuum photons and it is unlikely that there is as much as
  40 $\Msun$ neutral gas present directly outside the ionized shell because
  that would process a larger fraction of the stellar Lyman continuum flux
  into $\Ha$ emission. 

  Observational evidence has also been found for the existence of the shell of
  swept-up ambient ISM around the MS bubble. Using highly resolved {\it IRAS}
  images \citet{marston95} discovered an elliptical shell with average radius
  of 19 pc (assuming a distance to HD 192163 of 1.45 kpc) and mass of
  $\approx$8000~$\Msun$ (assuming a gas to dust mass ratio of 100).
  However, this value should be used with some caution since it was obtained
  with the same techniques that have been applied to produce the controversial
  result for the amount of neutral mass contained in the W-R bubble.

  NGC 6888 was the first SWB that has been observed in X-rays
  \citep{bochkarev88, wrigge94}. The latter authors examined the X-ray
  emission from NGC 6888 using the PSPC detector of the {\it ROSAT} satellite
  and found that it is filamentary and concentrated in the brightest optical
  features of the nebula. 70\% of the total X-ray emission originates from
  only ${\approx}1-2\%$ of the bubble volume. With their assumed distance to
  NGC 6888 of 1.8 kpc the X-ray luminosity is estimated to be
  $1.6 \times 10^{34}~\mathrm{ergs~s^{-1}}$ in the energy band $0.07-2.41$ keV
  and the plasma temperature to be $2 \times 10^6~\mathrm{K}$.  
  \citet{wrigge02} also used the HRI detector on board the {\it ROSAT}
  satellite to obtain X-ray emission maps with higher spatial resolution.
  They conclude that approximately half of the total X-ray emission originates
  from small filaments with typical size of a few tenths of a pc and typical
  luminosity in the {\it ROSAT} HRI band of several
  $10^{31}~\mathrm{ergs~s^{-1}}$, corresponding to a hydrogen number density
  in the filaments of a few $\mathrm{cm^{-3}}$. \citet{wrigge02} also examined
  the possibility that the emission of these filaments is produced by clumps
  of dense gas which evaporate in the bubble of hot gas. They find that this
  explanation could only be consistent with the observations if the bubble gas
  is sufficiently hot ($5-10 \times 10^6~\mathrm{K}$).

  S308 surrounding the WN4 star HD 50896 (= EZ CMa = WR 6) is the second SWB
  that has been observed in X-rays. The optically visible nebula is
  characterized by an almost spherically symmetric shell with a remarkable
  protrusion in the northwest quadrant but otherwise with no hints on the
  formation of pronounced instabilities. The distance to S308 and thus all
  quantities which scale with the distance are somewhat uncertain.
  \citet{chu82b} give $D = 1.5~\mathrm{kpc}$. \citet{hamann88} found that the
  range from 0.9 to a few kpc is in agreement with the results of their
  spectral analysis, with a preferred value of 2 kpc. \citet{howarth95}
  estimate $D = 1.8~\mathrm{kpc}$ with an uncertainty of 15\% based on their
  high-resolution observations of the interstellar \NaI D lines in the spectra
  of HD 50896 and several nearby stars. Measurements with the {\it Hipparcos}
  satellite indicated $D = 0.6^{+0.4}_{-0.2}~\mathrm{kpc}$ \citep{perryman97},
  which seems to be unreasonably close, when compared with the other
  distance-estimate techniques. The photometric distance to HD 50896 is
  $D = 1.0 \pm 0.2~\mathrm{kpc}$ \citep{vanderHucht01} and the kinematic
  distance based on radial velocity information (of S308 and neighboring
  \HII regions) and Galactic rotation is $D = 1.5 \pm 0.2~\mathrm{kpc}$
  \citep{chu03}. We will subsequently use a distance of 1.5 kpc and all the
  distance-depending values that we quote from other papers are scaled
  accordingly. 

  The radius of the optically visible shell is 9 pc and the shell expands with
  $v = 63~\mathrm{km~s^{-1}}$ \citep{chu03}. The ionized mass of the shell is
  54 $\Msun$ \citep[][scaled according to the values of radius and expansion
  velocity used here]{chu82b}. The derived average hydrogen number density of
  the ambient gas that has been swept up is thus
  $\approx$0.6~$\mathrm{cm^{-3}}$.
  An interesting implication of the distance and position on the sky is that
  S308 lies approximately 260 pc above the galactic plane, an unusual place
  for a massive star. Since HD 50896 also has a compact companion, possibly
  the stellar remnant of a SN explosion, it has been proposed that it is
  a runaway star \citep{firmani79}. However, because the location of HD 50896
  is almost central within S308 the transverse velocity of HD 50896 with
  respect to its shell cannot be very high. For an adopted dynamical age of
  the nebula of 0.14 Myr \citep{chu03} the star can reach its projected
  displacement from the center of the bubble with a velocity of less than
  $7~\mathrm{km~s^{-1}}$ \citep{chu82b}. There is no morphological evidence
  for supersonic transverse motion of the star + shell system with respect to
  its local ISM.

  \citet{wrigge99} used the {\it ROSAT} PSPC detector to examine the X-ray
  emission from S308. Despite some technical difficulties he found that the
  observed spectrum can be fitted by a two-temperature emission model,
  one component at $T= 1.5 \times 10^6~\mathrm{K}$ and the second
  at $T= 2.8 \times 10^7~\mathrm{K}$. A total luminosity of
  $2.1 \times 10^{33}~\mathrm{ergs~s^{-1}}$ (scaled to the distance of
  $D = 1.5~\mathrm{kpc}$) in the energy band between 0.1 keV and 2.04 keV 
  has been determined. The emitting volume is probably a thick shell
  with a ratio of inner to outer radius
  $\approx$0.5. 

  One of the major conclusions of \citet{wrigge99} regarding the theoretical
  models is that S308 cannot be described by either the \citet{weaver77}
  constant ambient density model or by the two-wind model of \citet{garcia95a}.
  A comparable bubble produced within the two-wind framework would have an
  X-ray luminosity of $9 \times 10^{33}~\mathrm{ergs~s^{-1}}$, which is
  roughly in agreement with the observed value (bearing in mind the
  uncertainty in the determination of the observed X-ray luminosity).
  The problem is that the stellar wind luminosity which is necessary to
  reproduce the bubble in the model is only $\approx$3\% of the stellar wind
  luminosity that \citet{hamann93} derived for HD 50896
  ($v_w = 1700~\mathrm{km~s^{-1}}$,
  $\dot{M}_w = 5.4 \times 10^{-5}~\Msun~\mathrm{yr^{-1}}$, scaled to the
  distance of $D = 1.5~\mathrm{kpc}$ used here).
  Even if the lower clumping-corrected mass-loss rate of \citet{nugis98} is
  used to derive the mechanical wind luminosity, the value from the
  model is still almost an order of magnitude lower. This is called the
  ``missing wind problem''. Moreover, the observed X-ray surface
  brightness profile of S308 is limb brightened while the theoretical profile
  is centrally filled. The result for the classical bubble model according to
  \citet{weaver77} is similar: Although the stellar wind luminosity needed to
  reproduce the observed bubble kinematics is higher than in the two-wind
  model of \citet{garcia95a}, it is still only $\approx$8\% of the
  observed value ($\approx$32\% for the clumping-corrected mass-loss rate). 

  \citet{chu03} observed the X-ray emission from the northwest quadrant
  of S308 using the EPIC CCD cameras of the {\it XMM-Newton} satellite.
  They found that the X-ray emission is completely interior to the optical
  shell and reconfirmed the limb brightening. The brightest X-ray emitting
  regions are linked to bright optical filaments. The total X-ray luminosity
  of S308, extrapolated from the observed flux from the northwest quadrant, is
  $\le (1.2 \pm 0.5) \times 10^{34}~\mathrm{ergs~s^{-1}}$ in the energy band
  between 0.25 keV and 1.5 keV. The observed spectrum can be fitted with the
  emission of an optically thin, nitrogen-enriched plasma of temperature
  $T = 1.1 \times 10^6~\mathrm{K}$. This is quite ``cool'' compared to the
  postshock temperature of order $10^8$ K which is expected for a rare stellar
  wind with terminal velocity of a few $1000~\mathrm{km~s^{-1}}$ and might
  indicate that the shocked W-R wind has mixed with cold gas by the processes
  of thermal evaporation and/or dynamic ablation. The spectrum is very soft
  so that there is basically no emission beyond 1 keV. The existence of a
  high-temperature gas component that substantially contributes to the
  observed emission and that has been claimed to be detected by
  \citet{wrigge99} is ruled out. Less than 6\% of the observed X-ray flux
  (which corresponds to 1.5\% of the unabsorbed X-ray flux) can be attributed
  to the emission of a hotter gas component. The reason for these
  contradicting results is probably the low signal-to-noise ratio in the
  {\it ROSAT} PSPC data that \citet{wrigge99} obtained together with a
  number of point sources which have not been resolved by the PSPC detector.
  Since the \OIII $\lambda 5007$ emission of S308 has a sharp rim,
  the nebular shell is probably still surrounded by unaffected
  RSG wind material. This RSG material cannot extend too much farther out,
  if the protrusion in the northwest quadrant is interpreted as a first
  blowout. There is a gap between the outer rim of the \OIII $\lambda 5007$
  emission and the outer edge of the X-ray emission of between 90 arcsec to
  over 200 arcsec, corresponding to $0.7-1.5$ pc. \citet{chu03} interpret this
  gap as being filled by the W-R shell and probably a transition layer. This
  is also indicated by the detection of an interstellar \NV absorption line
  toward HD 50896, which \citet{boroson97} attribute to the shell of S308.

  The electron density in the X-ray emitting gas is estimated to be
  $n_e = 0.28 \pm 0.04~\mathrm{cm^{-3}}$ or
  $n_e = 0.63 \pm 0.09~\mathrm{cm^{-3}}$ for an assumed hot gas volume filling
  factor of 0.5 or 0.1, respectively. The filling factor is expected to
  be closer to 0.5 since the limb-brightened X-ray emission profile suggests
  emission from a thick shell. The resulting mass of the X-ray emitting gas is
  $11 \pm 5~\Msun$ or $5 \pm 3~\Msun$ for the two volume filling factors.
  Assuming that the optical emission which has been used to determine the
  expansion velocity comes from the W-R shell, the age of the W-R phase can
  be estimated using equation \ref{eq_t_dyn}. For
  $R_{\mathrm{shell}}$ = 9 pc, $v_{\mathrm{shell}}$ = $63~\mathrm{km~s^{-1}}$,
  and $\eta = 1.0$ this yields $t_{\mathrm{dyn}}$ = 0.14 Myr.
  For a clumping-corrected mass-loss rate of HD 50896 of
  $1.4 \times 10^{-5}~\Msun~\mathrm{yr^{-1}}$
  \citep[][scaled to the distance used here]{nugis98}
  the mass blown into the bubble by the W-R wind is 2 $\Msun$.
  This is less than the X-ray emitting gas mass, a fact which in turn supports
  the idea that RSG material is mixed with the shocked W-R wind in the
  bubble or that the main source of X-rays is different from the shocked W-R
  wind. Observations of S308 in the optical waveband furthermore show that
  besides photoionization there is also shock heating present, indicated by
  the high \OIII $\lambda 5007$/$\Hb$ ratio of 20 \citep{esteban92b}.
  Similar to the case of NGC 6888, the \OIII $\lambda 5007$ emission leads the
  $\Ha$ emission by 16-20 arcsec \citep{gruendl00} corresponding to
  $0.12-0.15$ pc. Furthermore, S308 is located in an \HI cavity
  swept free by the MS wind of HD 50896 \citep{arnal96}.
  
\section{Numerical Method}
\label{sec_numerics}

\subsection{Radiation Hydrodynamics Scheme}
\label{sec_num_rad_hydro}

  The numerical code used to obtain the results presented in this paper is
  described in \citeauthor{freyer03}. The hydrodynamical equations are solved
  together with the transfer of H-ionizing photons on a two-dimensional
  cylindrical grid. The time dependent ionization and recombination of
  hydrogen is calculated each time step and we carefully take stock of all the
  important energy exchange processes in the system. For a closer description
  of the algorithm we refer the reader to \citet{yorke95}, \citet{yorke96},
  and \citeauthor{freyer03}.

\subsection{Initial Conditions}
\label{subsec_ini_cond}

  We use the same undisturbed background gas layer with hydrogen number
  density $n_0 = 20~\mathrm{cm^{-3}}$ and temperature $T_0 = 200~\mathrm{K}$
  for the reasons described in \citeauthor{freyer03} and start our
  calculations with the sudden turn-on of the zero-age main-sequence (ZAMS)
  stellar radiation field and stellar wind. After $t = 700~\mathrm{yr}$ the
  spherical one-dimensional solution is used as initial model for the
  two-dimensional calculations.

\subsection{Boundary Conditions}
\label{subsec_bound_cond}

  The basic difference between the 60 $\Msun$ case described in
  \citeauthor{freyer03} and the 35 $\Msun$ case is the set of time-dependent
  stellar parameters. The mass-loss rate ($\dot{M}_w$), terminal velocity
  of the wind ($v_w$), effective temperature ($T_{\mathrm{eff}}$), and
  photon luminosity in the Lyman continuum ($L_{\mathrm{LyC}}$) drive the
  evolution of the circumstellar gas. For the 35 $\Msun$ star we adopt the
  stellar parameters given by \citet[][their version with the ``fast''
  $75~\mathrm{km~s^{-1}}$ RSG wind]{garcia96b} shown in
  Figure \ref{35Msun_input4apj_up.eps}. Here, the star undergoes the
  following evolution: MS O star $\rightarrow$ RSG star $\rightarrow$ W-R star
  $\rightarrow$ SN. The MS phase lasts for about $4.52~\mathrm{Myr}$, the
  RSG phase $0.234~\mathrm{Myr}$ and the final W-R phase $0.191~\mathrm{Myr}$.
  After $4.945~\mathrm{Myr}$ the calculation stops and the star 
  explodes as a SN.  

  During the star's MS phase the mass-loss rate rises approximately from
  $3 \times 10^{-7}~\Msun~\mathrm{yr^{-1}}$ to
  $10^{-6}~\Msun~\mathrm{yr^{-1}}$. With the onset of the
  RSG phase the mass-loss rate jumps up to almost
  $10^{-4}~\Msun~\mathrm{yr^{-1}}$. In the final W-R phase it
  is of the order $(2-3) \times 10^{-5}~\Msun~\mathrm{yr^{-1}}$.
  The terminal velocity of the stellar wind on the MS is between
  $2 \times 10^{3}~\mathrm{km~s^{-1}}$ and
  $4 \times 10^{3}~\mathrm{km~s^{-1}}$, during the
  RSG phase it drops below $100~\mathrm{km~s^{-1}}$ until it again reaches
  values of $(1-4) \times 10^{3}~\mathrm{km~s^{-1}}$ in the final W-R phase.
  The effective temperature during the MS phase lies in the range of
  $(3-4) \times 10^4~\mathrm{K}$, falls down to a few thousand Kelvin in the
  RSG phase and temporarily reaches more than $10^5~\mathrm{K}$ during the
  final W-R phase. While the total photon luminosity of the star is relatively
  constant with values of $(7-14) \times 10^{38}~\mathrm{ergs~s^{-1}}$ during
  the whole evolution of the star, the photon luminosity in the Lyman
  continuum and the mechanical luminosity of the wind reflect the strong
  changes of mass-loss rate, terminal velocity of the wind and effective
  temperature. The Lyman continuum luminosity during the MS is in the range of
  $(2-3) \times 10^{38}~\mathrm{ergs~s^{-1}}$. In the course of the RSG phase
  it is insignificant due to the strong decrease of the effective temperature.
  The inverse situation occurs during the subsequent W-R phase:
  Because of the high effective temperature, most of the radiative power
  is emitted above the Lyman edge and the Lyman continuum luminosity reaches
  almost $10^{39}~\mathrm{ergs~s^{-1}}$. The mechanical wind luminosity
  ($\dot{M}_w v_w^2 / 2$) during the MS phase is
  $(1-2) \times 10^{36}~\mathrm{ergs~s^{-1}}$, drops by approximately one
  order of magnitude during the RSG phase, and reaches the highest values
  between $10^{37}~\mathrm{ergs~s^{-1}}$ and $10^{38}~\mathrm{ergs~s^{-1}}$
  in the final W-R phase.

  Here, the radius of the ``wind generator region'' is
  $2.16 \times 10^{17}~\mathrm{cm}$. The grid organization scheme as well as
  the other boundary conditions are unchanged compared to the 60 $\Msun$ case
  presented in \citeauthor{freyer03}.

\subsection{Geometry and Resolution}
\label{subsec_geometry}

  The size of the coarsest mesh, representing the size of the 
  computational domain, is
  $r_{\mathrm{max}} = z_{\mathrm{max}} = 50~\mathrm{pc}$. This value ensures
  that the coarsest grid covers the entire evolution during the lifetime
  of the star and thus prohibits that the ionization front or moving material
  can reach the outer boundary of the coarsest grid. The size is smaller than
  for the 60 $\Msun$ case since the size of the developing bubble is smaller
  as well. Six nested grids are employed within the coarsest grid, resulting
  in seven grid levels. On each grid level $125 \times 125$ cells (excluding
  ghost cells) are used resulting in a linear resolution that ranges from
  $6.25 \times 10^{-3}~\mathrm{pc}$ close to the star to 0.4 pc in the
  outermost parts of the coarsest grid.

\section{Results and Discussion}
\label{sec_results}

\subsection{A Resolution Study}
\label{subsec_validity}

  Similar to \citeauthor{freyer03} we performed a resolution study for the
  35 $\Msun$ case to ensure that the spatial resolution is high enough and
  that the numerical results are trustworthy. The same selection of parameters
  was calculated with three different resolutions,
  medium (125 cells in each dimension), low (61 cells in each dimension),
  and high (253 cells in each dimension).  Again, we compare only the
  first Myr of the evolution since the high-resolution model would require
  too much CPU time for the entire lifetime of the star.
  Figure \ref{comp_resol_228_229_230apj_up.eps} shows the results of this
  resolution study. 

  The quality of the results is comparable to that of the
  60 $\Msun$ case presented in \citeauthor{freyer03}. The variation of the
  thermal energy with resolution is $\lesssim$ 0.05 dex at almost any time
  during the first Myr. For ionization energy and kinetic energy of bulk
  motion the deviation between low and high resolution is $\lesssim$ 0.05 dex
  for $0.5~\mathrm{Myr} \le t \le 1.0~\mathrm{Myr}$ and $\lesssim$ 0.1 dex for
  $t \le 0.5~\mathrm{Myr}$, the latter case with a considerably smaller shift
  between medium and high resolution than between low and medium resolution,
  indicating that these values are already close to the actual limit.
  These results show that for the 35 $\Msun$ case discussed here
  the errors in our energetic analysis due to resolution effects are within
  an acceptable range, while there are morphological details which remain to
  be explored in future simulations with even higher resolution.

\subsection{Evolution of the 35 $\Msun$ Case}
\label{subsec_35msun_model}

  Figures \ref{ion_uchii_apj230.001.001.3.med.mono.eps} to 
  \ref{ion_uchii_apj229.967.008.1.med.mono.eps} depict the evolution of the
  gas in the vicinity of and under the influence of the 35 $\Msun$ model star.
  The data are plotted in the same manner as in \citeauthor{freyer03} for
  the 60 $\Msun$ case. Once again, we begin our discussion with the initial
  model that has been set up from the one-dimensional solution. For the
  35 $\Msun$ case this is done after 700 yr because --- due to the lower
  mechanical wind luminosity of the star --- the pressure in the hot bubble is
  lower. Thus, the forward shock is weaker and the shell of swept-up
  material is less heated and collapses earlier.

  As expected, the basic structure of the SWB/\HII region seen in Figure
  \ref{ion_uchii_apj230.001.001.3.med.mono.eps} is the same as for the
  60 $\Msun$ case except for the length and time scales. The stellar wind flows
  with the terminal velocity of nearly $4000~\mathrm{km~s^{-1}}$ freely out to
  $r \approx 0.08~\mathrm{pc}$, the position of the reverse shock, where it is
  heated up to about $10^8~\mathrm{K}$.  The forward shock that sweeps up the
  \HII region is at $r \approx 0.23~\mathrm{pc}$ and moves with some
  $130~\mathrm{km~s^{-1}}$. Density and temperature immediately behind this
  shock front are $\rho \approx 1.4 \times 10^{-22}~\mathrm{g~cm^{-3}}$ and
  $T \approx 2.5 \times 10^5~\mathrm{K}$, respectively, which is in good
  agreement with the jump conditions for a strong shock moving into 
  photoionized gas with $T \approx 8000~\mathrm{K}$. The ionization front
  is still weak R-type at 4.7 pc.

  After $5 \times 10^4~\mathrm{yr}$
  (Figure \ref{ion_uchii_apj230.017.001.3.med.mono.eps}) the
  hot bubble extends out to about 2.3 pc. The shell of swept-up \HII region
  expands with some $24~\mathrm{km~s^{-1}}$. Density knots have been produced
  in the thin shell, similar to those seen in the 60 $\Msun$ case, altering
  the optical depth along different radial lines of sight. The \HII region has
  begun to expand with almost $10~\mathrm{km~s^{-1}}$, sweeping up
  the ambient neutral gas, but the ionization front has started to retreat
  at the places where the clumps in the stellar wind shell cast shadows into
  the \HII region. The basic morphological 
  structure is still comparable to the 60 $\Msun$ case.

  Figure \ref{ion_uchii_apj230.030.001.3.med.mono.eps} shows the evolutionary
  state of the 35 $\Msun$ case after 0.2 Myr. The radius of the \HII region
  is approximately 9.5 pc and it expands into the ambient medium at several
  $\mathrm{km~s^{-1}}$. (When neutral shadows and ionized
  extensions appear, we define the radius of the \HII region as the
  distance from the star to the ``undisturbed'' ionization front which is
  neither extended along the fingers nor shortened by the shadows.)
  The hot bubble has grown to $r \approx 4.8~\mathrm{pc}$ and
  the expansion velocity of the stellar wind shell has decelerated to
  $\approx$13~$\mathrm{km~s^{-1}}$. This is comparable to the sound speed
  in the \HII region; i.e., the outermost side of the stellar wind shell is no
  longer bound by a shock front and has begun to expand into the \HII region
  as a result of the pressure gradient between the shell and the \HII region.
  (For clarity, we distinguish the stellar wind shell and the \HII region.
  Actually, the stellar wind shell is an important part of the \HII region
  because it is, at least at this point in time, photoionized by the star.)
  At $t = 0.2~\mathrm{Myr}$ the geometrical thickness of the stellar wind
  shell has already grown to 1 pc and the density in the shell is
  $(6-9) \times 10^{-23}~\mathrm{g~cm^{-3}}$.

  As a result of the growth of the geometrical shell thickness, the thin-shell
  instability that triggered the formation of density clumps in the shell
  ceased and the clumps dissolved; the period of time they were
  present in the stellar wind shell was relatively short. Thus, the shadows
  in the \HII region are much less pronounced and the finger-like extensions
  of the \HII region are shorter and less numerous than for the 60 $\Msun$
  case. Nevertheless, density fluctuations in the \HII region of more than
  an order of magnitude have been produced. (This may be a lower limit because
  of restrictions in resolution.) Maximum densities are around
  $\rho \approx 5 \times 10^{-23}~\mathrm{g~cm^{-3}}$ and minima around
  $\rho \approx 3 \times 10^{-24}~\mathrm{g~cm^{-3}}$.
  At these densities, temperatures, and masses the clumps are not
  gravitationally bound.

  The dissolution of the stellar wind shell can impressively be seen in Figure
  \ref{ion_uchii_apj230.041.001.3.med.mono.eps} at $t = 0.3~\mathrm{Myr}$.
  The expansion of the stellar wind shell has become a 2 pc broad belt of
  outflow with density $(2-5) \times 10^{-23}~\mathrm{g~cm^{-3}}$ from the
  contact discontinuity at $r \approx 6~\mathrm{pc}$ into the \HII region
  with about sound speed.    

  Additional processes are triggered by the rarefaction of the gas that
  previously belonged to the stellar wind shell: Because of the increase of
  hydrogen recombination time with lower density, excess photons are
  generated which reionize the shadowed regions (see lower panel of
  Figure \ref{ion_uchii_apj230.041.001.3.med.mono.eps}) and advance the
  ionization front even further, evaporating additional material from the
  \HII region shell of swept-up ambient medium ahead. This evaporation is
  visible in the density increase and the disturbed velocity field in the
  outer parts of the \HII region close to the ionization front.
  As a transient phenomenon, another shell with a density of
  $(3-6) \times 10^{-23}~\mathrm{g~cm^{-3}}$ develops at
  $t \approx 0.35~\mathrm{Myr}$ when the former stellar wind shell gas
  flowing into the \HII region collides with the photoevaporated
  material from this swept-up ambient shell.

  Together with the relics of denser fragments in the \HII region, the
  larger opacity in this dense shell causes once again the formation of
  small ripples in the ionization front at $t \approx 0.4~\mathrm{Myr}$
  (Figure \ref{ion_uchii_apj230.048.002.3.med.mono.eps}). At this time the
  radius of the hot bubble is about 7.5 pc and the geometrical thickness of
  the whole \HII region is about 4 pc. However, the newly formed shell soon
  dissolves because of its overpressure with respect to the rest of the
  \HII region, and within certain limits it can thus be said that the
  dissolution of the stellar wind shell leads to a ``rehomogenization'' of
  the \HII region.

  If we continue our analysis of this case to $t = 0.6~\mathrm{Myr}$
  (Figure \ref{ion_uchii_apj230.063.002.2.med.mono.eps}, please note the
  different scale), we see that the hot bubble has grown to almost 10 pc in
  radius. The shell-like \HII region, still expanding into the ambient medium
  at about $10~\mathrm{km~s^{-1}}$, becomes more and more homogeneous.
  Nevertheless, remaining density fluctuations from the previous formation
  and destruction processes are still in the range from
  $7 \times 10^{-24}~\mathrm{g~cm^{-3}}$ to
  $4 \times 10^{-23}~\mathrm{g~cm^{-3}}$.

  We compare the morphological characteristics of the circumstellar gas
  which result from our runs with different resolution after 1 Myr in
  Figure \ref{ion_uchii_apj230.092.002.2.med.mono.eps} (high resolution) and
  Figure \ref{ion_uchii_apj229.404.001.2.med.mono.eps} (medium resolution).
  As can be expected the overall structure in both runs is the same.
  The radius of the shell of swept-up ambient gas around the \HII region is
  $\approx$17 pc and it expands at almost $10~\mathrm{km~s^{-1}}$.
  The geometrical thickness of the \HII region itself is slightly larger in
  the high-resolution run, the velocity field in the \HII region is stronger
  perturbed than in the medium-resolution run, and some fragments of the
  photoionized gas protrude and get mixed into the hot gas. All the latter
  deviations can be understood in terms of the finer substructures in the
  \HII region, which were able to form in the high-resolution run.
  However, the trend toward a rehomogenization of the \HII region continues:
  The density fluctuations in the \HII region are only in the range of
  $9 \times 10^{-24}~\mathrm{g~cm^{-3}}$ to
  $2 \times 10^{-23}~\mathrm{g~cm^{-3}}$, except for very small regions at
  the inner side of the shell of swept-up ambient gas. Here the density is
  reduced after neutral clumps from the shell have been evaporated
  ``explosively''. Subsequently for $t > 1~\mathrm{Myr}$ we consider only
  the medium-resolution model.

\begin{figure}
  \plotone{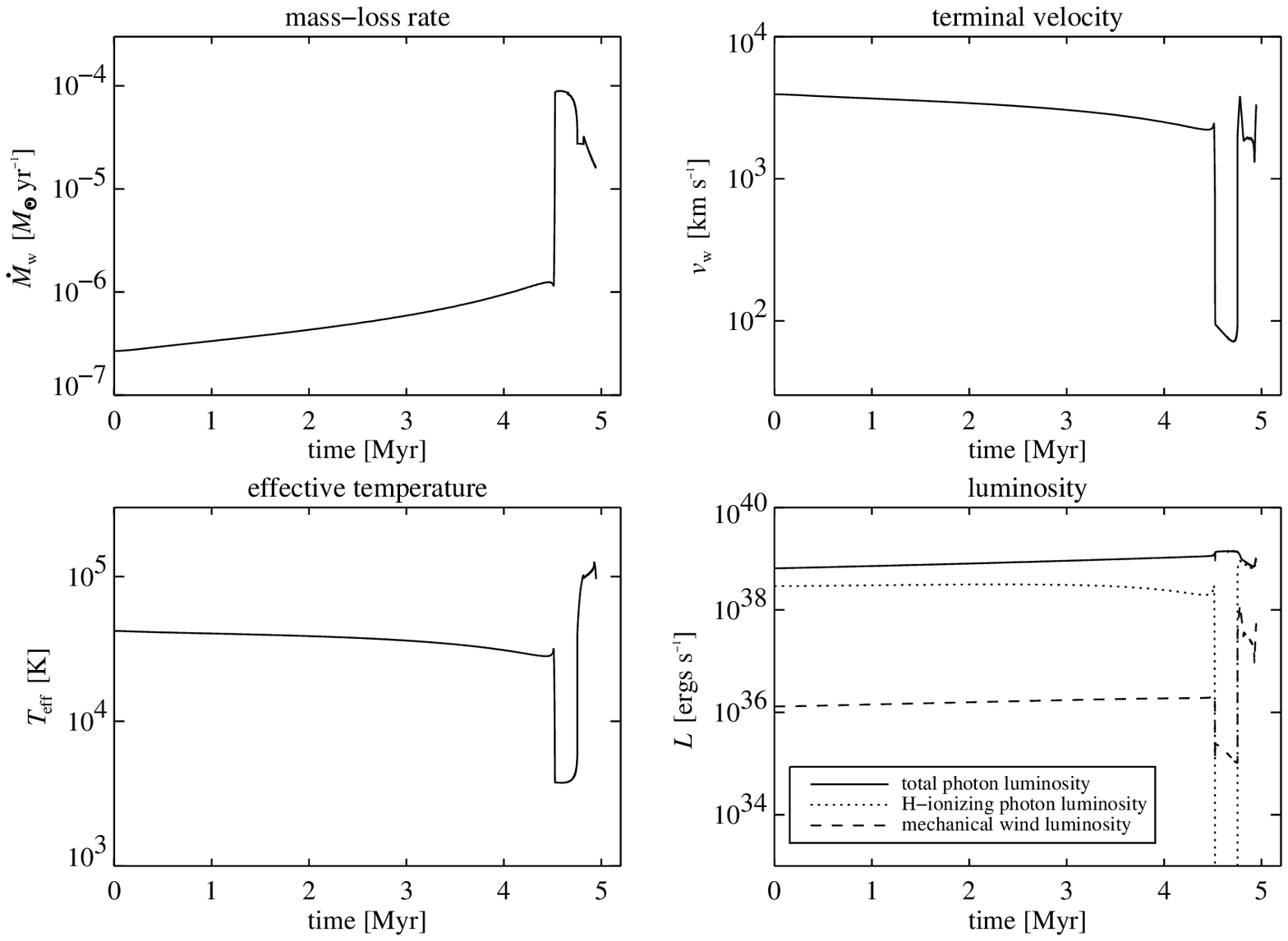}
  \caption{Time-dependent stellar parameters used as boundary conditions for
           the calculation of the 35 $\Msun$ case: mass-loss rate
           ({\it{top left}}), terminal velocity of the wind ({\it{top right}}),
           effective temperature ({\it{bottom left}}), and luminosity
           ({\it{bottom right}}). All parameters are adopted from
           \citet{garcia96b}.
           \label{35Msun_input4apj_up.eps}
          }
\end{figure}

\begin{figure}
  \plotone{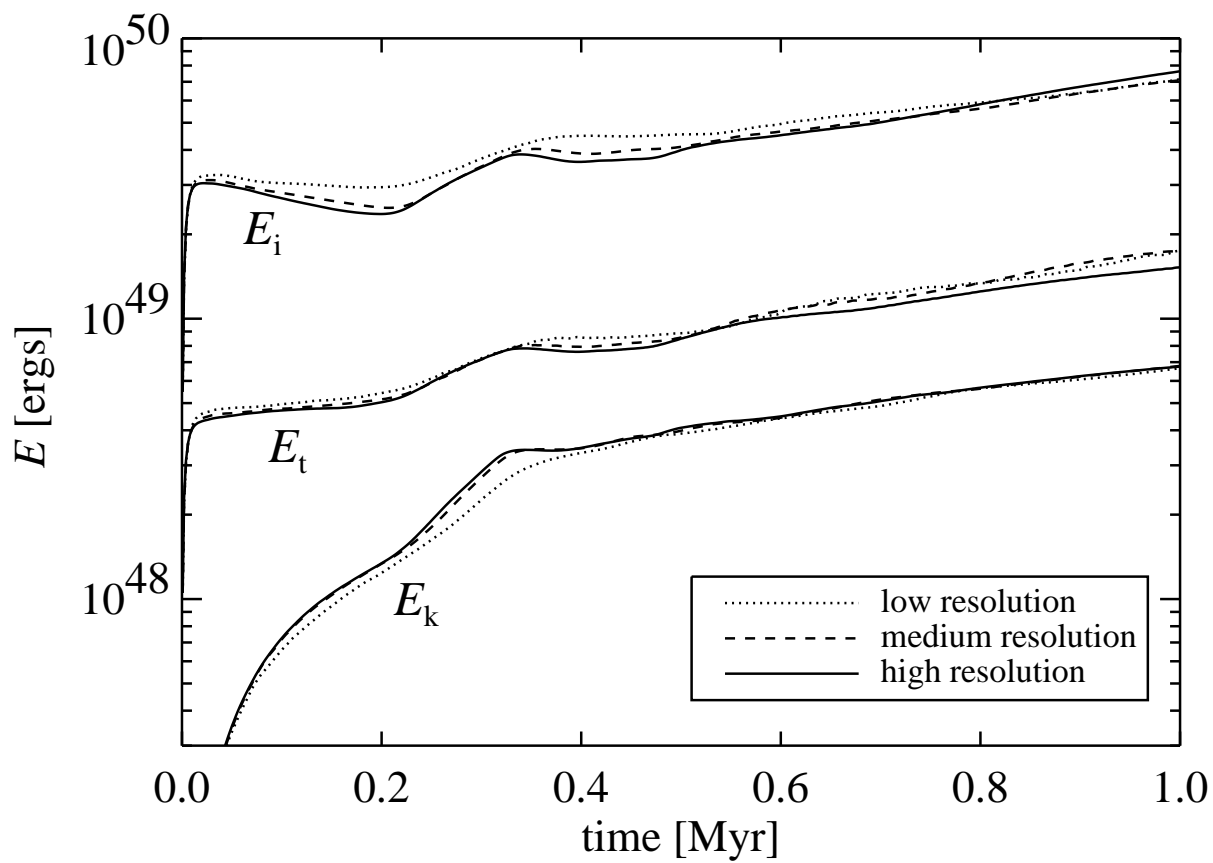}
  \caption{Resolution study for the 35 $\Msun$ case. $E_k$ is the total
           kinetic energy of bulk motion in the system, $E_t$ the thermal
           energy, and $E_i$ the ionization energy (13.6 eV per ionized
           hydrogen atom). 
           \label{comp_resol_228_229_230apj_up.eps}
          }
\end{figure}

\begin{figure}
  \epsscale{0.50}
  \plotone{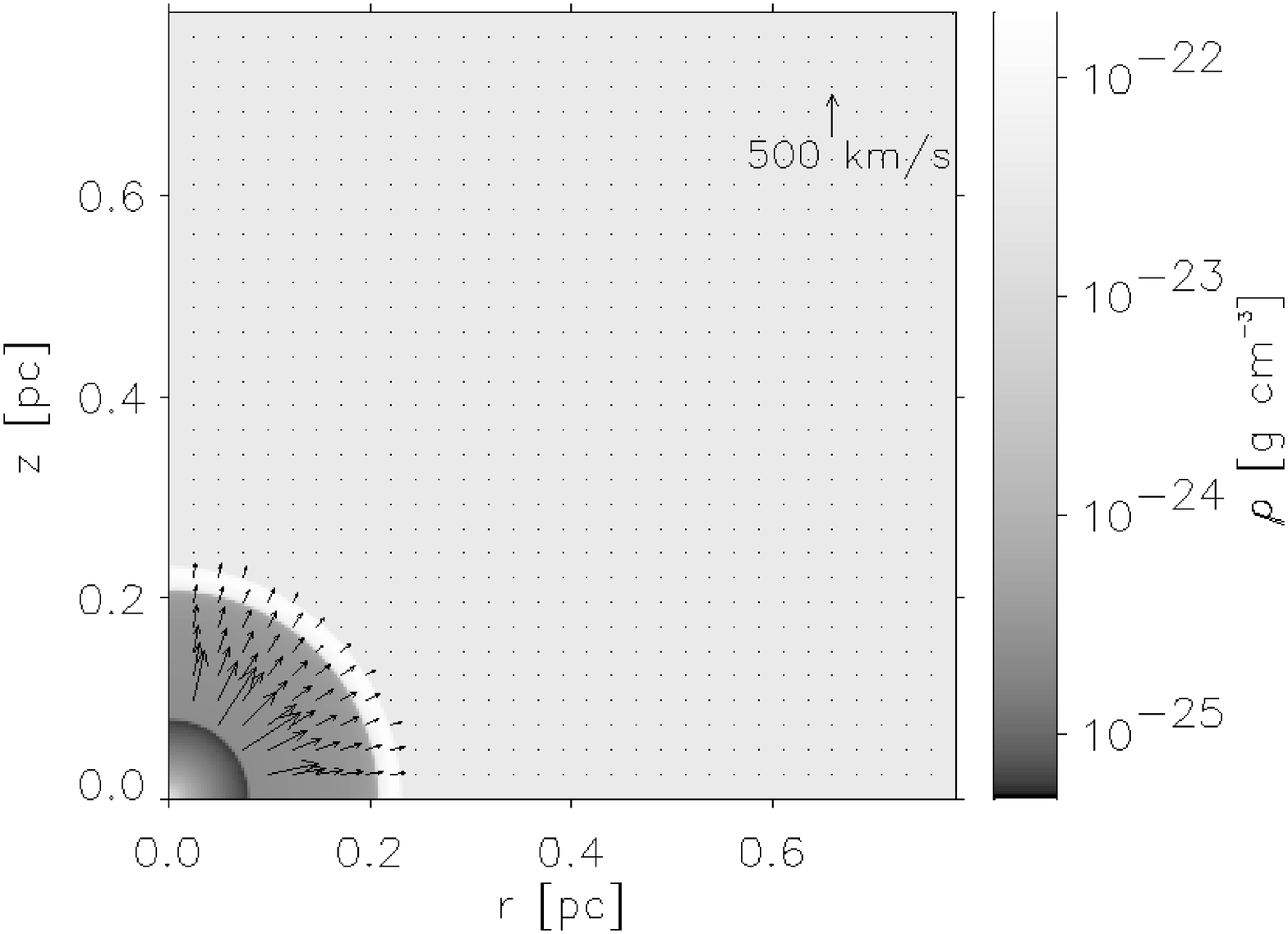}
  \plotone{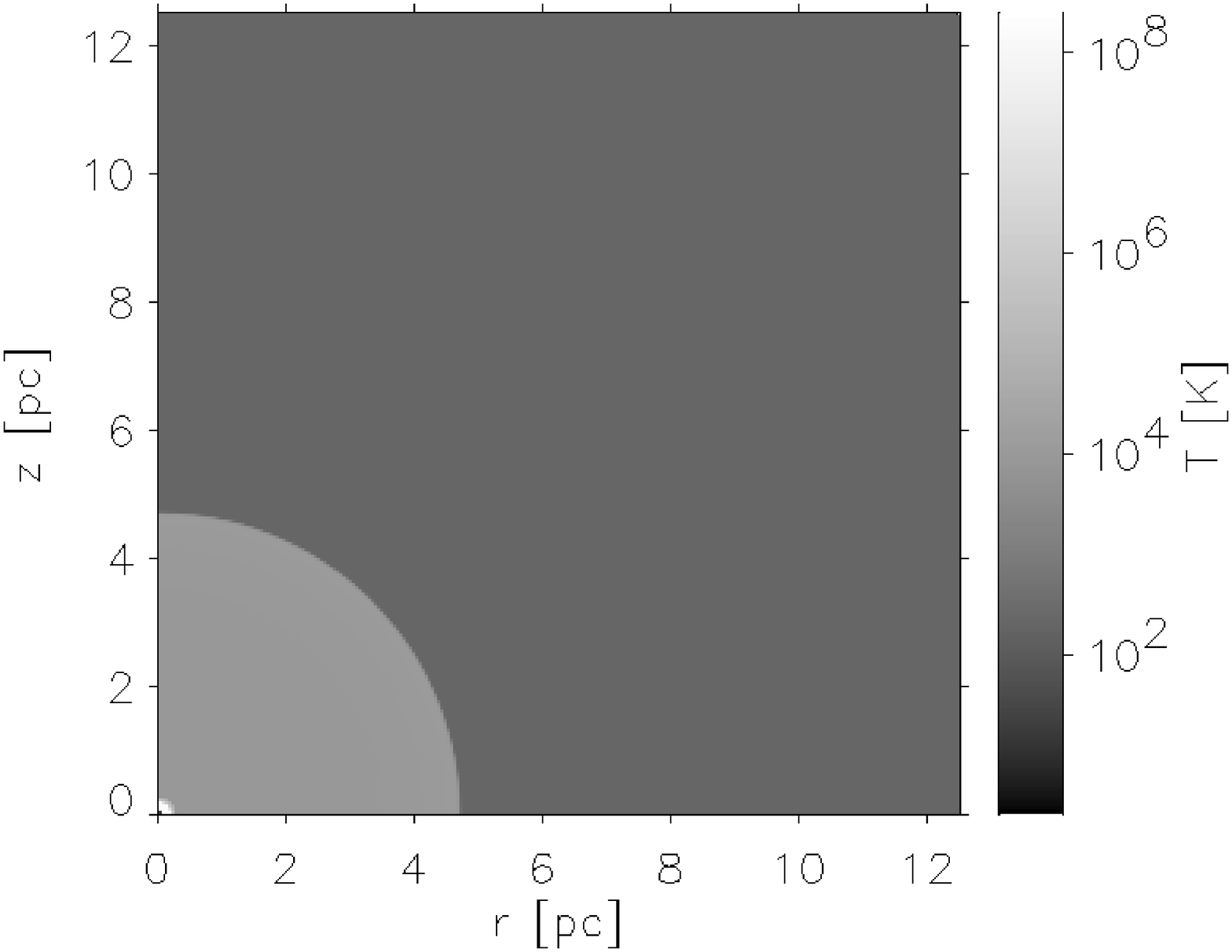}
  \plotone{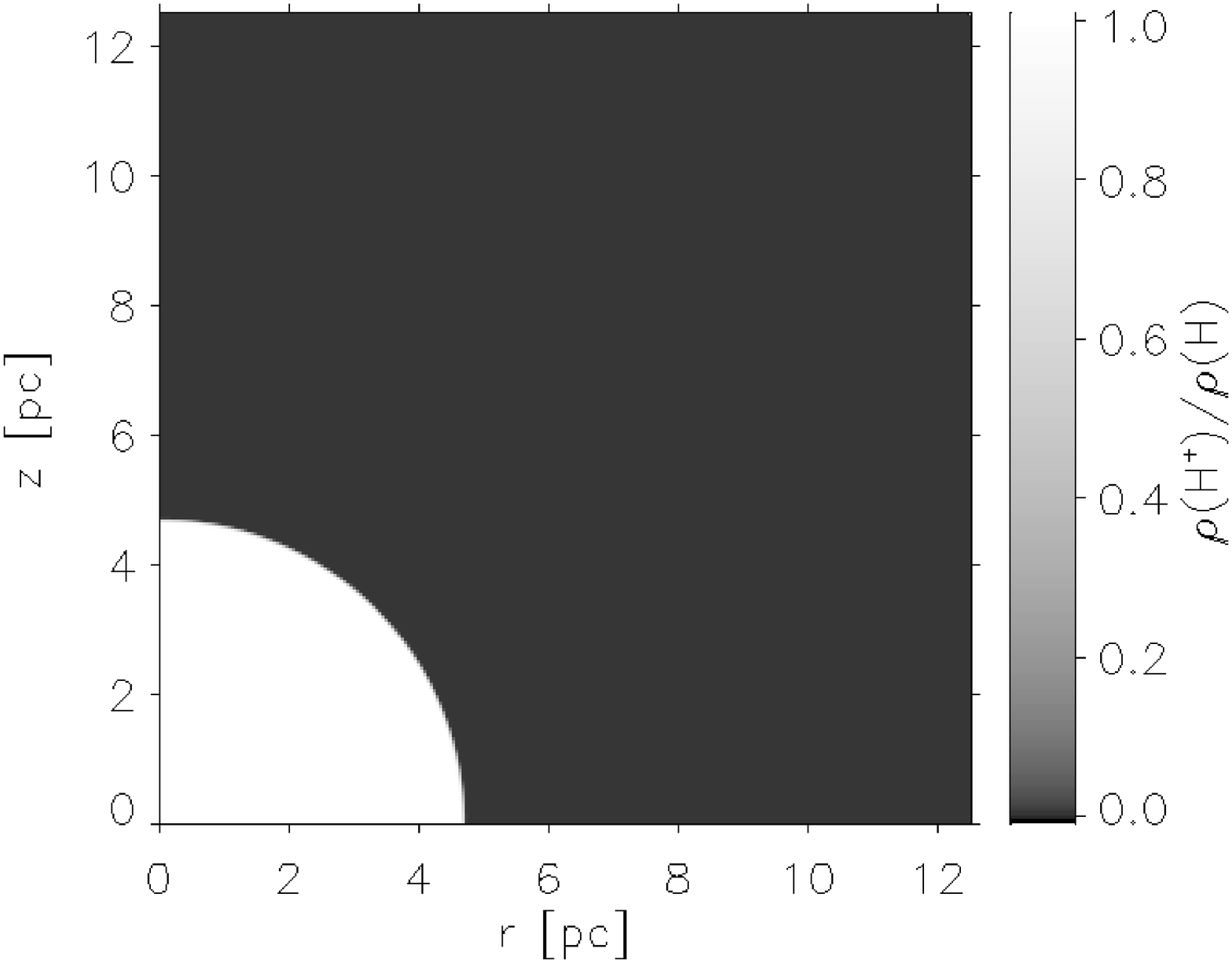}
  \caption{Circumstellar mass density and velocity field ({\it{top}}),
           temperature ({\it{middle}}), and degree of hydrogen ionization
           ({\it{bottom}}) for the 35 $\Msun$ case at evolutionary age 700 yr
           (high-resolution run). The velocity arrows in the free-flowing
           wind zone have been omitted to prevent confusion. The star is
           located in the center of the coordinate system. Note the
           different length scales.
           \label{ion_uchii_apj230.001.001.3.med.mono.eps}
          }
\end{figure}

\begin{figure}
  \epsscale{0.50}
  \plotone{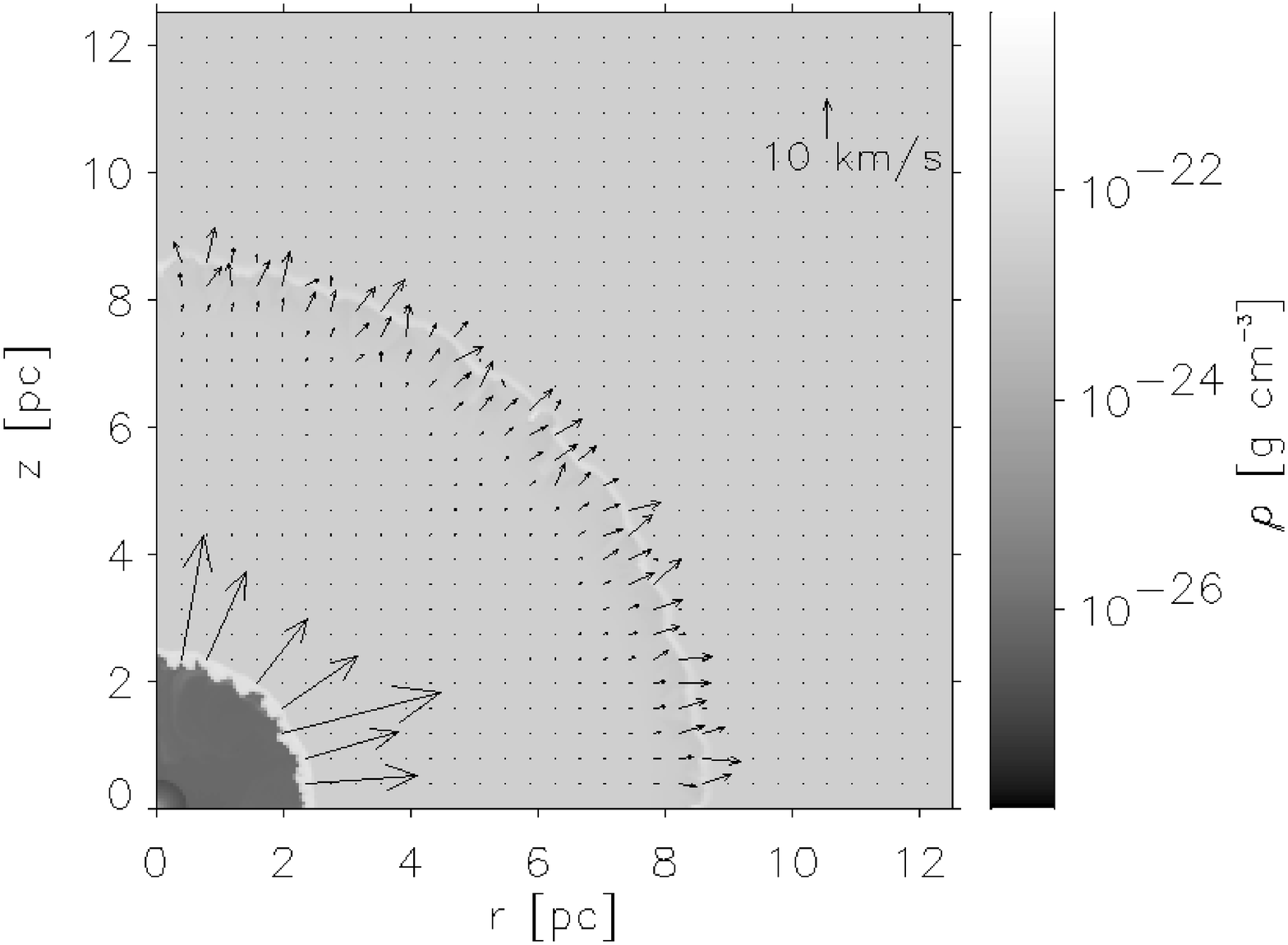}
  \plotone{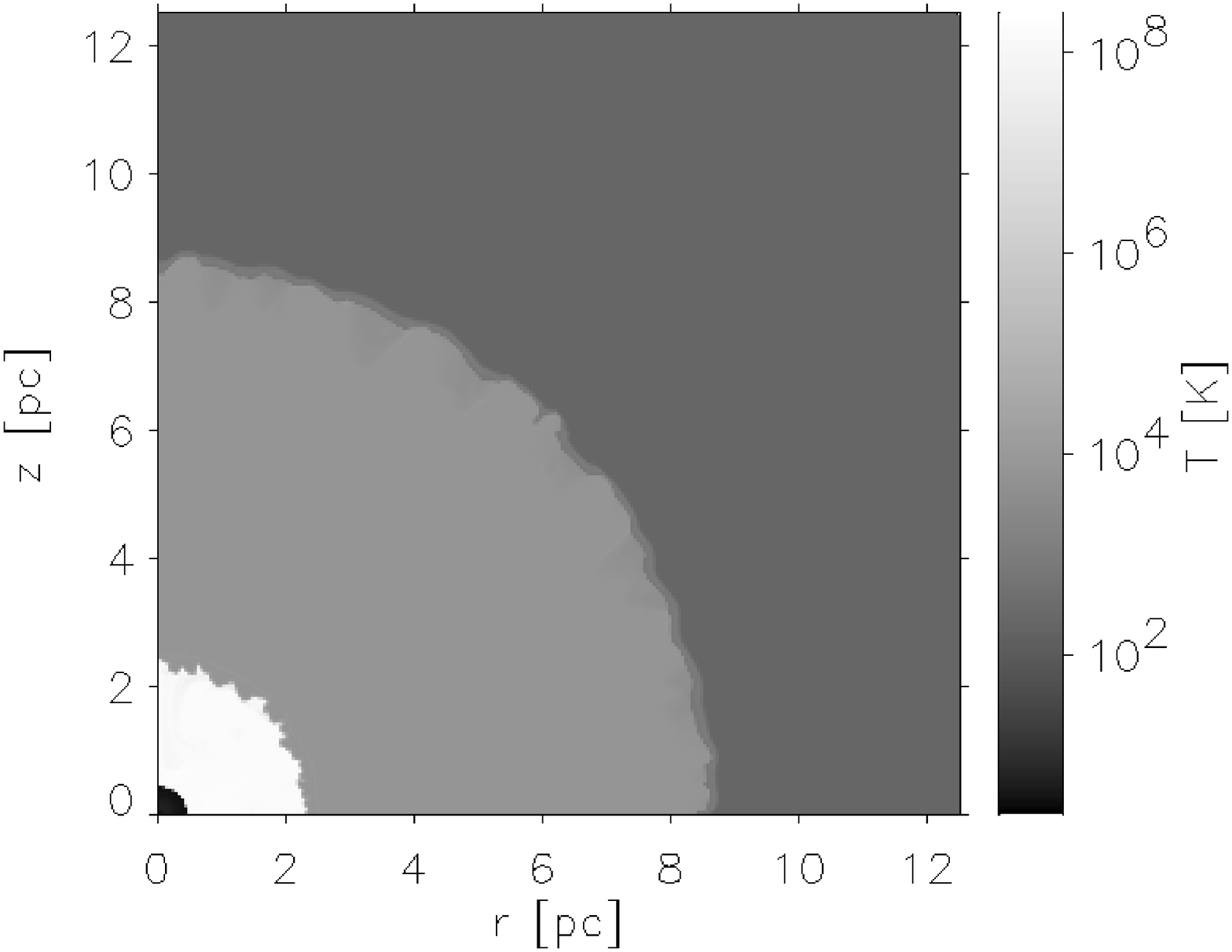}
  \plotone{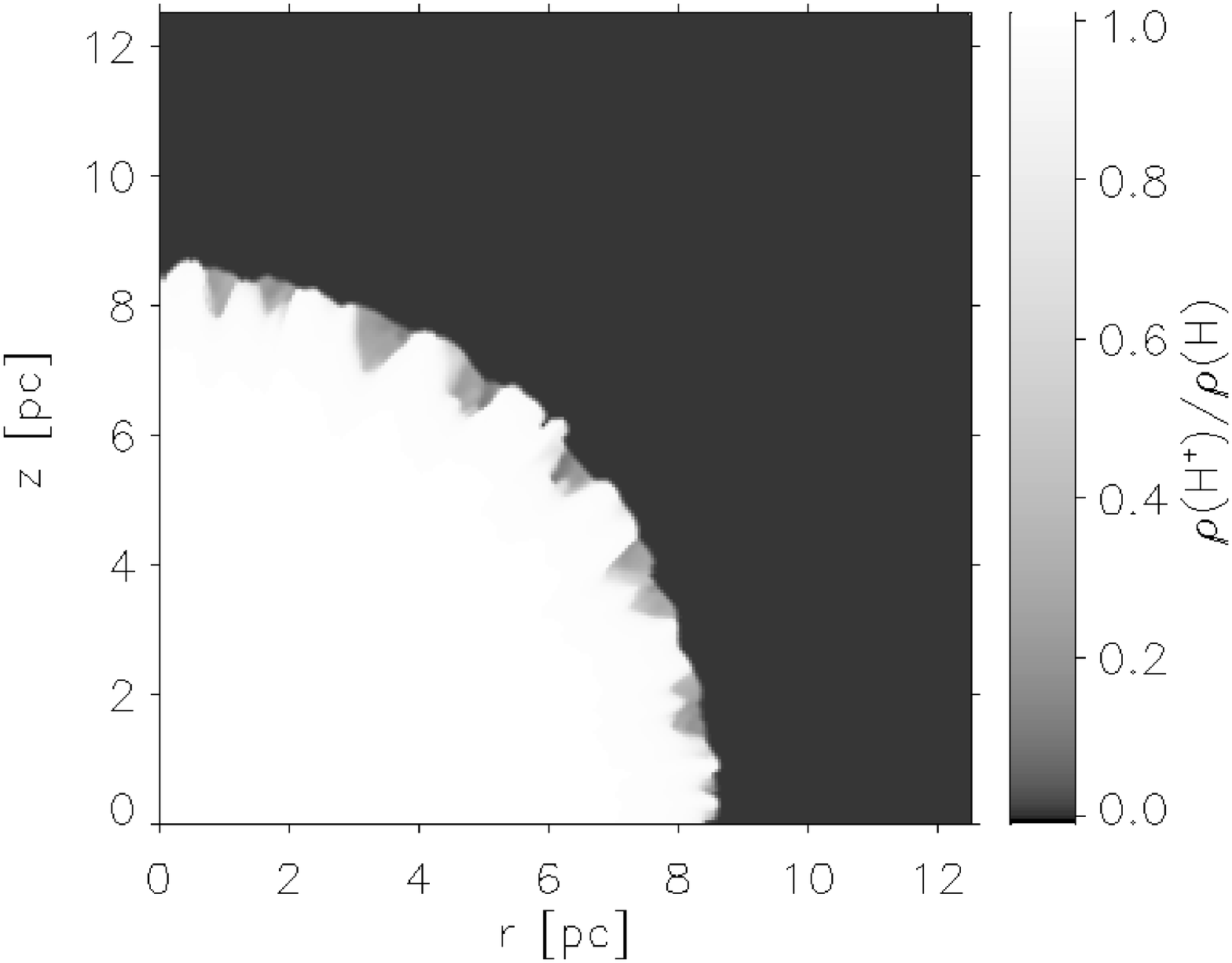}
  \caption{Same as Fig.~\ref{ion_uchii_apj230.001.001.3.med.mono.eps}, but at
           age $5 \times 10^4~\mathrm{yr}$. The velocity arrows in the
           free-flowing wind zone and in the hot bubble have been omitted
           to prevent confusion.
           \label{ion_uchii_apj230.017.001.3.med.mono.eps}
          }
\end{figure}

\begin{figure}
  \epsscale{0.50}
  \plotone{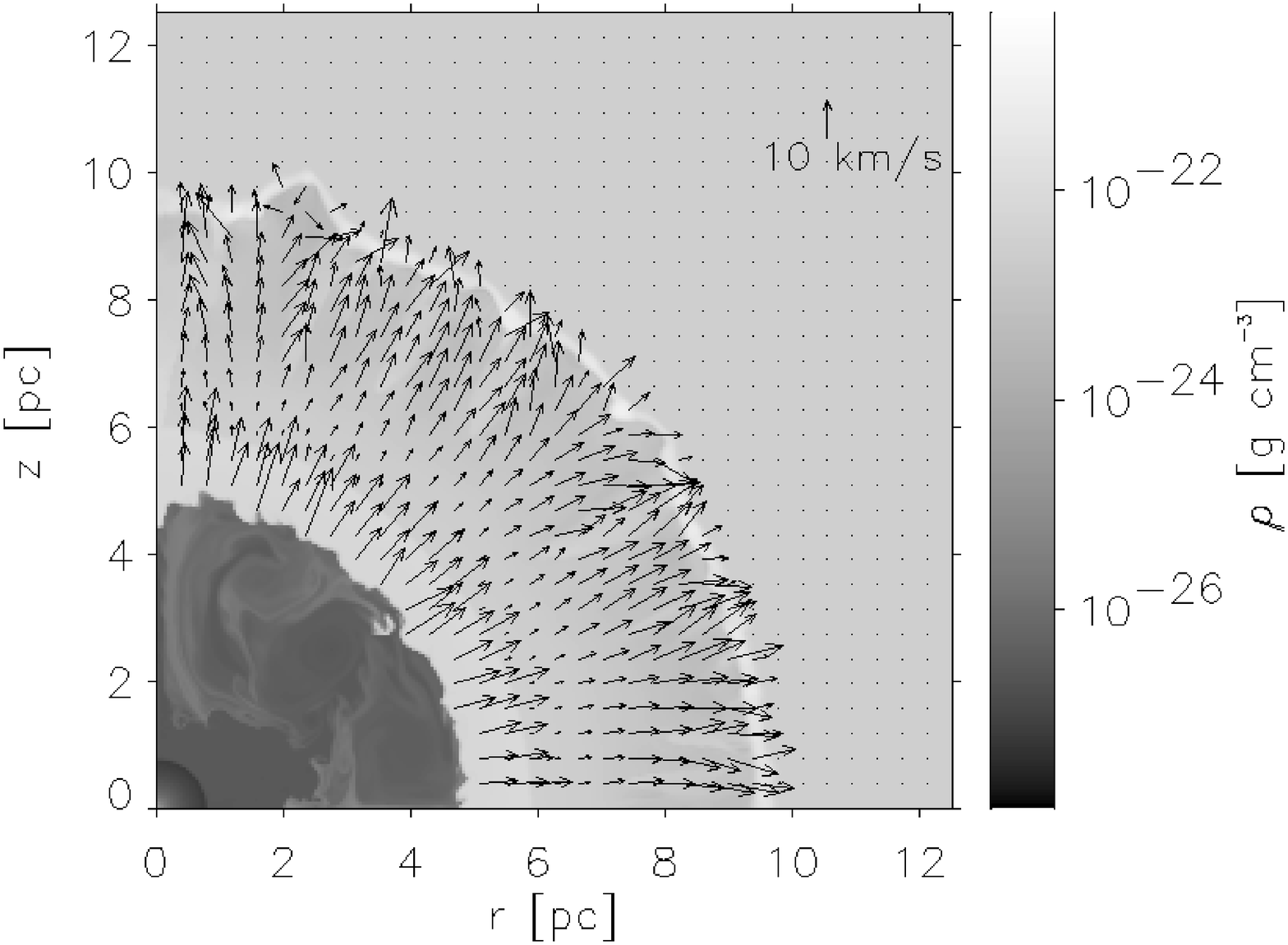}
  \plotone{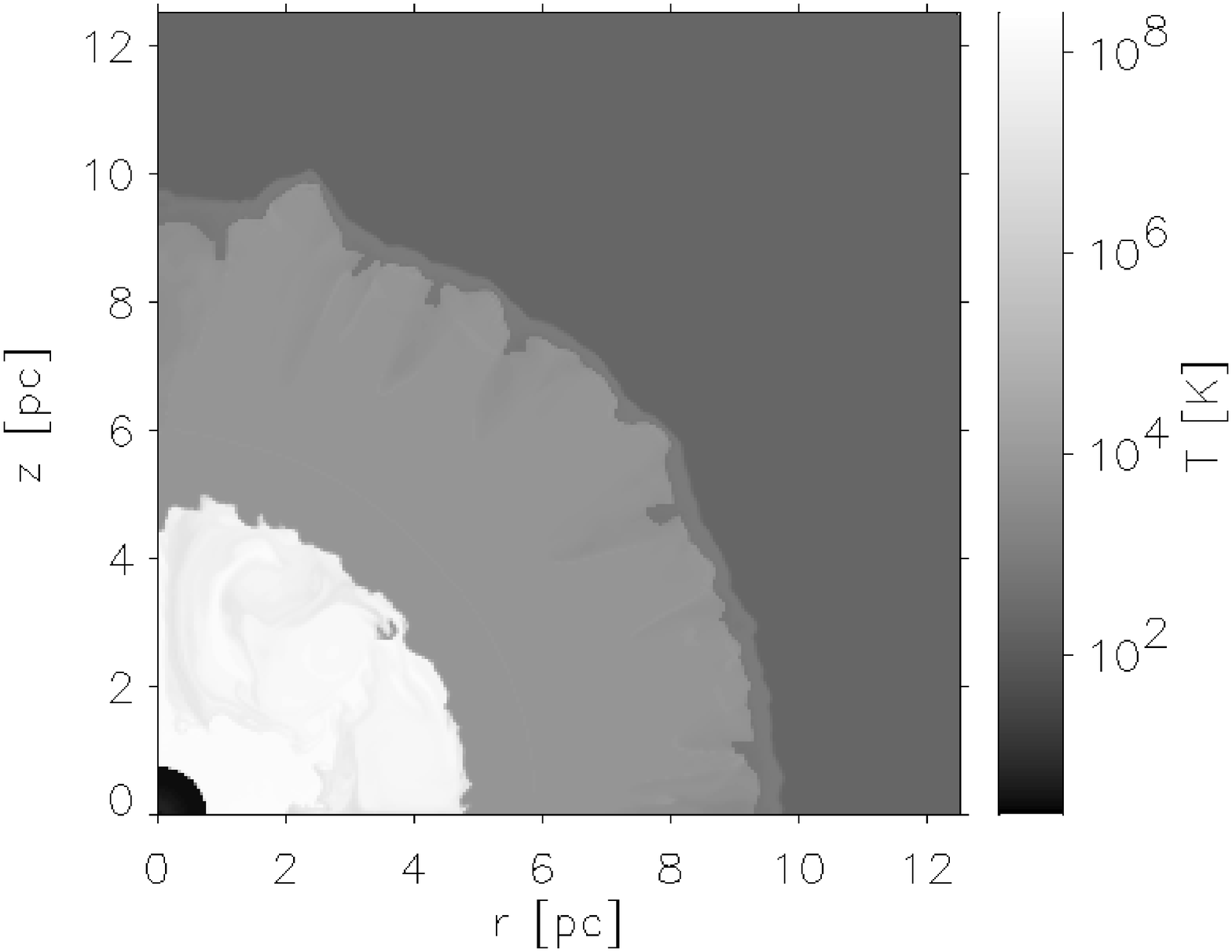}
  \plotone{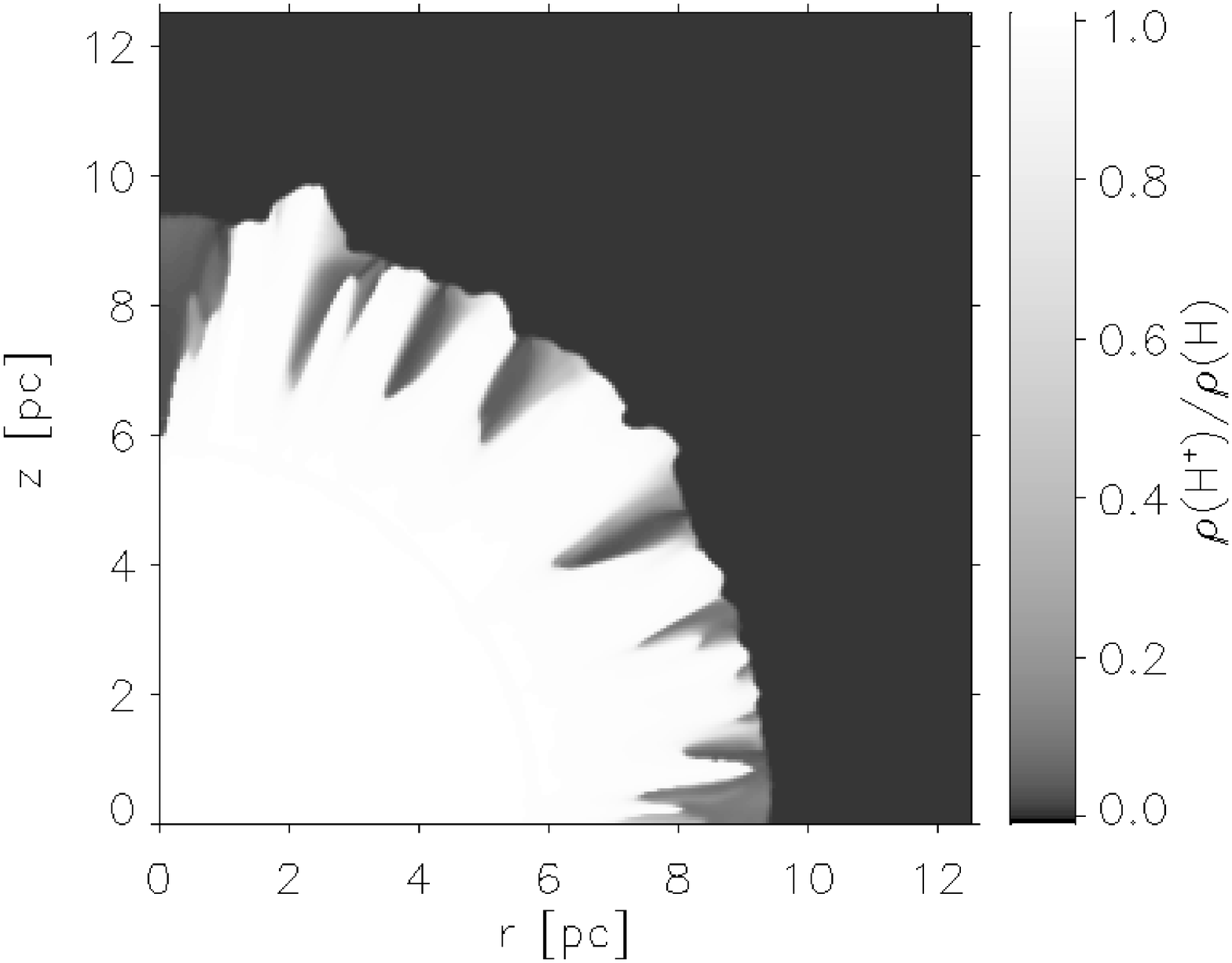}
  \caption{Same as Fig.~\ref{ion_uchii_apj230.017.001.3.med.mono.eps},
           but at age 0.2 Myr
           \label{ion_uchii_apj230.030.001.3.med.mono.eps}
          }
\end{figure}

\begin{figure}
  \epsscale{0.50}
  \plotone{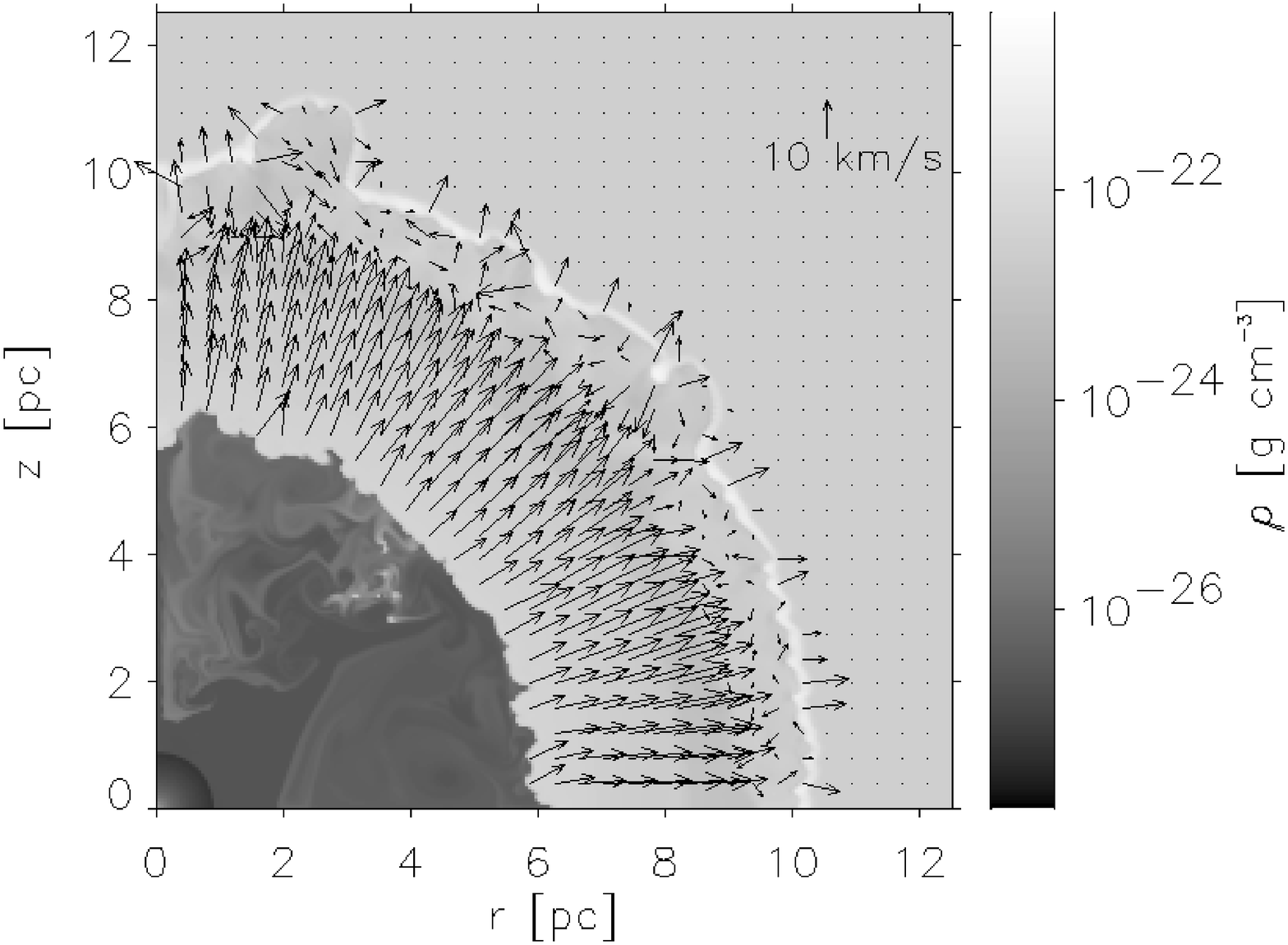}
  \plotone{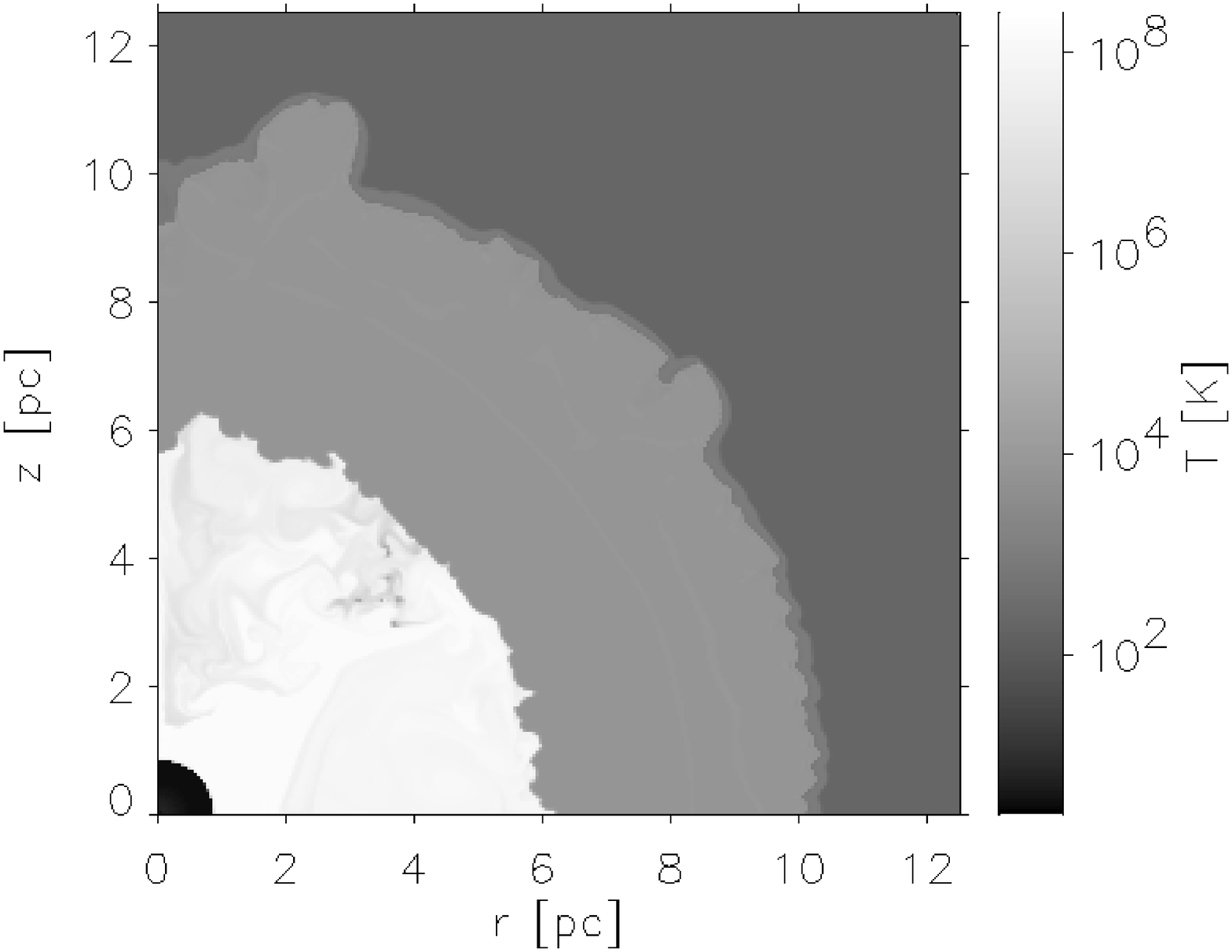}
  \plotone{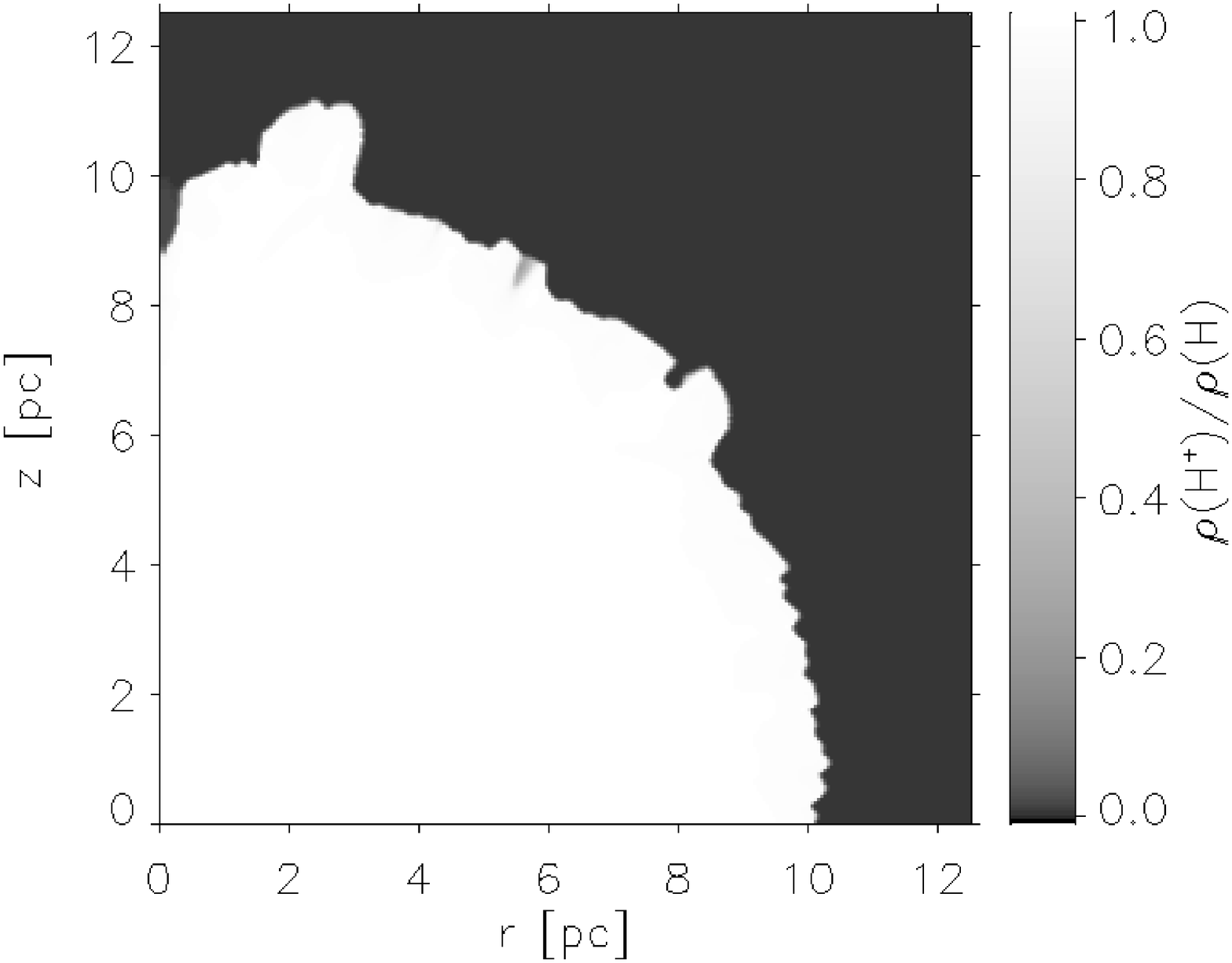}
  \caption{Same as Fig.~\ref{ion_uchii_apj230.017.001.3.med.mono.eps},
           but at age 0.3 Myr
           \label{ion_uchii_apj230.041.001.3.med.mono.eps}
          }
\end{figure}

\begin{figure}
  \epsscale{0.50}
  \plotone{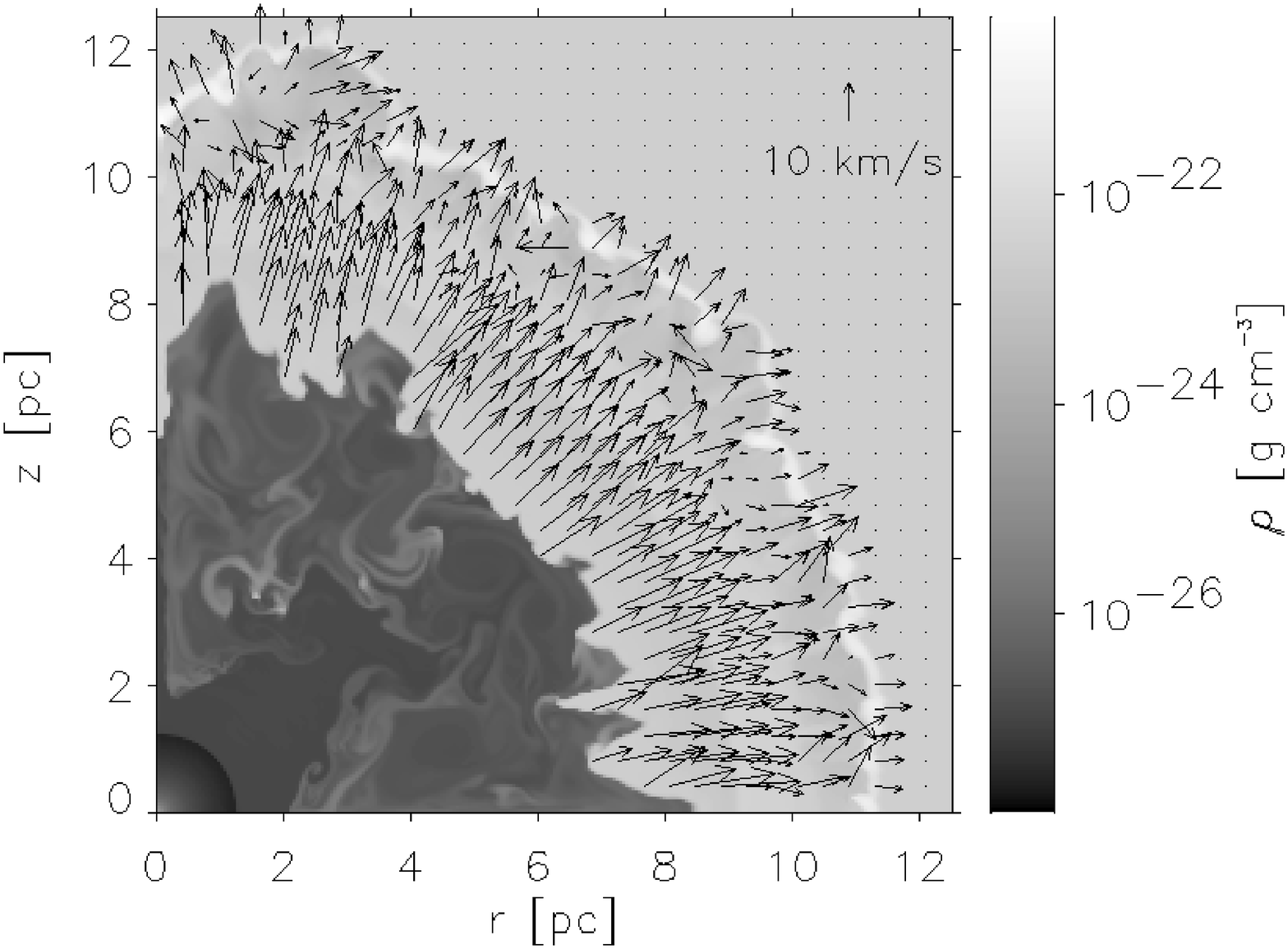}
  \plotone{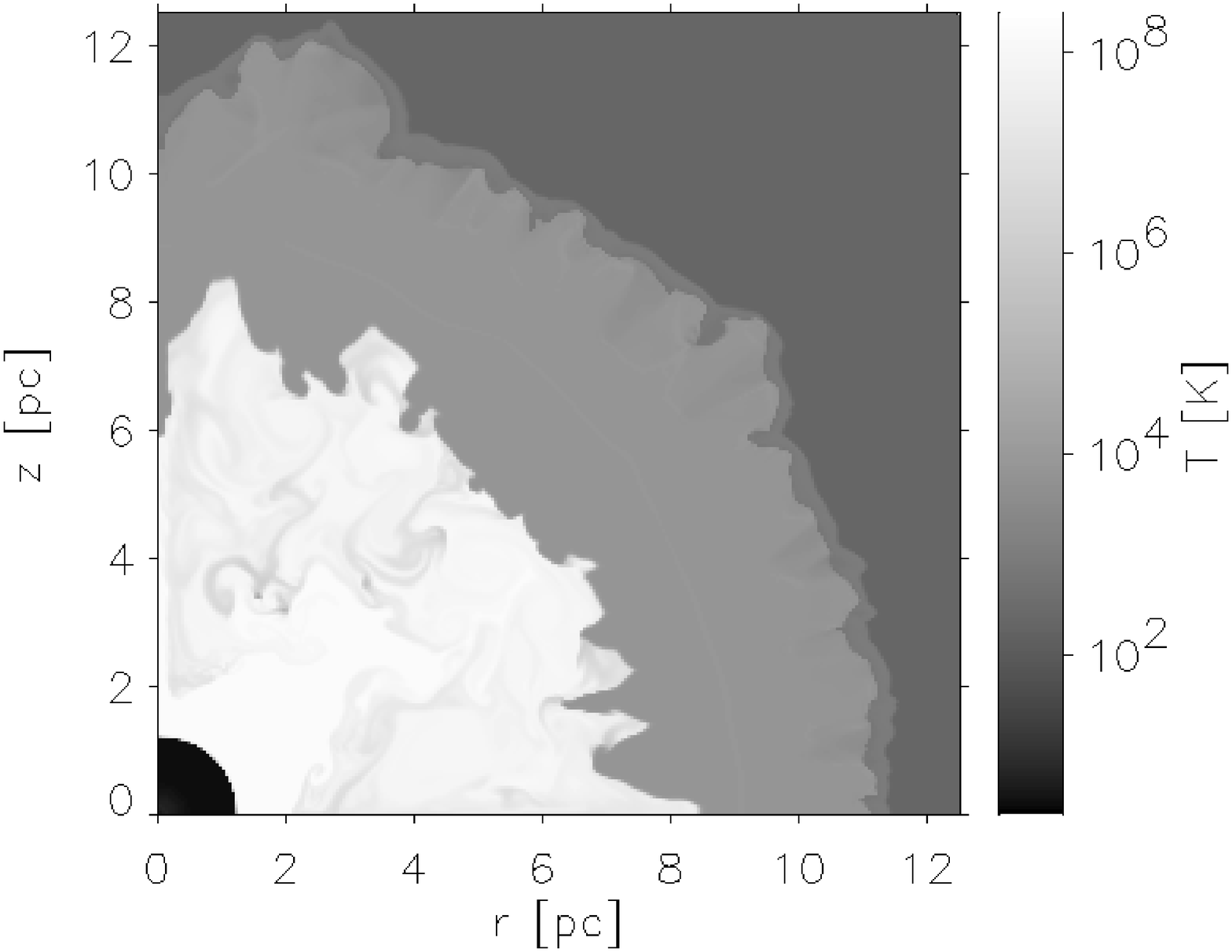}
  \plotone{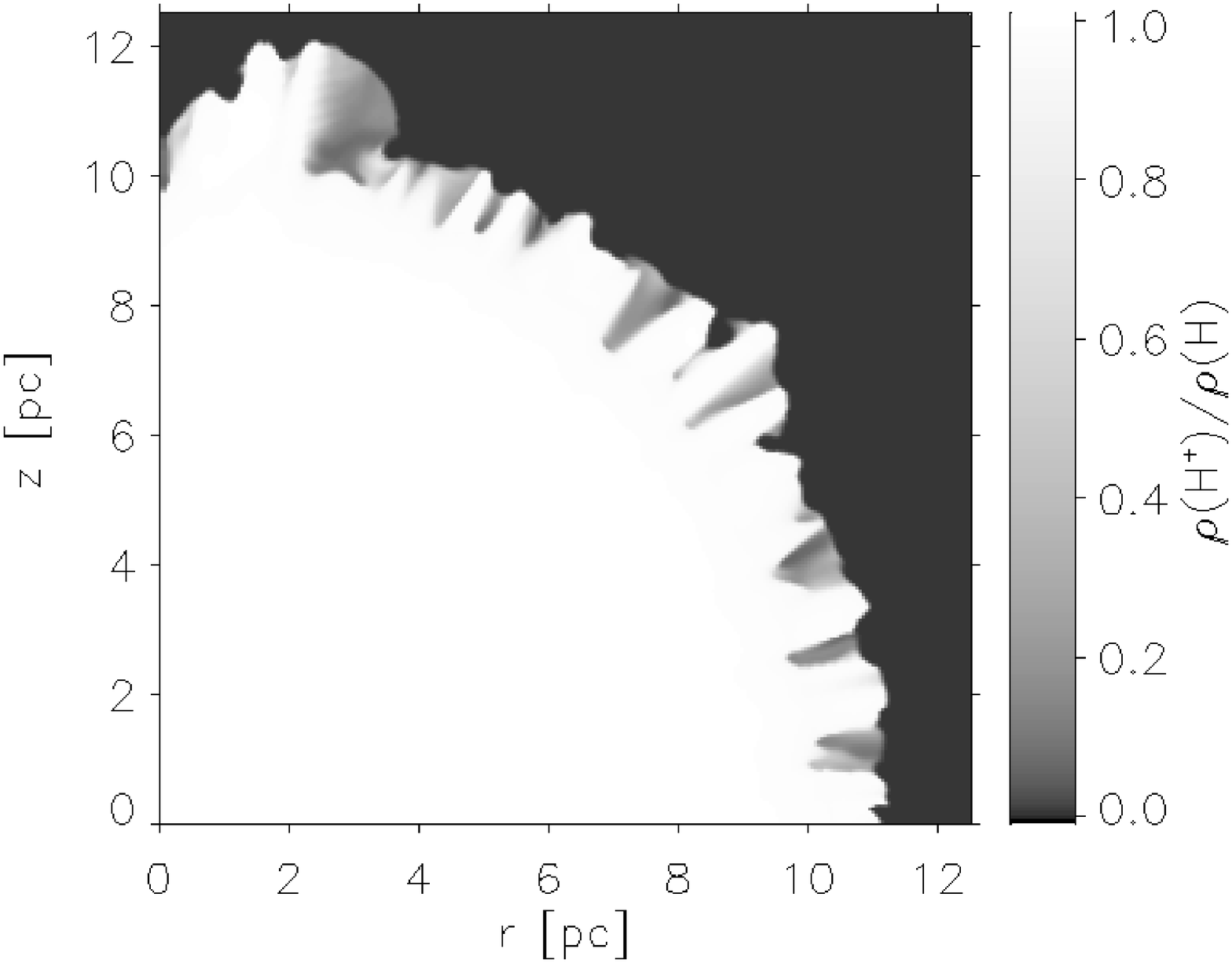}
  \caption{Same as Fig.~\ref{ion_uchii_apj230.017.001.3.med.mono.eps},
           but at age 0.4 Myr
           \label{ion_uchii_apj230.048.002.3.med.mono.eps}
          }
\end{figure}

\begin{figure}
  \epsscale{0.50}
  \plotone{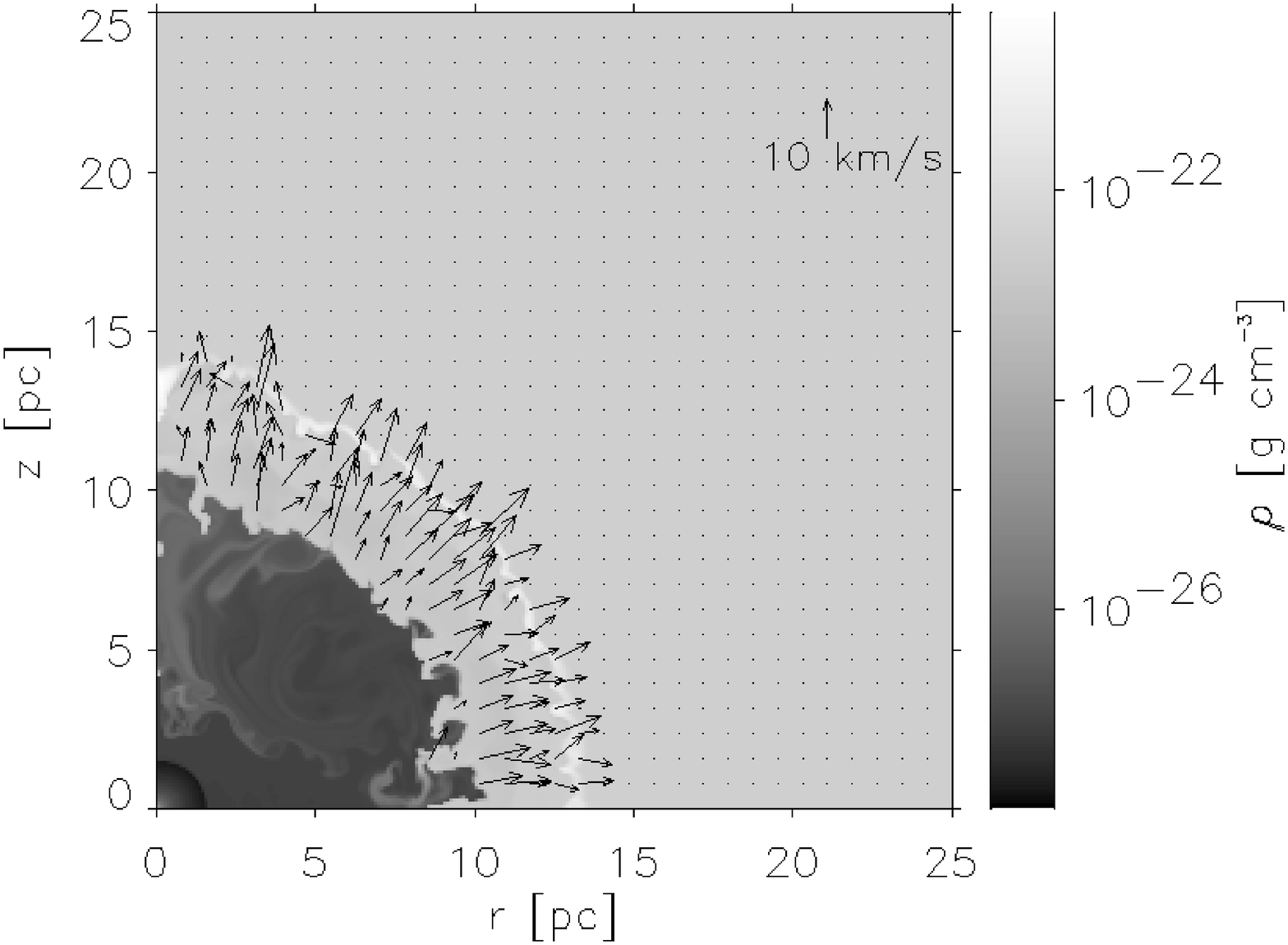}
  \plotone{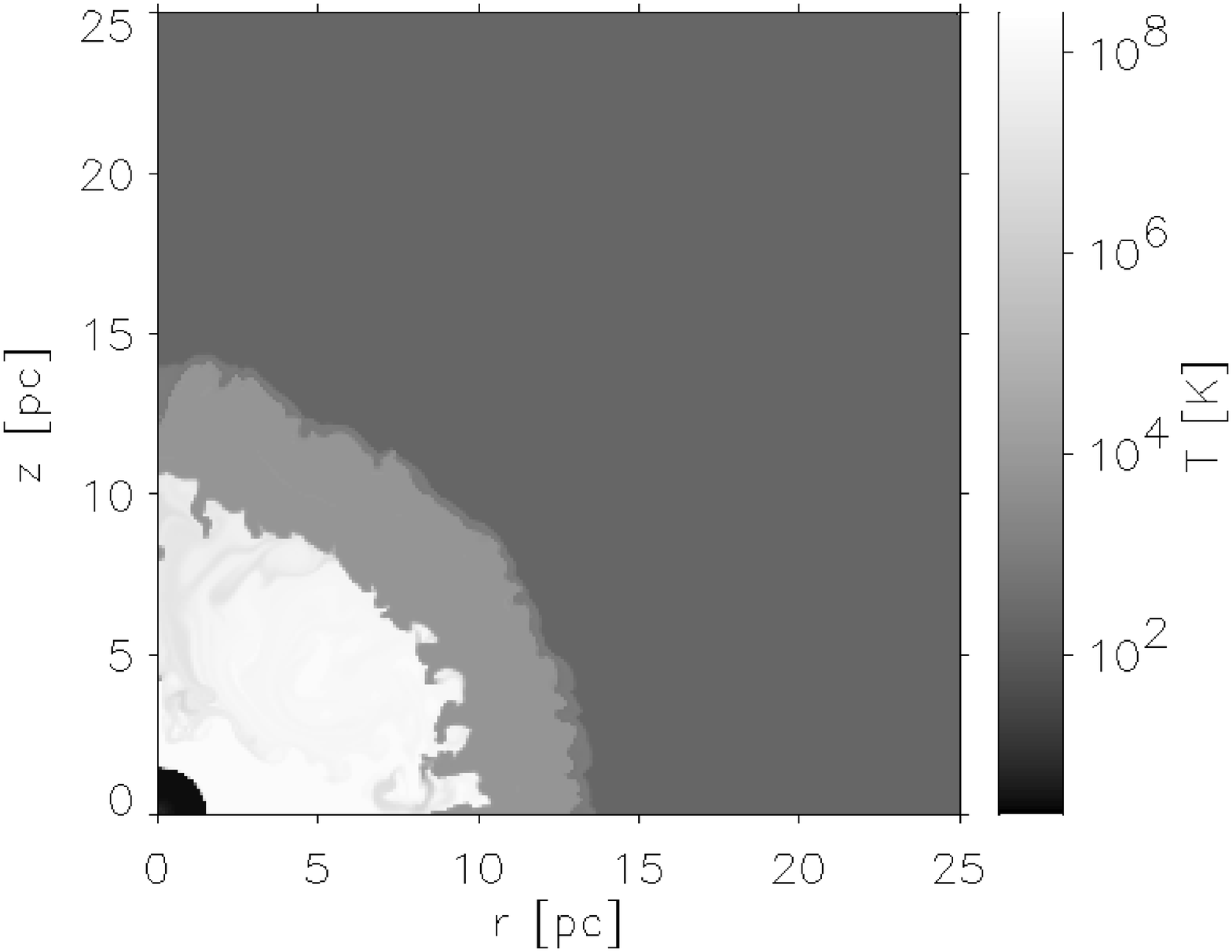}
  \plotone{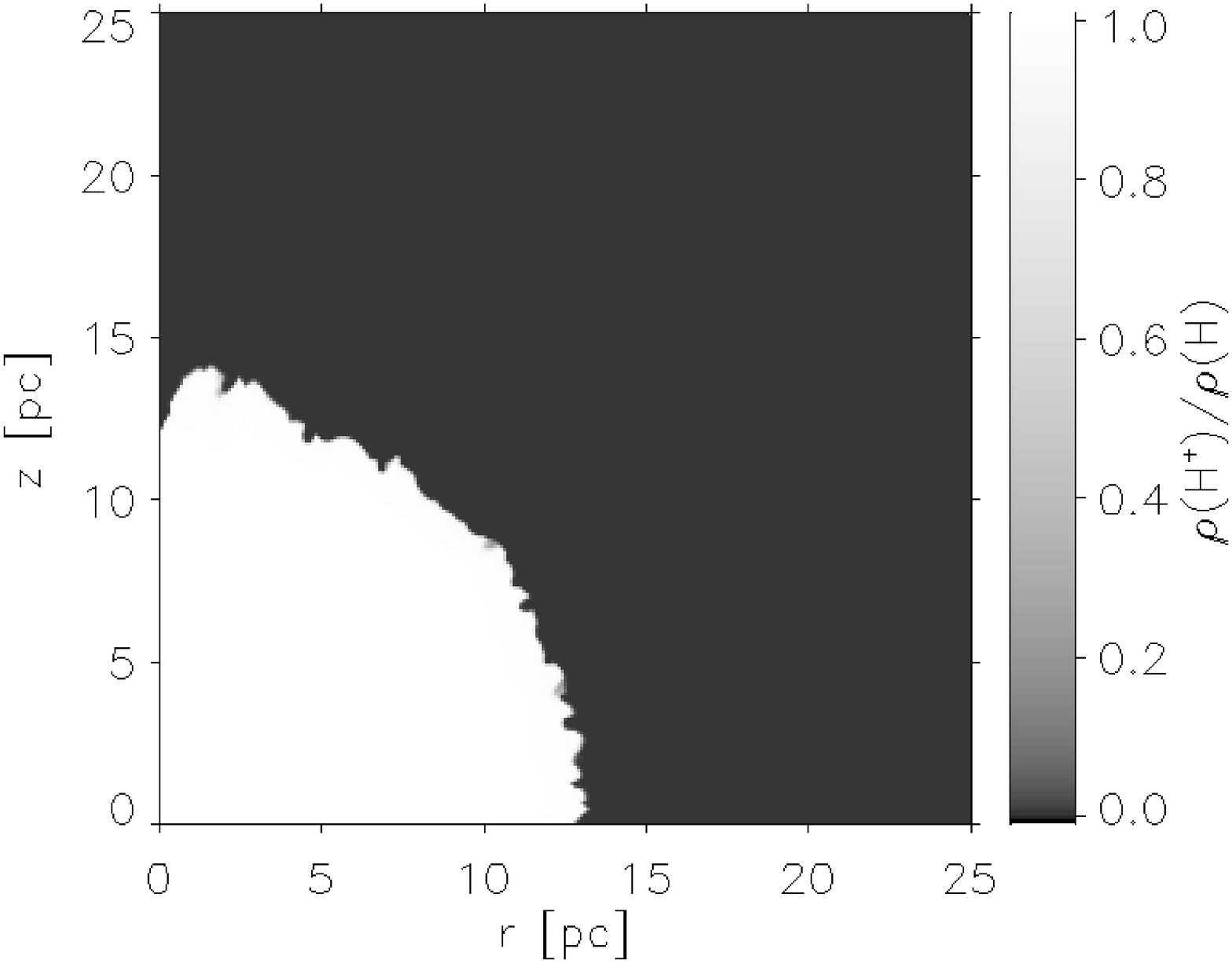}
  \caption{Same as Fig.~\ref{ion_uchii_apj230.017.001.3.med.mono.eps},
           but at age 0.6 Myr. Because of the bubble expansion,
           a larger volume is shown.
           \label{ion_uchii_apj230.063.002.2.med.mono.eps}
          }
\end{figure}

\begin{figure}
  \epsscale{0.50}
  \plotone{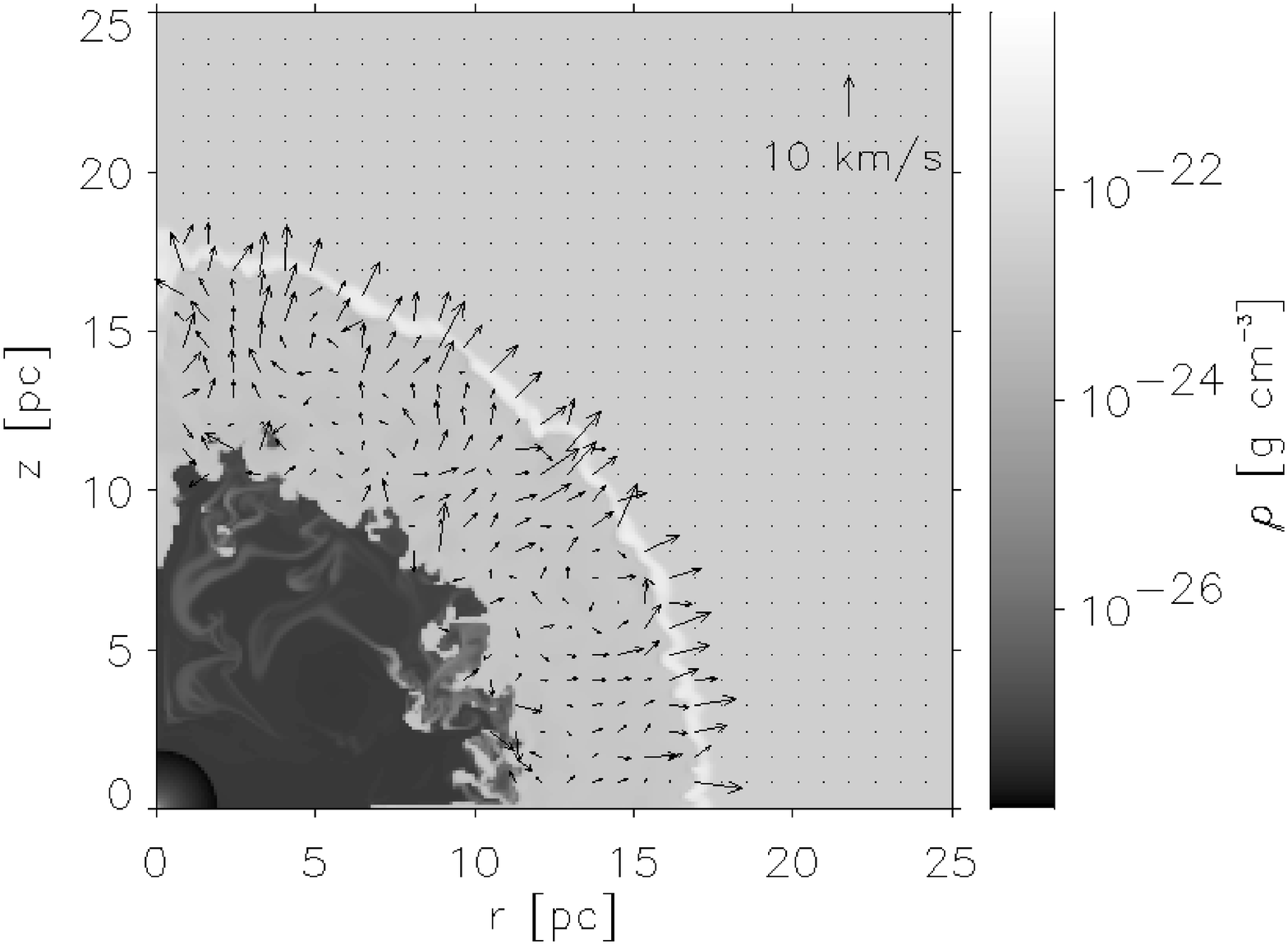}
  \plotone{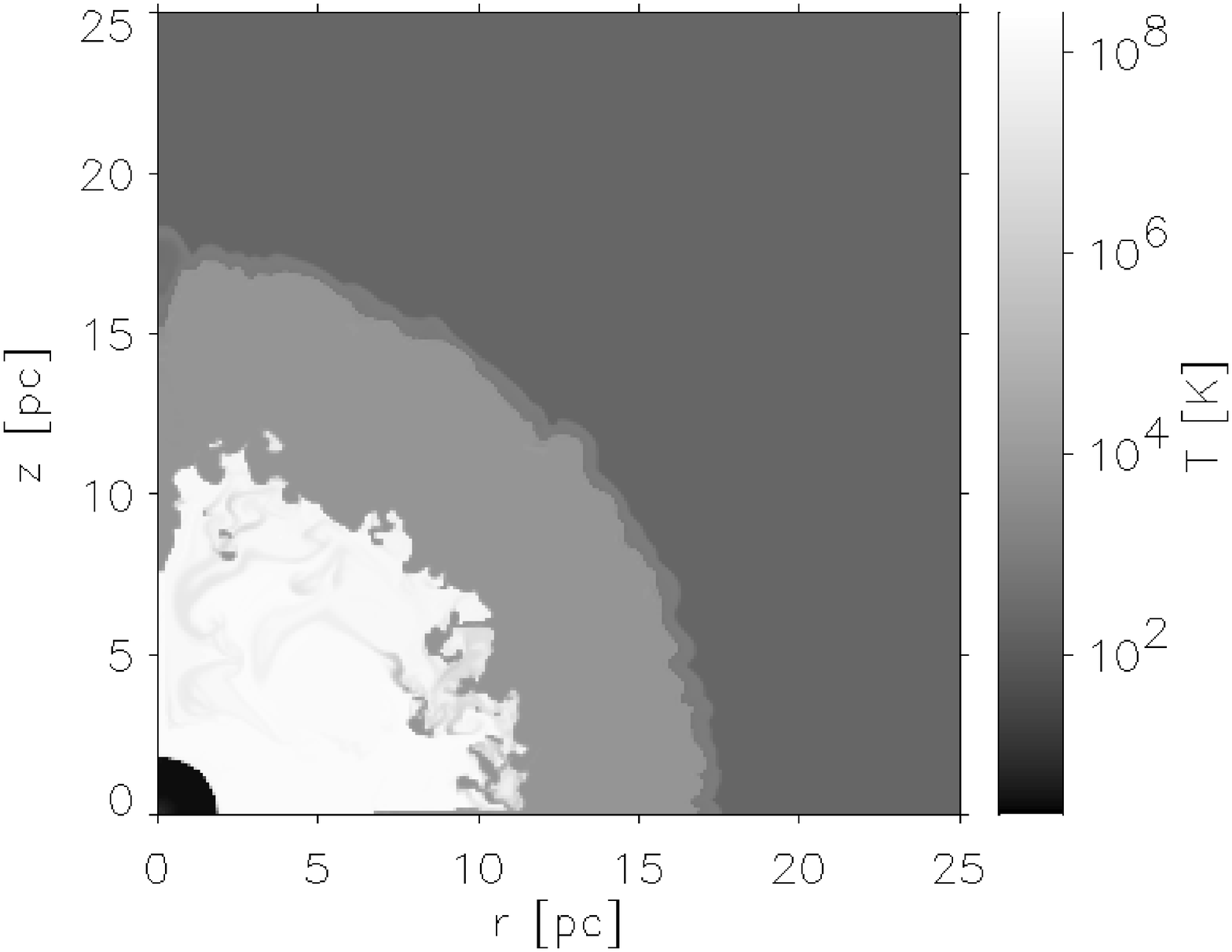}
  \plotone{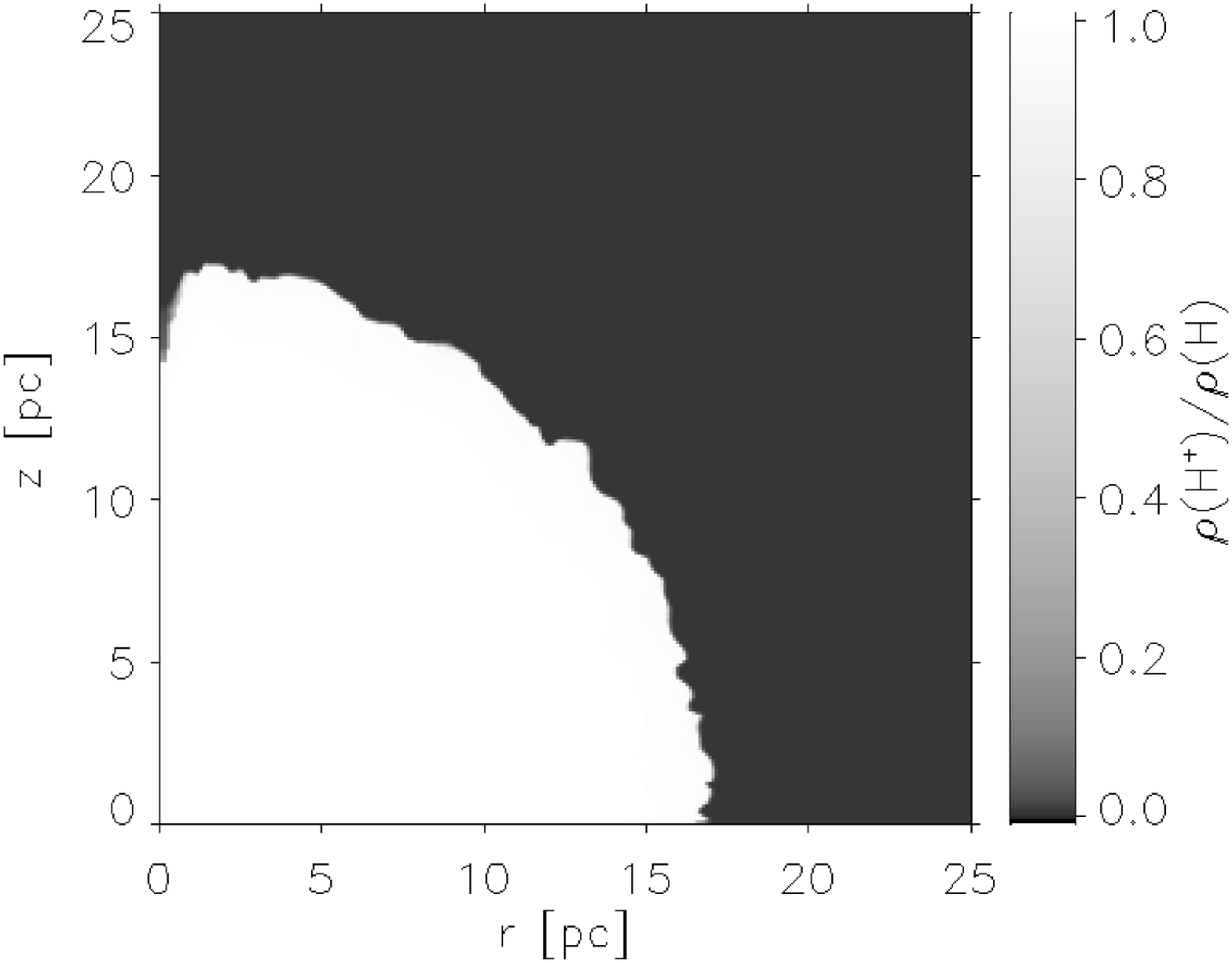}
  \caption{Same as Fig.~\ref{ion_uchii_apj230.063.002.2.med.mono.eps},
           but at age 1.0 Myr
           \label{ion_uchii_apj230.092.002.2.med.mono.eps}
          }
\end{figure}

\begin{figure}
  \epsscale{0.50}
  \plotone{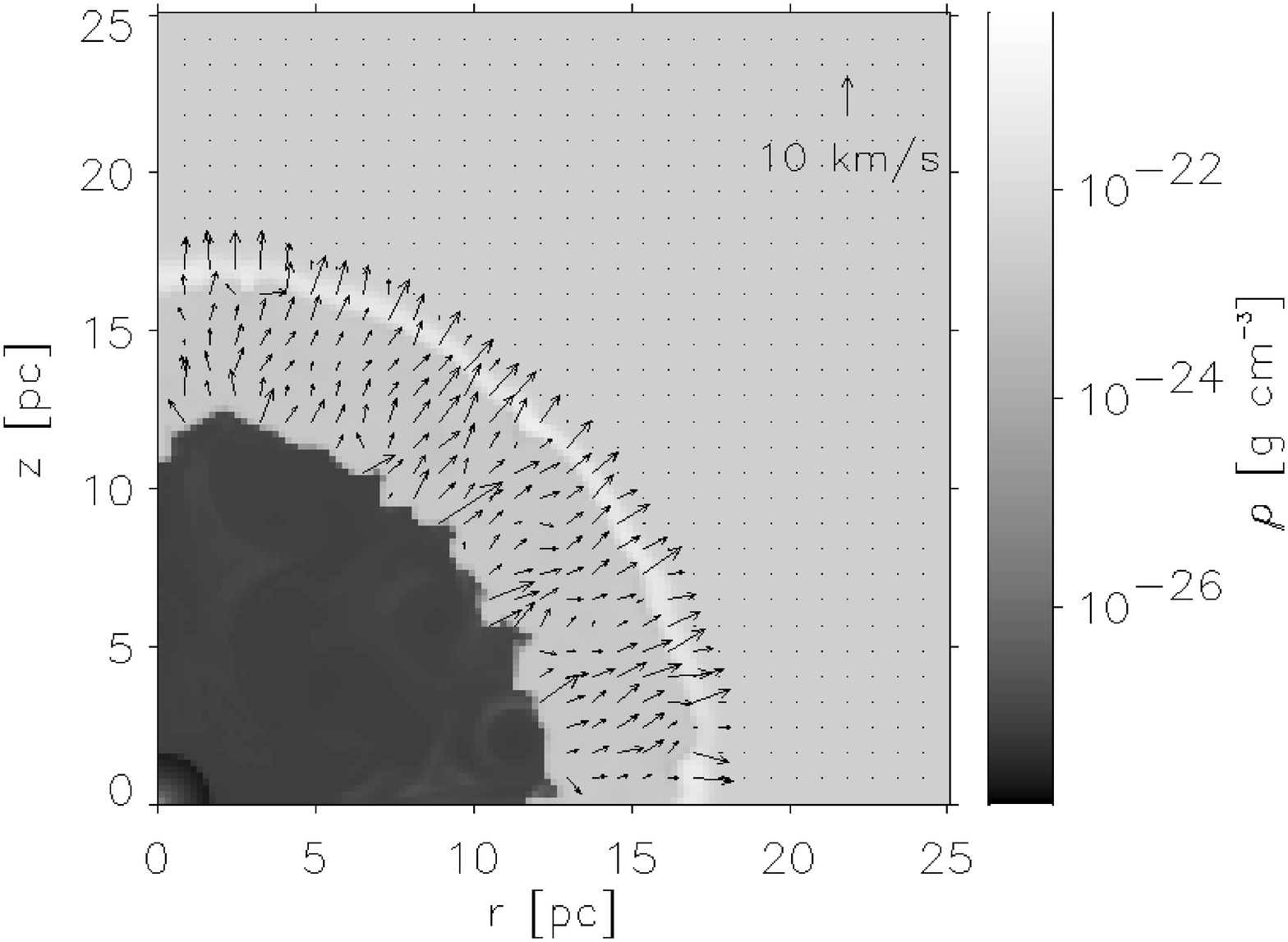}
  \plotone{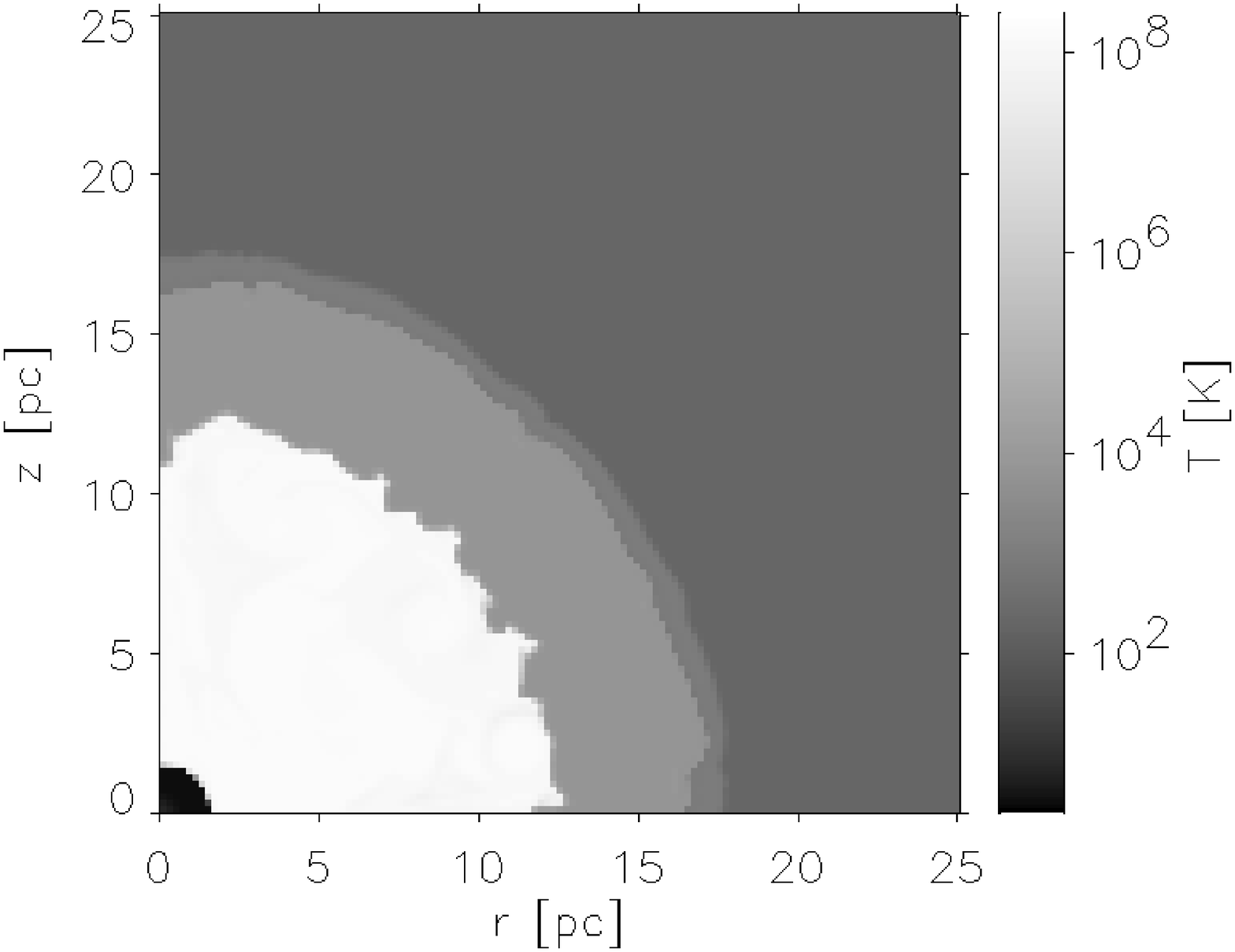}
  \plotone{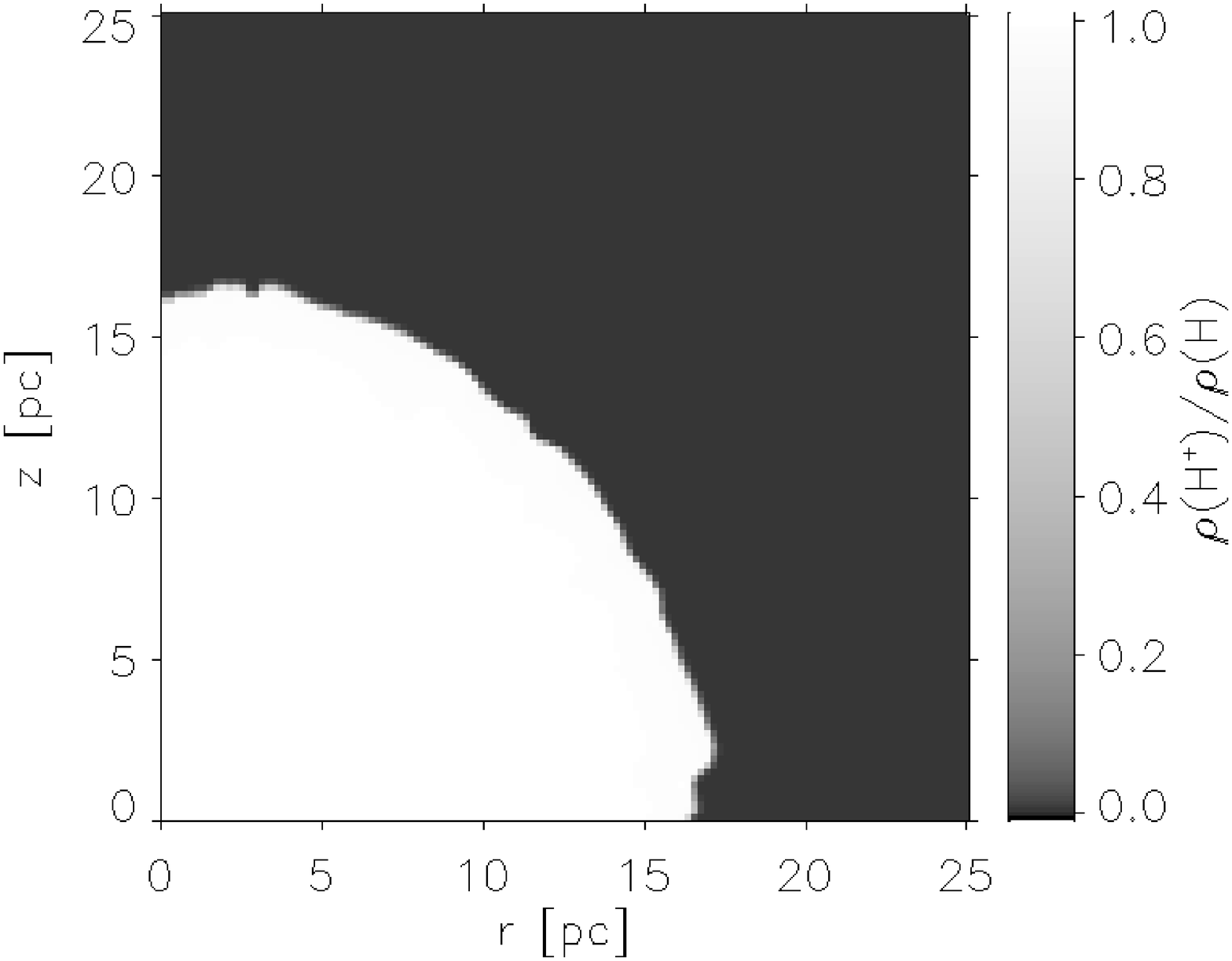}
  \caption{Same as Fig.~\ref{ion_uchii_apj230.092.002.2.med.mono.eps},
           but for the medium-resolution run
           \label{ion_uchii_apj229.404.001.2.med.mono.eps}
          }
\end{figure}

\begin{figure}
  \epsscale{0.50}
  \plotone{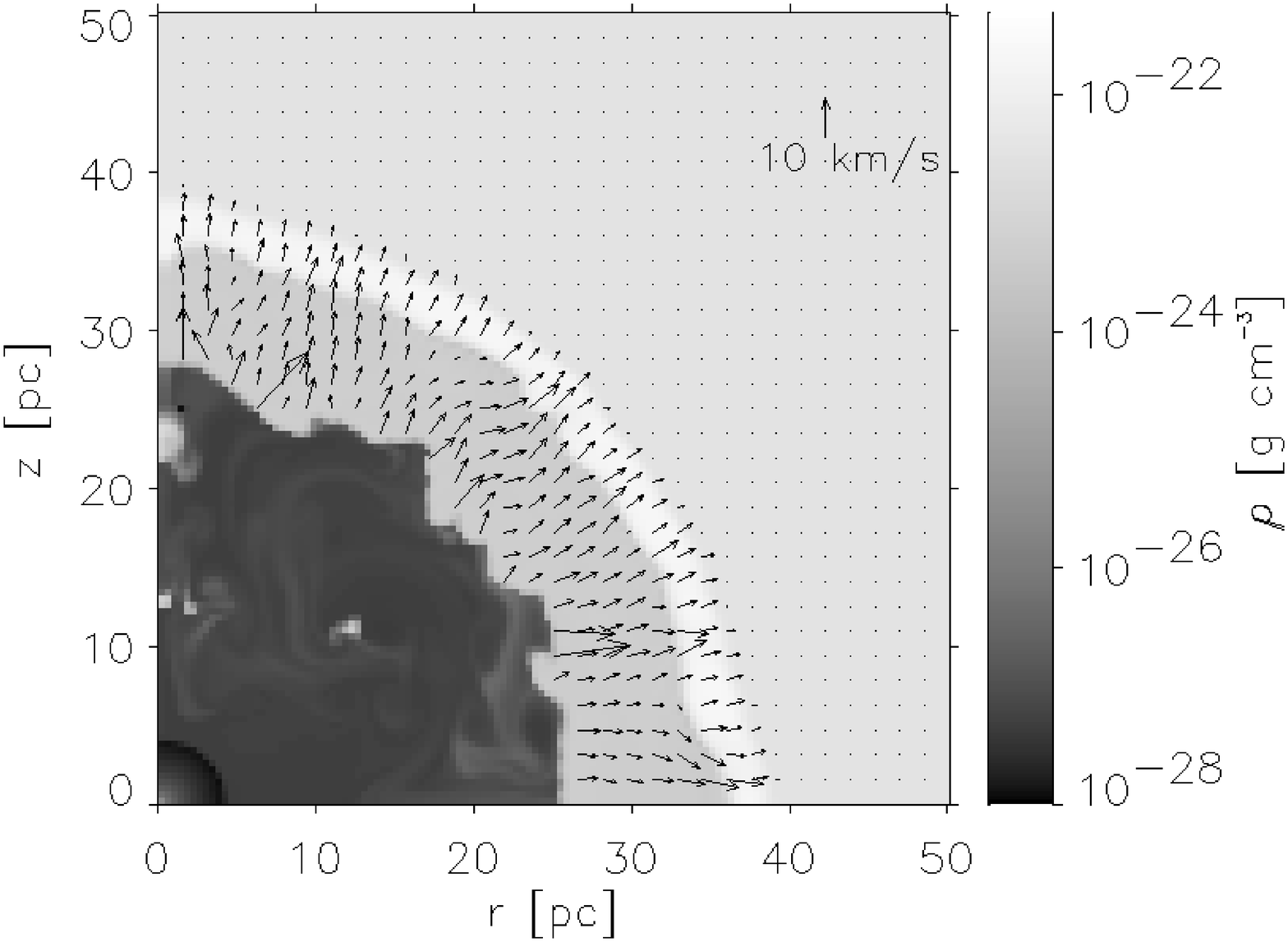}
  \plotone{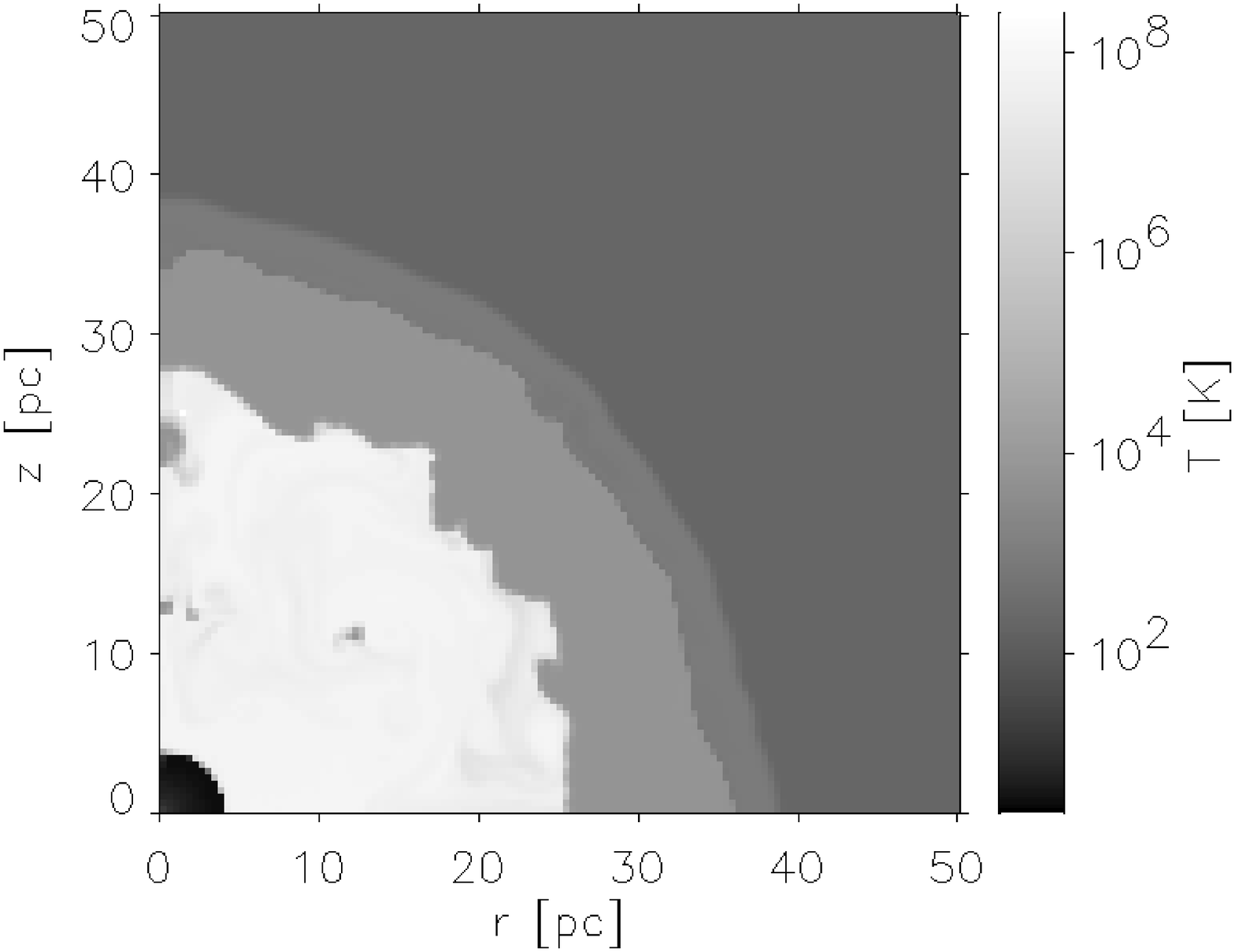}
  \plotone{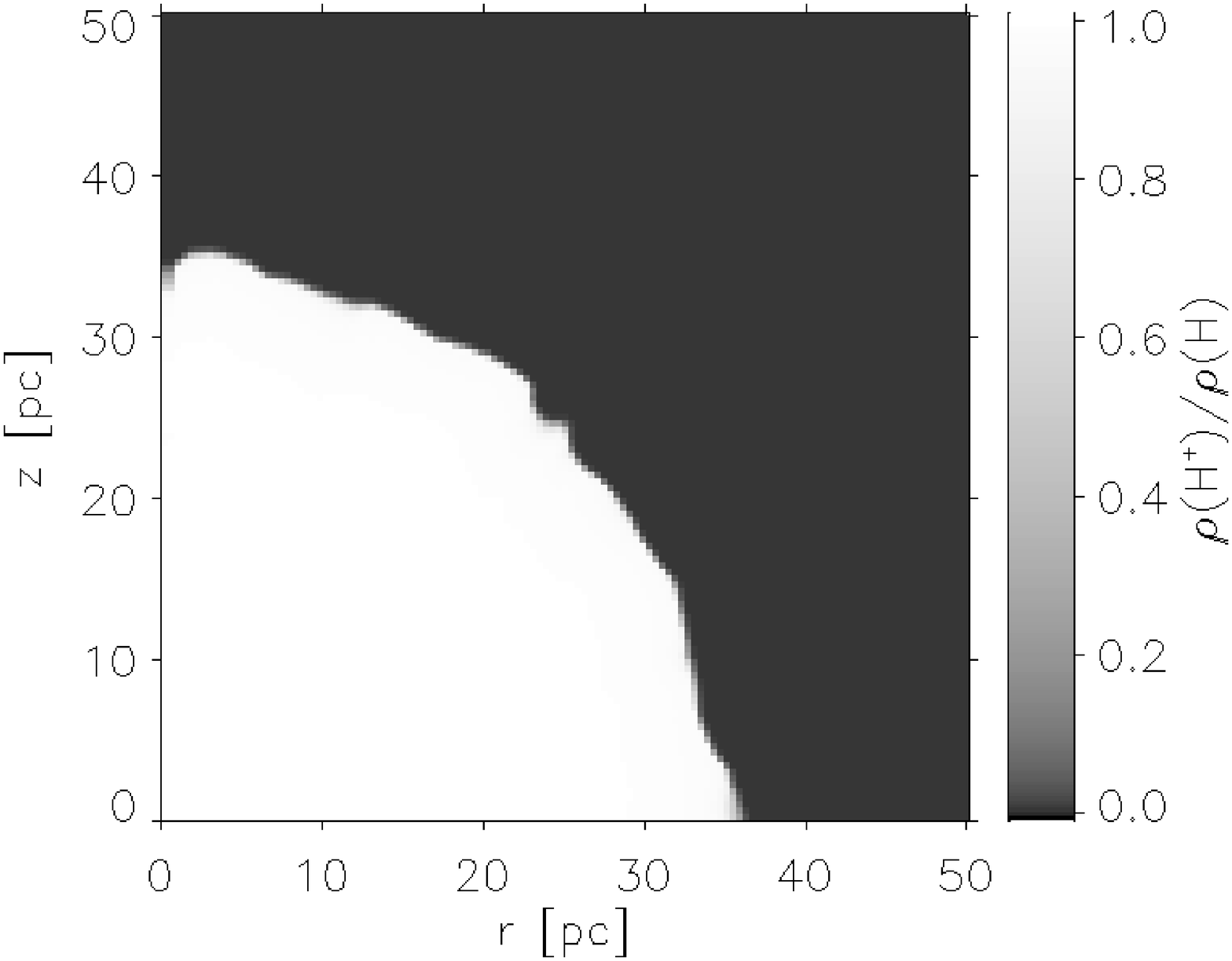}
  \caption{Same as Fig.~\ref{ion_uchii_apj229.404.001.2.med.mono.eps},
           but at age 4.0 Myr, and the displayed area is enlarged once again
           \label{ion_uchii_apj229.942.003.1.med.mono.eps}
          }
\end{figure}

\begin{figure}
  \epsscale{0.50}
  \plotone{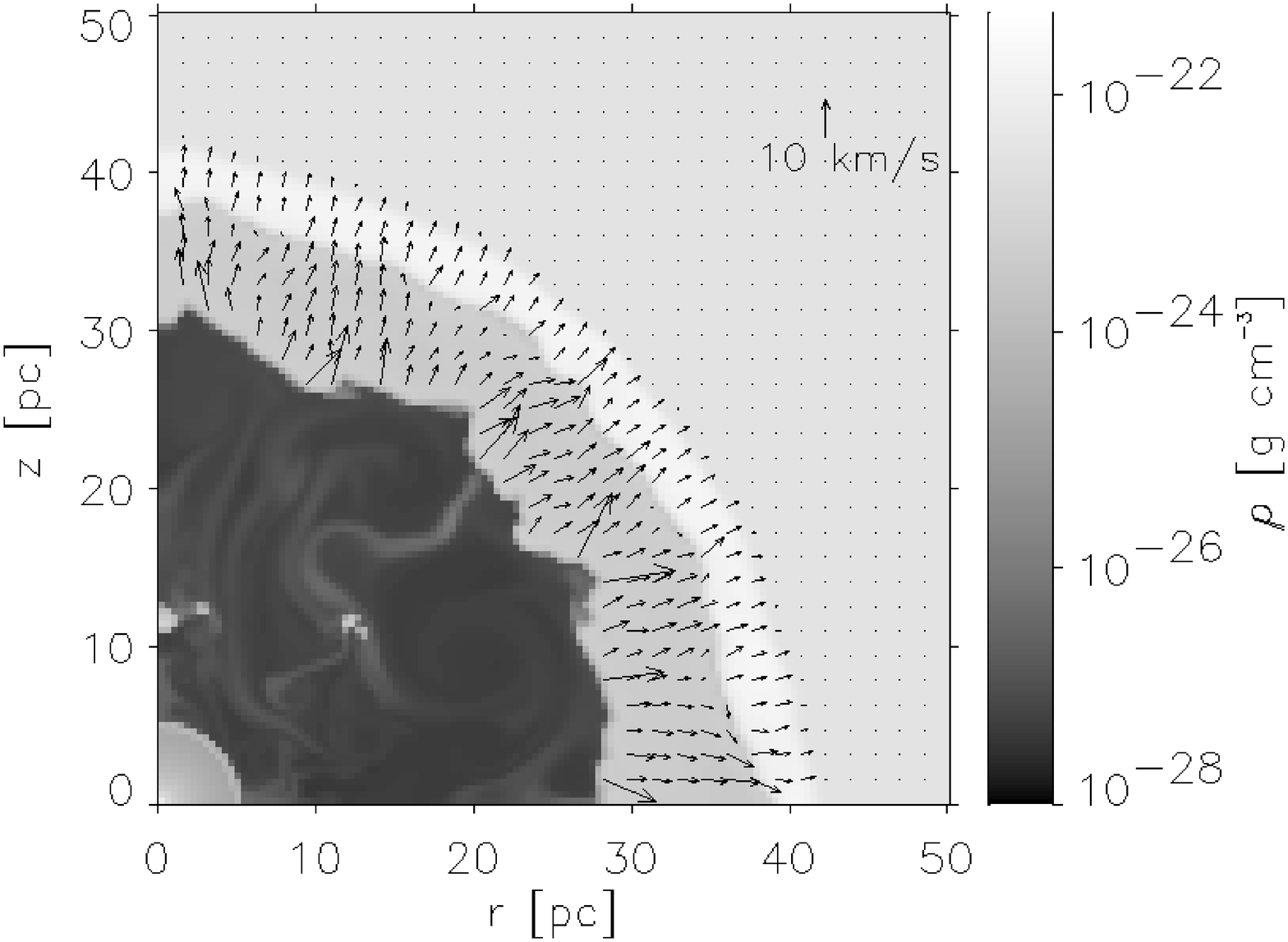}
  \plotone{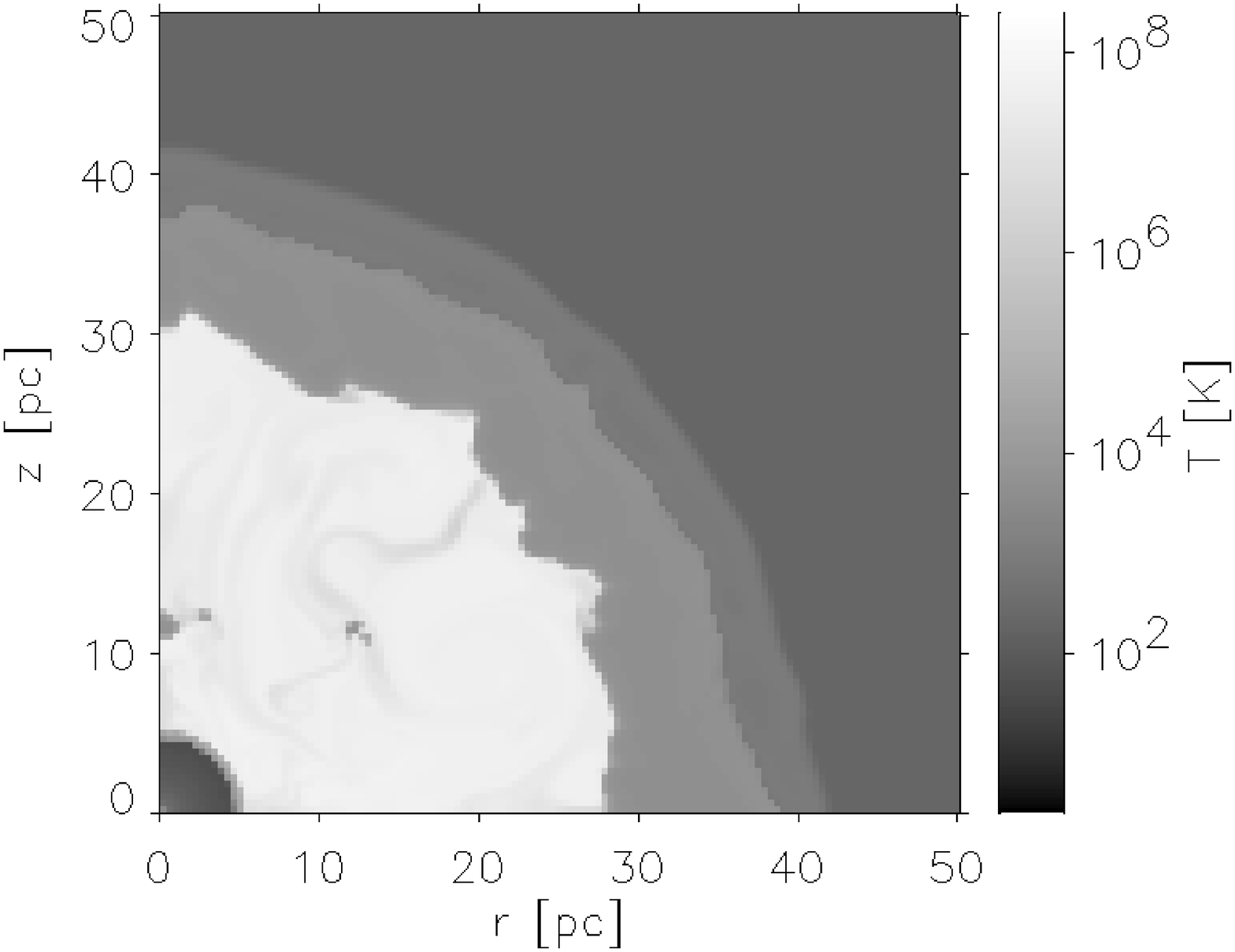}
  \plotone{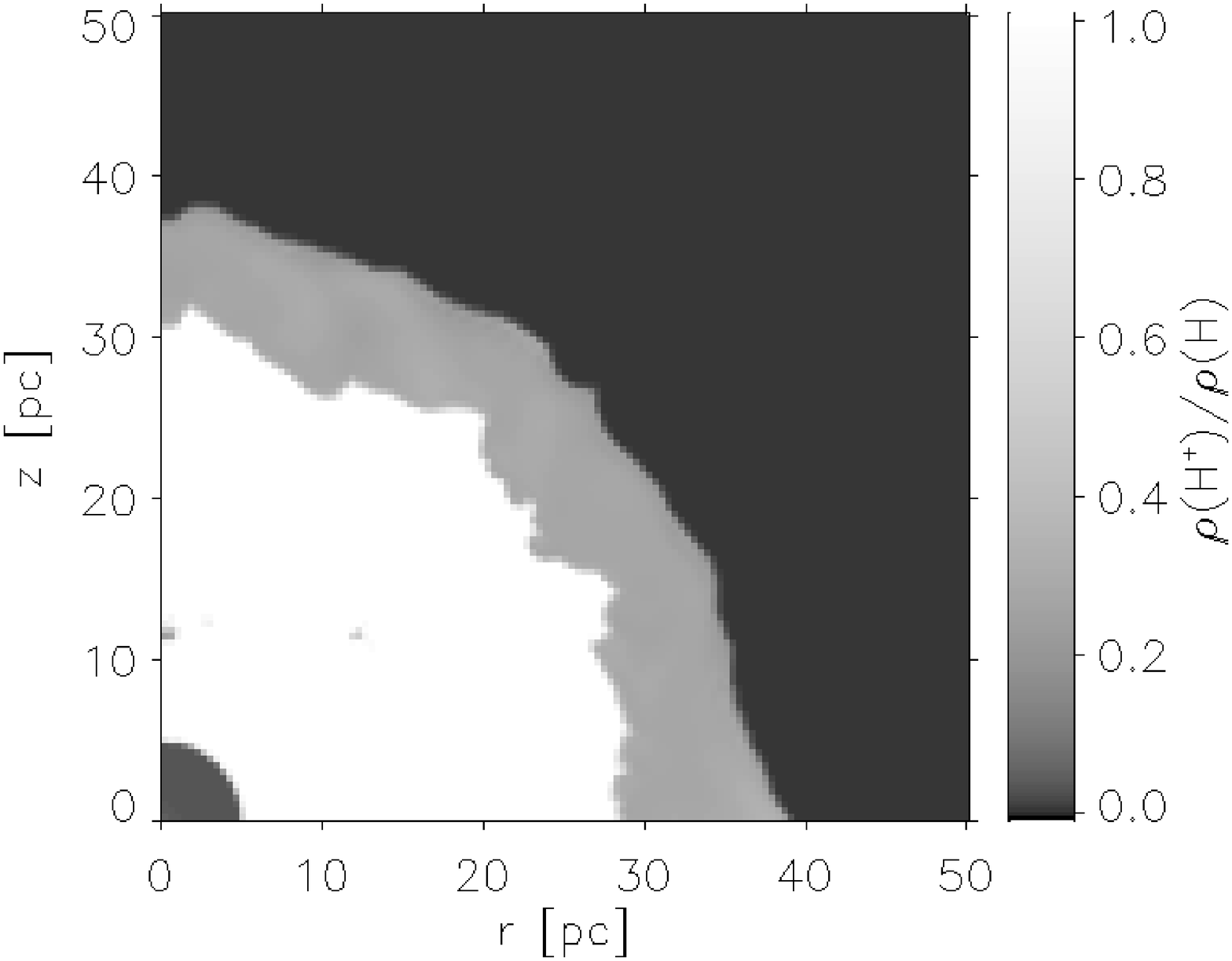}
  \caption{Same as Fig.~\ref{ion_uchii_apj229.942.003.1.med.mono.eps},
           but at age 4.59 Myr
           \label{ion_uchii_apj229.960.017.1.med.mono.eps}
          }
\end{figure}

\begin{figure}
  \epsscale{0.50}
  \plotone{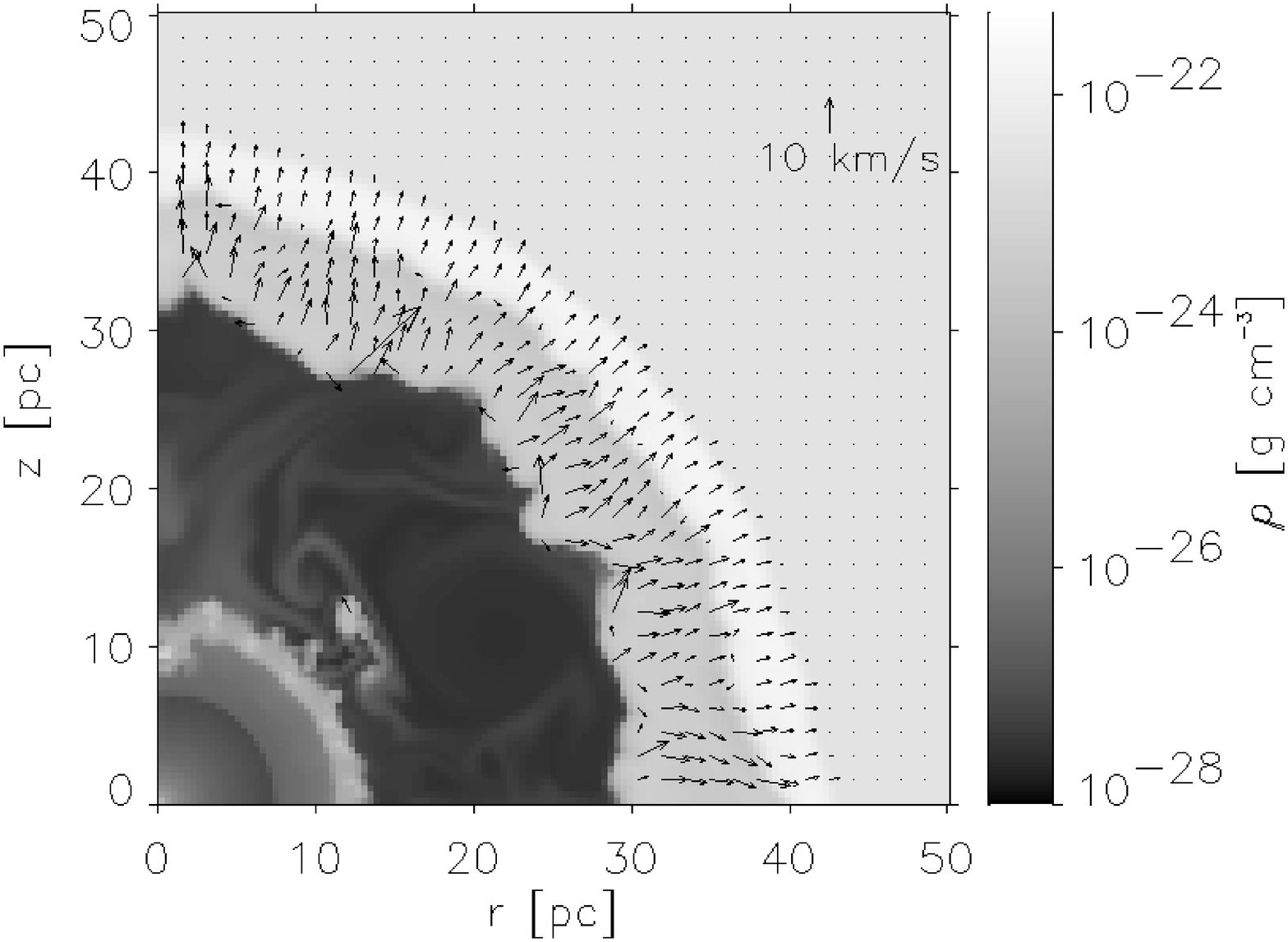}
  \plotone{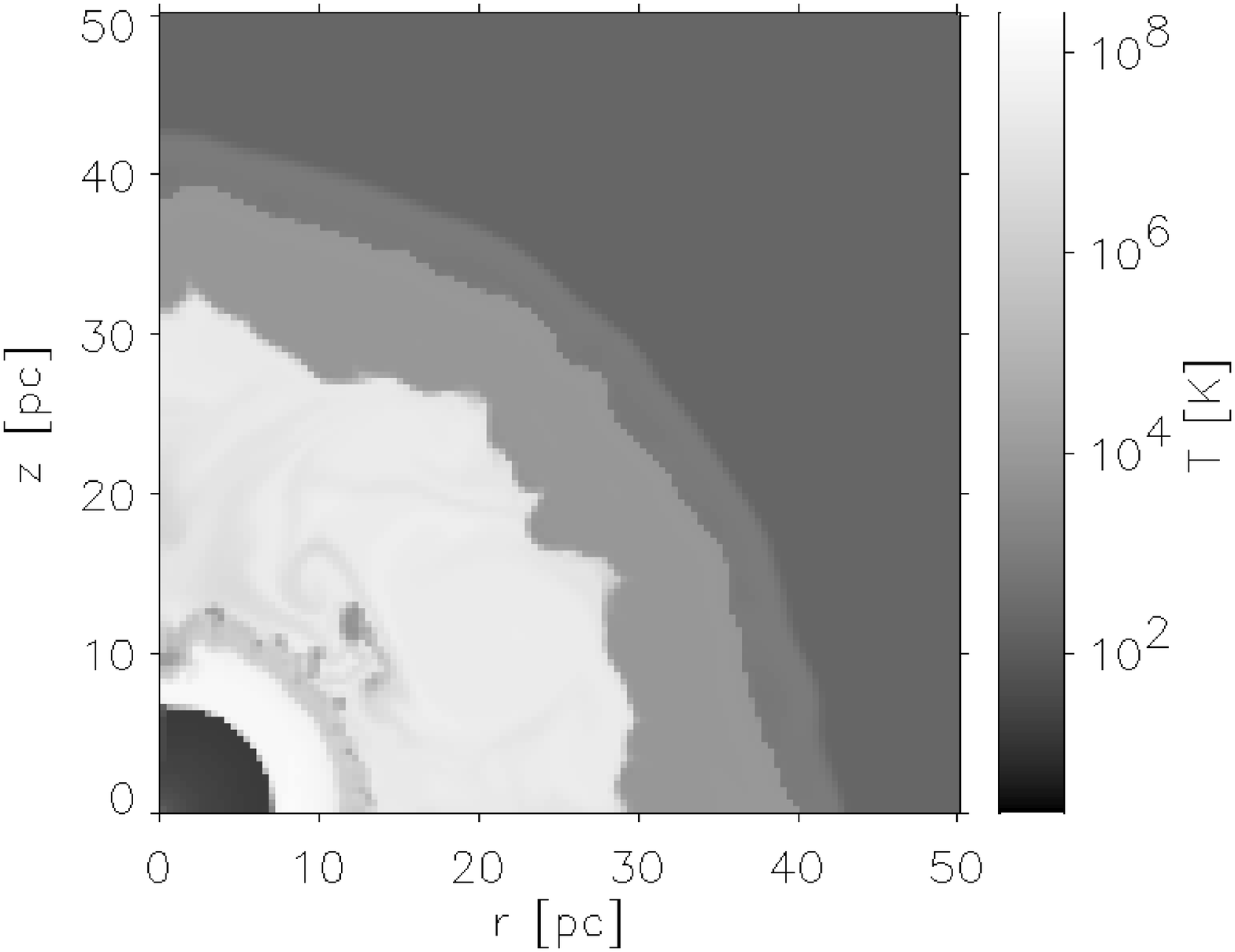}
  \plotone{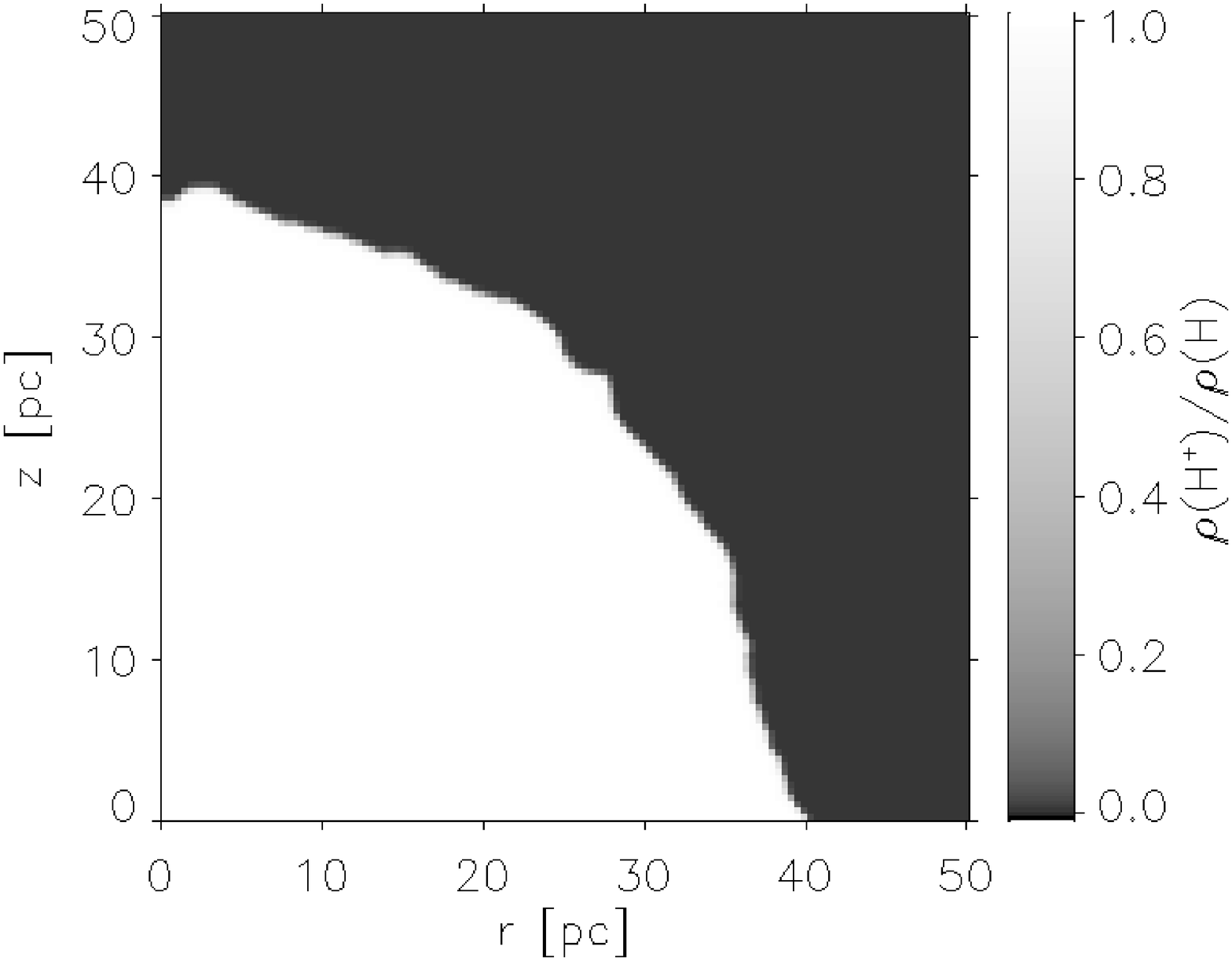}
  \caption{Same as Fig.~\ref{ion_uchii_apj229.942.003.1.med.mono.eps},
           but at age 4.78 Myr
           \label{ion_uchii_apj229.962.006.1.med.mono.eps}
          }
\end{figure}

\begin{figure}
  \epsscale{0.50}
  \plotone{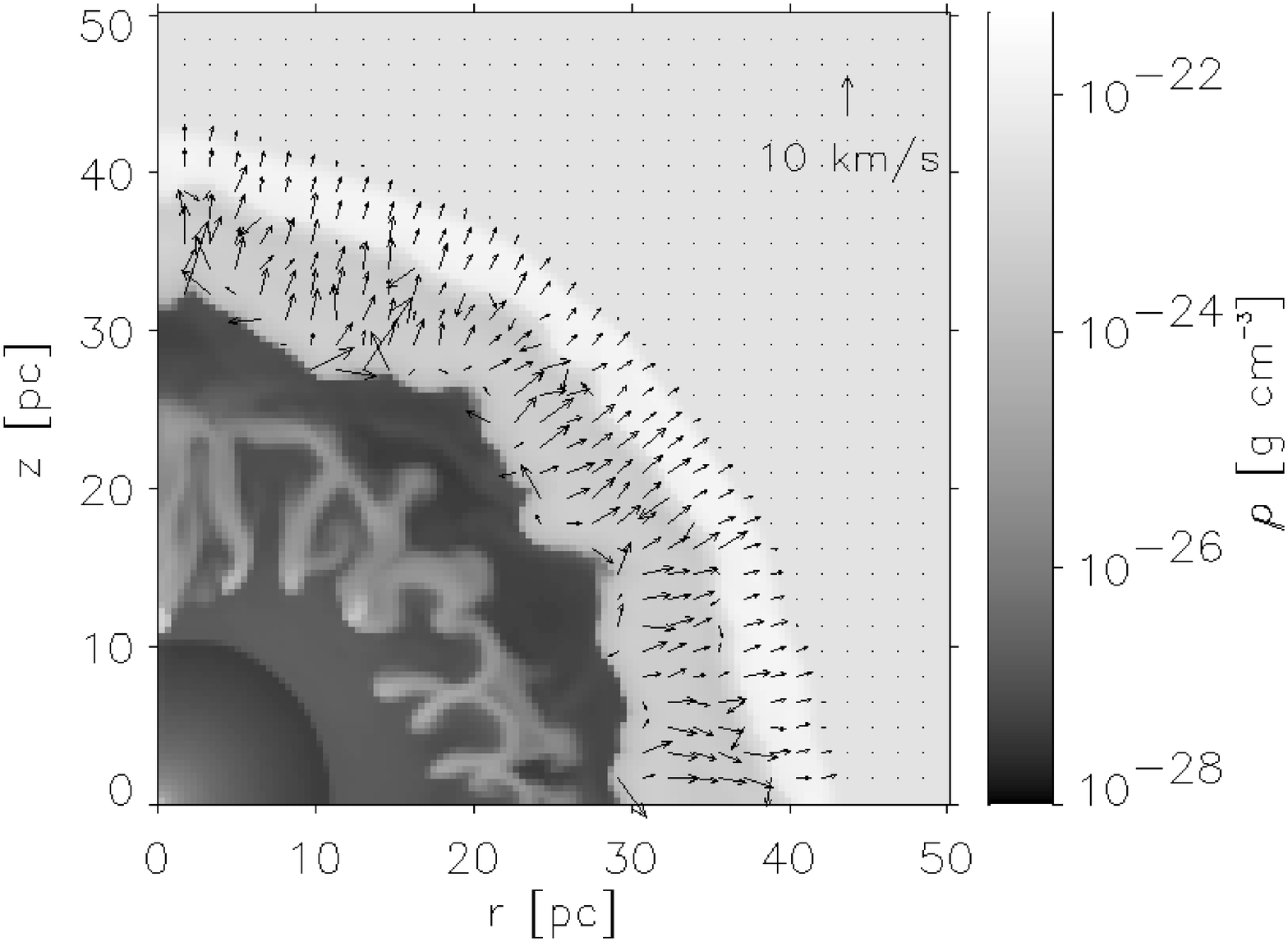}
  \plotone{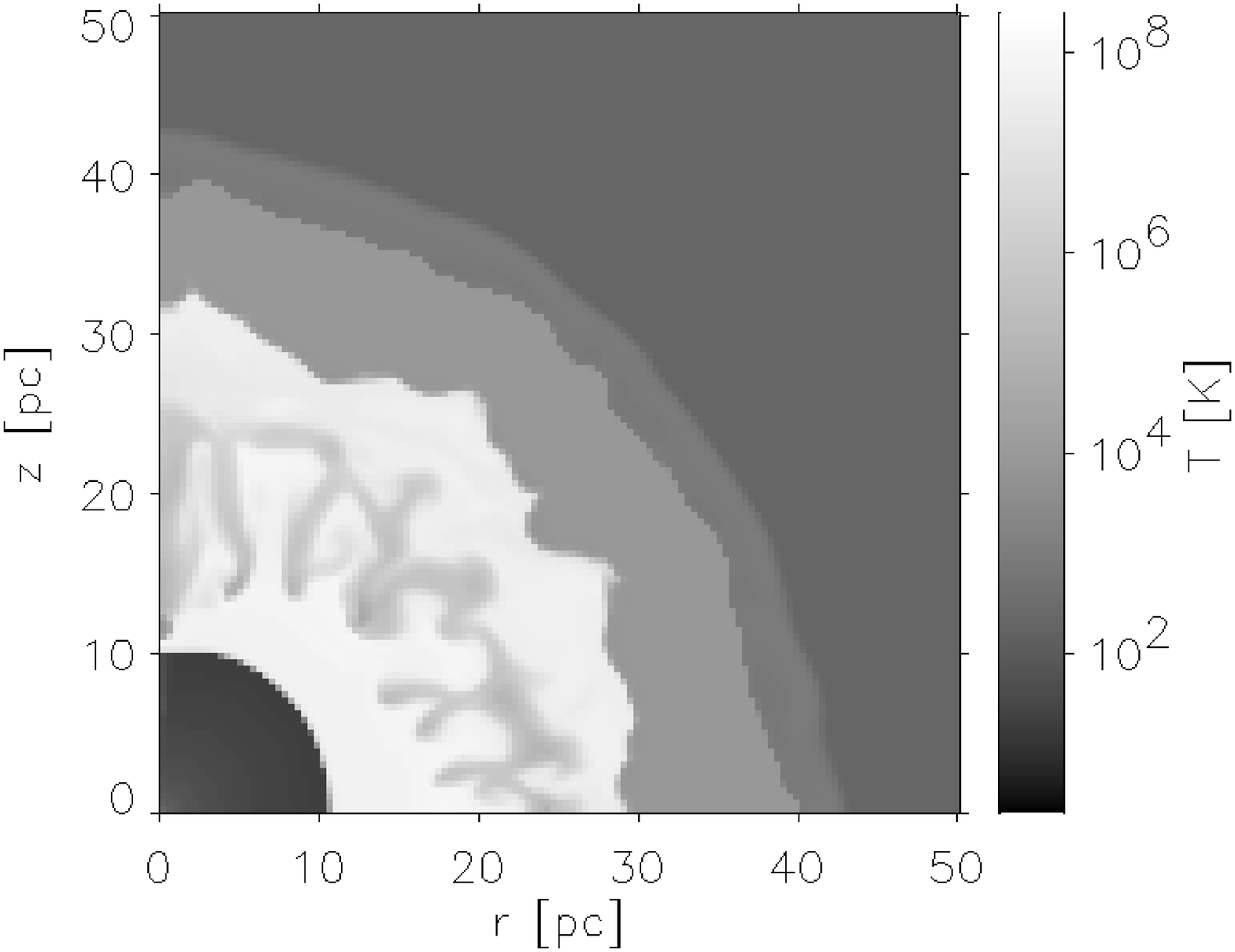}
  \plotone{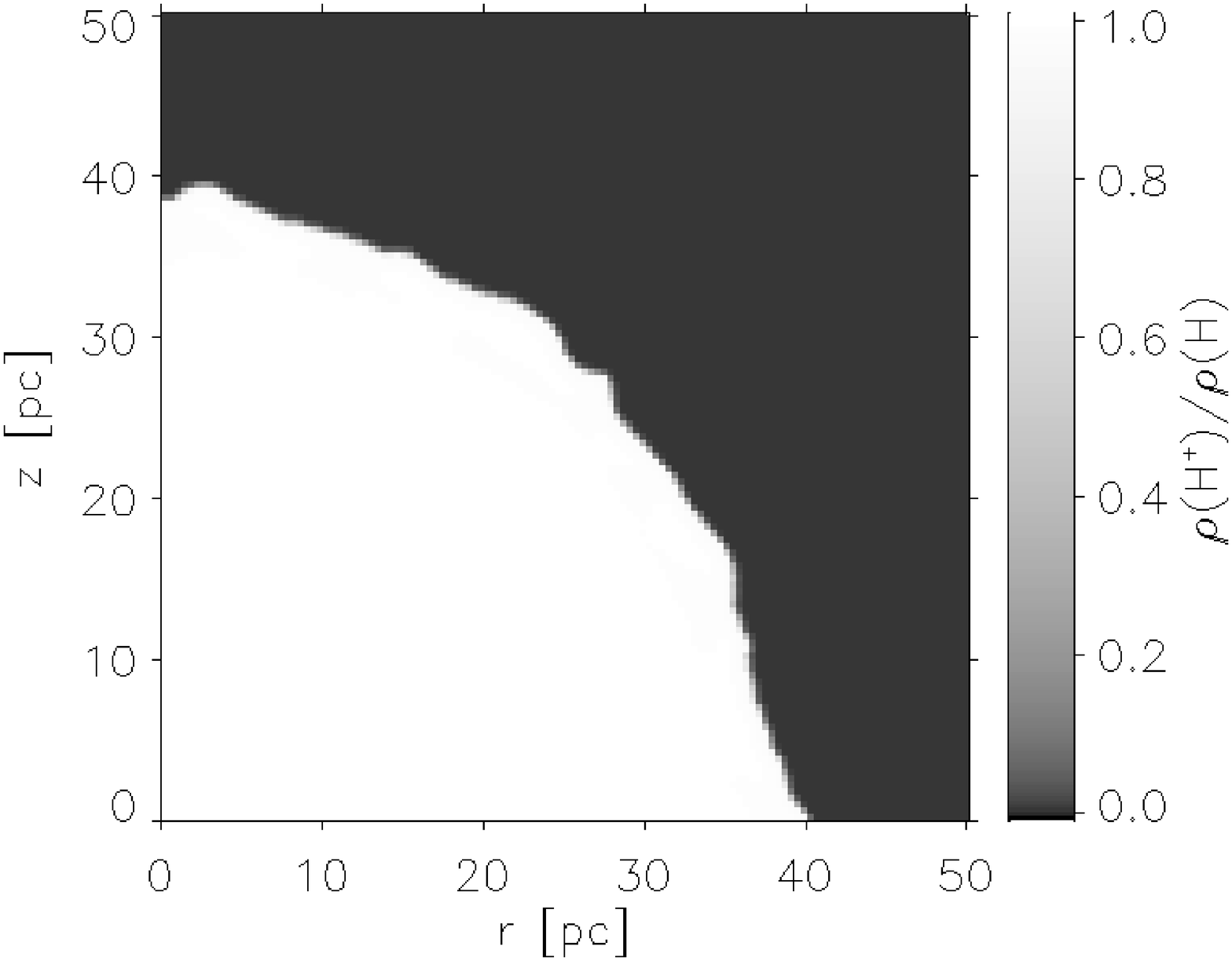}
  \caption{Same as Fig.~\ref{ion_uchii_apj229.942.003.1.med.mono.eps},
           but at age 4.80 Myr
           \label{ion_uchii_apj229.963.004.1.med.mono.eps}
          }
\end{figure}

\begin{figure}
  \epsscale{0.50}
  \plotone{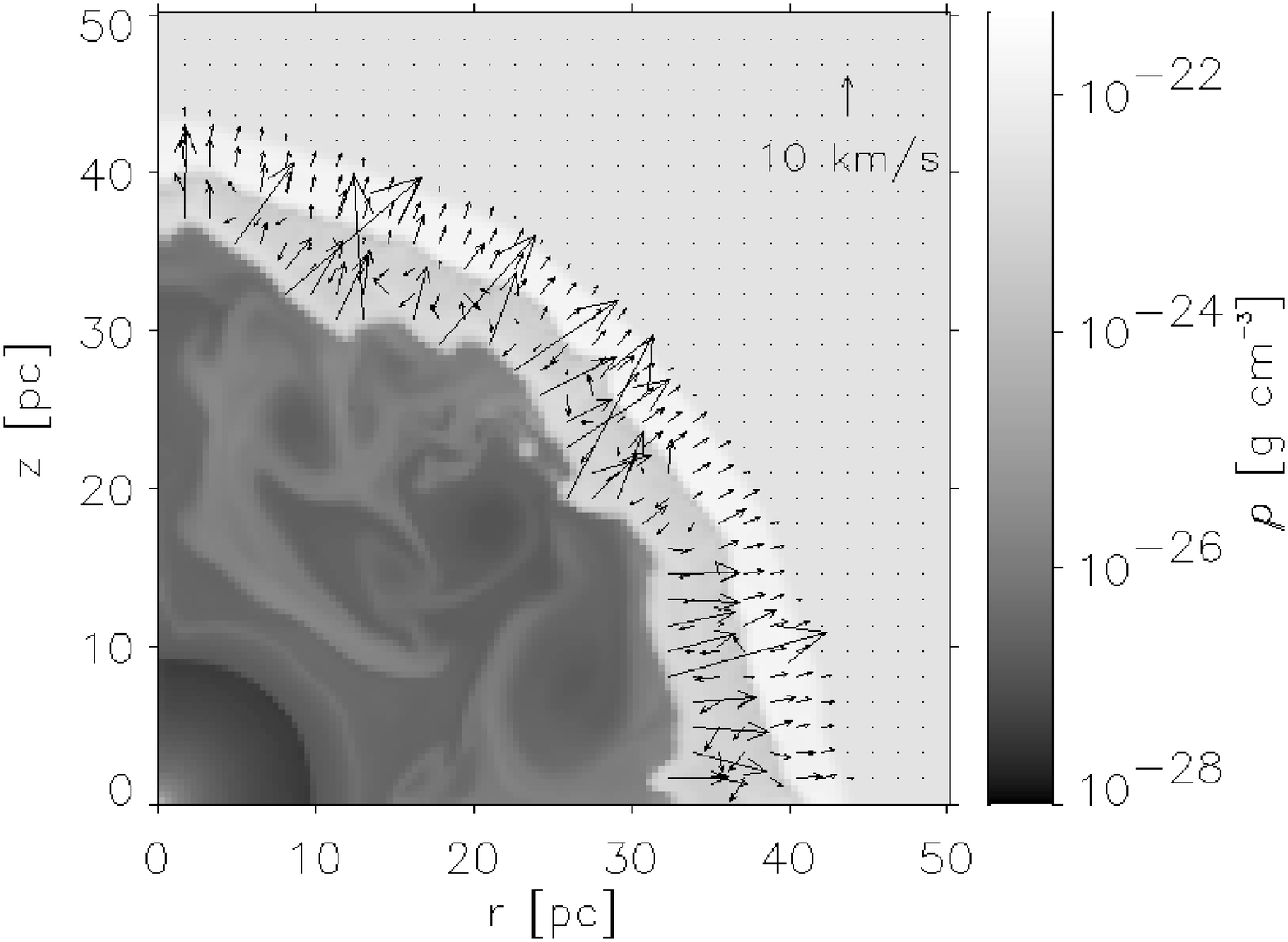}
  \plotone{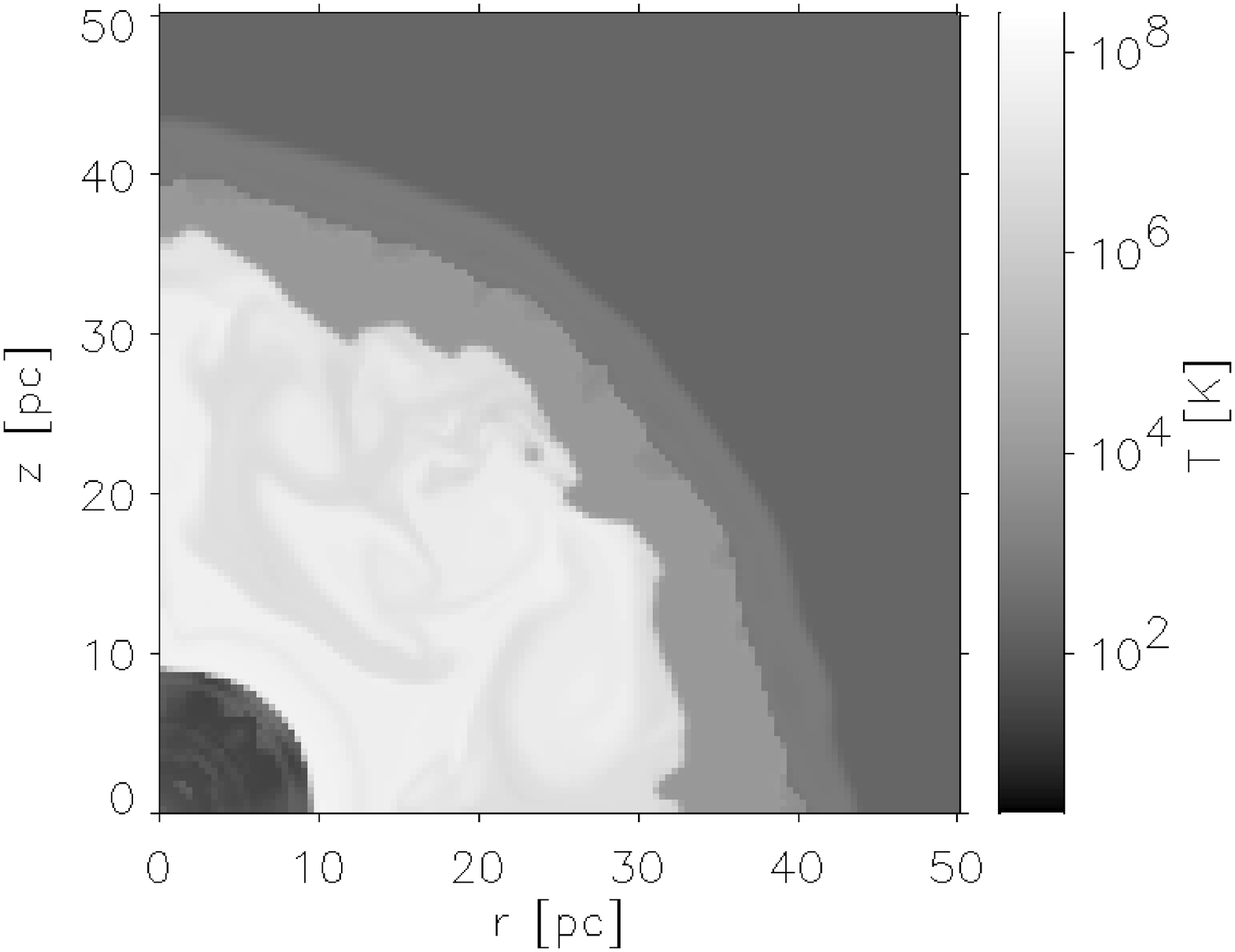}
  \plotone{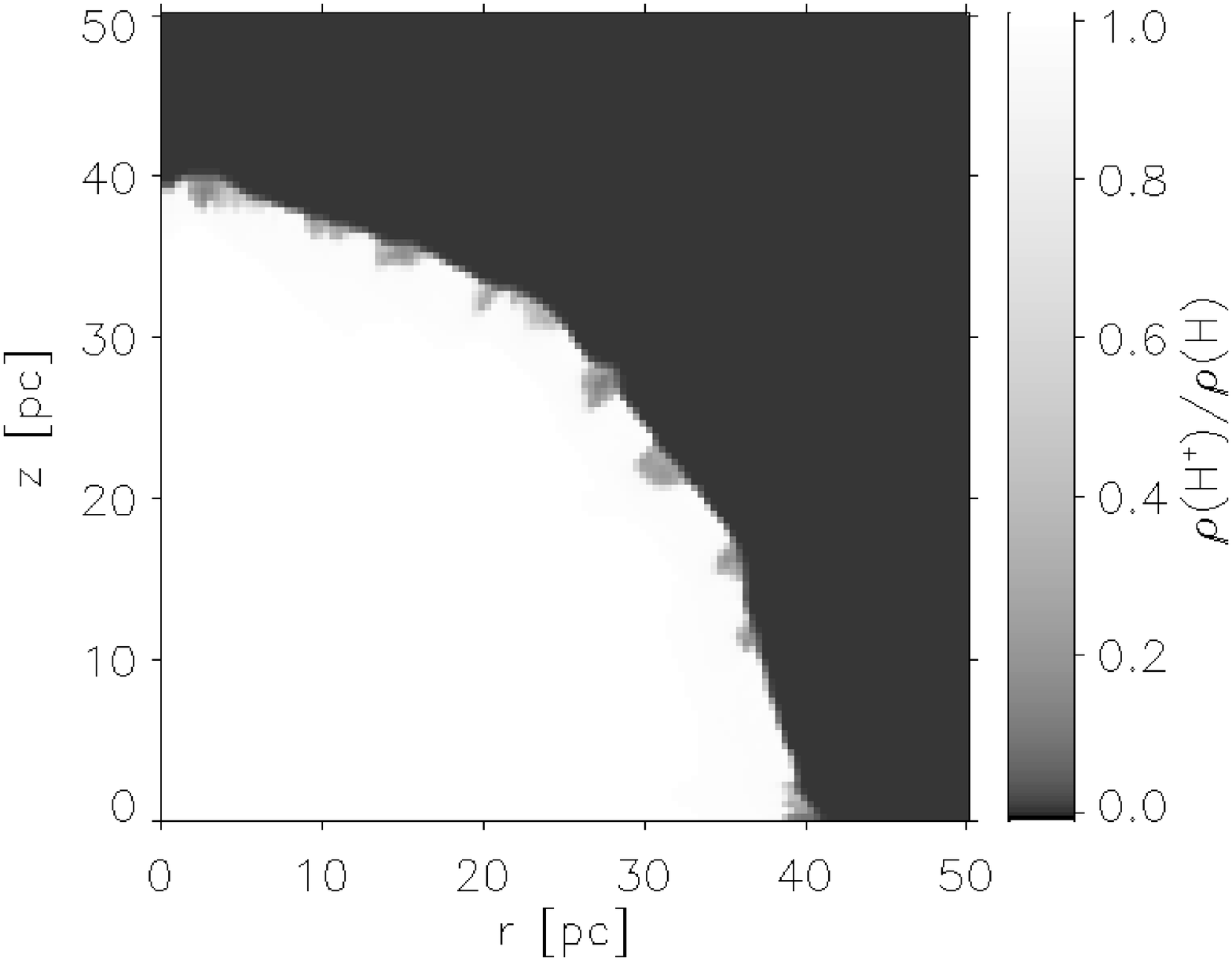}
  \caption{Same as Fig.~\ref{ion_uchii_apj229.942.003.1.med.mono.eps},
           but at age 4.945 Myr
           \label{ion_uchii_apj229.967.008.1.med.mono.eps}
          }
\end{figure}

  Since the stellar parameters vary only very gradually during the MS phase
  until the star enters the RSG stage
  (see Figure \ref{35Msun_input4apj_up.eps}), the basic structure of the
  SWB/\HII region remains the same during the next 3 Myr except that it
  continues to grow. Thus, we can proceed with our description of the evolution
  to Figure \ref{ion_uchii_apj229.942.003.1.med.mono.eps}, which shows the
  bubble at the age of 4 Myr (please note once again the larger scale). The
  radius of the hot bubble is now about 27 pc and the \HII region extends out
  to $\approx$35 pc. The hot bubble and the photoionized \HII region are in
  pressure equilibrium; as the pressure of the hot bubble decreases through
  expansion, the pressure of the \HII region drops, too. Because the
  temperature in the \HII region is nearly constant in time, the density has
  decreased to $(3.9-5.5) \times 10^{-24}~\mathrm{g~cm^{-3}}$. A few dense
  clumps are visible in the hot bubble which have detached from the
  \HII region. The thermal pressure of the SWB/\HII region is still
  $\approx$20 times higher than that of the ambient medium and the whole
  structure still expands at about $4-5~\mathrm{km~s^{-1}}$. As a result of
  the additionally swept-up material, the reduced pressure in the \HII region,
  and the deceleration of the expansion, the geometrical thickness of the
  outer shell has increased to almost 3 pc while its density has decreased
  to $\rho \approx (1-2) \times 10^{-22}~\mathrm{g~cm^{-3}}$. Thus, the mass
  collected in this shell is about $2 \times 10^{38}~\mathrm{g}$. This result
  is in good agreement with the picture that most of the ambient gas that has
  been swept up during the expansion since the ionization front turned to
  D-type is still stored in the outer shell and that only a minor fraction
  has been evaporated into the \HII region.

  At $t \approx 4.52~\mathrm{Myr}$ the star enters the RSG phase. The effective
  temperature of the star decreases to a few thousand Kelvin and the Lyman
  continuum flux drops by many orders of magnitude. The mass-loss rate strongly
  increases and the terminal velocity of the wind decreases; i.e., the star
  blows a dense and slow wind into the hot bubble. This can be seen in Figure  
  \ref{ion_uchii_apj229.960.017.1.med.mono.eps}, which depicts the state of the
  circumstellar gas at $t \approx 4.59~\mathrm{Myr}$. At this time the 
  dense RSG wind fills the inner 5 pc of the volume and expands at about
  $90~\mathrm{km~s^{-1}}$ into the hot bubble extending out to about 30 pc.
  Although the wind speed during the RSG phase is relatively low (compared to
  the MS and W-R phases), the material becomes shocked when it is decelerated
  by the pressure of the MS bubble because the sound speed in the RSG wind
  is also low. (The gas is cold and neutral, or even molecular, because the
  soft stellar radiation field in the RSG phase is incapable of ionizing and
  heating it.) Since the density of the RSG wind is fairly high, this reverse
  shock is radiative and forms the very thin and dense RSG shell, which is
  not completely resolved in our calculation
  \citep[compare to][the case of the fast RSG wind]{garcia96b}. Because of the
  reduction of Lyman continuum flux, the hydrogen in the former \HII region
  between  $30~\mathrm{pc} \lesssim r \lesssim 38~\mathrm{pc}$ starts to
  recombine. The degree of hydrogen ionization drops to $\approx$ 0.3 as the
  gas cools from about 6600 K at $t = 4~\mathrm{Myr}$ to 4800 K. Thus, the
  thermal pressure in the (former) \HII region drops significantly, leading to
  a further broadening and slowing down of the outer shell of swept-up ambient
  medium, which at this time has a thickness of about 3 pc and a velocity of
  $\approx$4~$\mathrm{km~s^{-1}}$.

  The transition from the RSG phase of the star to the W-R phase occurs at
  $t \approx 4.754~\mathrm{Myr}$. In the W-R phase the mass-loss rate is
  somewhat lower than in the RSG stage (see Figure
  \ref{35Msun_input4apj_up.eps}), but the terminal velocity of the W-R wind
  is much higher (a few thousand $\mathrm{km~s^{-1}}$). The mechanical
  luminosity of the W-R wind is thus much higher than that of the RSG wind.
  Because of the increase of effective temperature to about $10^5~\mathrm{K}$,
  the stellar photon output occurs largely beyond the Lyman continuum limit;
  the star's Lyman continuum luminosity $\approx 10^{39}~\mathrm{ergs~s^{-1}}$
  is higher than during the MS phase of the star.
  Figure \ref{ion_uchii_apj229.962.006.1.med.mono.eps} shows the structure of
  the circumstellar gas at $t = 4.78~\mathrm{Myr}$, some
  $2.6 \times 10^4~\mathrm{yr}$ after the start of the W-R wind. The fast
  (and thus less dense) W-R wind reestablishes a non-radiative reverse shock
  that heats the W-R wind to about $10^8~\mathrm{K}$. The expansion of the hot
  gas sweeps up the slow RSG wind in the W-R shell. The expansion velocity of
  the W-R shell is high, almost $400~\mathrm{km~s^{-1}}$. Therefore, a strong
  shock ahead of the W-R shell heats the RSG wind material and the W-R shell
  is thick (of the order 1 pc) and hot (${\approx}10^6~\mathrm{K}$).

  Figure \ref{ion_uchii_apj229.962.006.1.med.mono.eps} pinpoints the moment
  when the W-R shell and the RSG shell collide so that only one shell of
  swept-up RSG wind is visible within the MS bubble \citep[see][]{garcia96b}.
  (The shell distortion at the $z$-axis is a result of the dense clump that
  has been there before.) Stellar Lyman continuum photons ionize the
  RSG wind gas (before it is shock heated by the W-R shell) and reionize
  the former \HII region, which can be seen by comparison of the lower panels
  of Figures \ref{ion_uchii_apj229.960.017.1.med.mono.eps} and
  \ref{ion_uchii_apj229.962.006.1.med.mono.eps}.

  As long as the W-R shell moves through the RSG wind material, the thermal
  pressure of the shocked W-R wind inside the W-R shell is counterbalanced
  by the ram pressure of the RSG material it sweeps up.
  When the shock passes over the boundary between the RSG and MS winds and
  moves into the lower density medium, it speeds up as a rarefaction wave
  travels back into the W-R shell. It is the overpressure of
  the shocked rarefied W-R gas which is accelerating the much denser W-R shell.
  This is the classical case of a Rayleigh-Taylor unstable configuration
  and the W-R shell is torn into long filaments as the shocked W-R wind
  breaks through it. This can be seen very impressively in Figure
  \ref{ion_uchii_apj229.963.004.1.med.mono.eps} at $t = 4.80~\mathrm{Myr}$.

  Figure \ref{ion_uchii_apj229.967.008.1.med.mono.eps} shows the final,
  pre-supernova state of the circumstellar gas after 4.945 Myr of evolution.
  The wind flows freely out to $9-10~\mathrm{pc}$. This value is somewhat
  uncertain, however, due to the very complex flow pattern. The freely flowing
  wind is surrounded by the hot bubble that extends out to $31-37$ pc
  (on average 34 pc). The gas density in the hot bubble is
  $(1-30) \times 10^{-27}~\mathrm{g~cm^{-3}}$. This is fairly high (compared
  to the density at the end of the MS phase of the star) because some of the
  RSG/W-R ejecta mixes with the hot bubble gas, enhancing the density
  significantly and lowering the temperature.

  The thickness of the shell-like \HII region has shrunk to $3-8$ pc with an
  average of 5.5 pc, after the RSG ejecta hit it and the pressure in the
  hot bubble has risen. This also enhances the density in the \HII region to
  $(2-20) \times 10^{-24}~\mathrm{g~cm^{-3}}$. The higher spread is a
  result of the higher W-R Lyman continuum flux (compared to the MS phase),
  leading to increased photoevaporation of the swept-up ambient material.
  Finally, the shell of swept-up ambient gas is $3-4$ pc thick and the gas
  density in the shell is $(1-2) \times 10^{-22}~\mathrm{g~cm^{-3}}$.
  The expansion velocity of this shell is $3-4~\mathrm{km~s^{-1}}$ and
  therefore still supersonic with respect to the cold ambient ISM.
  The outermost shock moving into the ISM heats the swept-up ISM gas to
  almost $10^3~\mathrm{K}$. Thus, the whole structure extends out to a
  distance of $43-44$ pc from the star at this evolutionary time.

\subsection{The Energy Balance in the Circumstellar Gas}
\label{subsec_e_balance}

\subsubsection{Numerical Results}

\begin{figure}
  \epsscale{1.00}
  \plotone{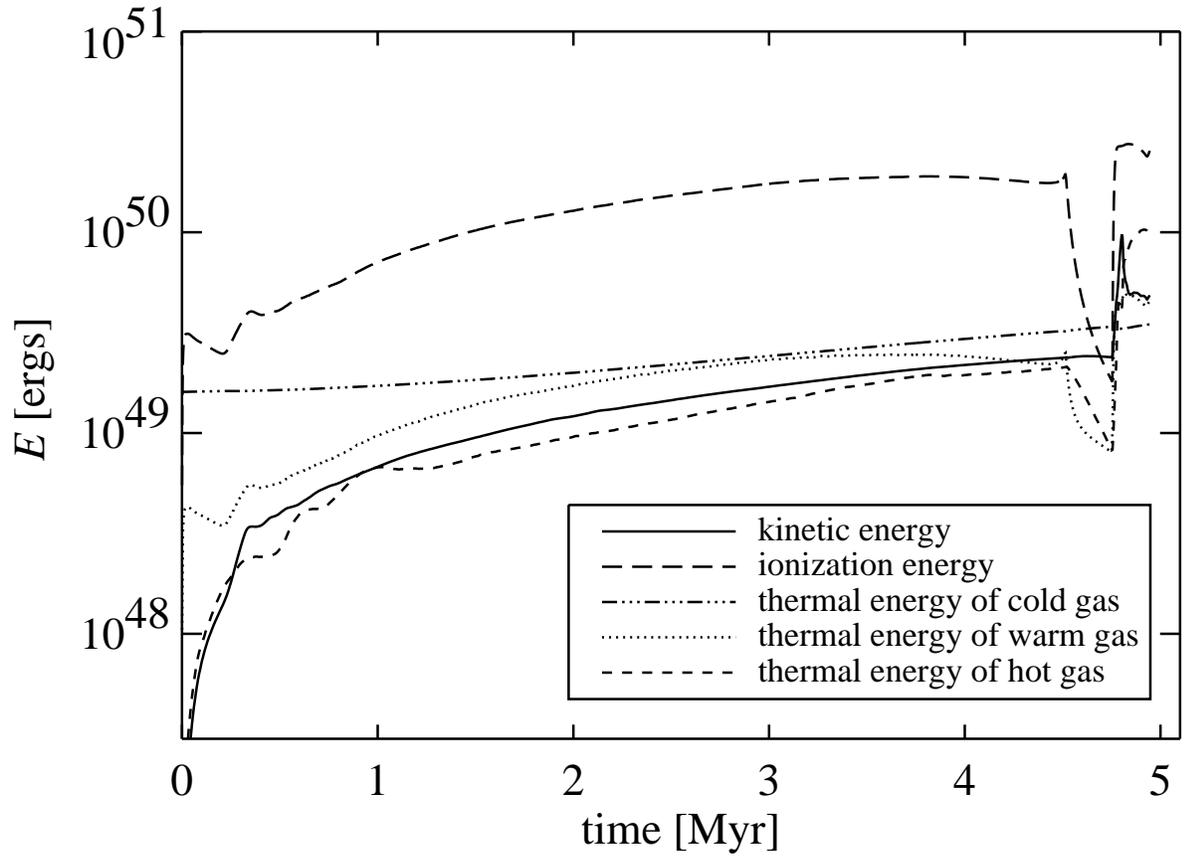}
  \caption{Temporal evolution of kinetic, thermal, and ionization energy
           for the 35 $\Msun$ case. For details see text.
           \label{e_distri229apj_up.eps}
          }
\end{figure}

  We discuss the energization of the circumstellar gas around the 35 $\Msun$
  star on the basis of Figure \ref{e_distri229apj_up.eps} and compare it with
  its counterpart for the 60 $\Msun$ case (Figure 17 in \citeauthor{freyer03}).
  The plot shows selected energy contributions in the circumstellar gas as a
  function of time, namely the kinetic energy of bulk motion in the whole
  computational domain, the ionization energy (13.6 eV per ionized hydrogen
  atom), the thermal energy of cold ($T \le 10^3~\mathrm{K}$), warm
  ($10^3~\mathrm{K} < T < 10^5~\mathrm{K}$), and hot ($T \ge 10^5~\mathrm{K}$)
  gas, respectively. 

  The ionization energy dominates after several thousand years, when the
  ionization front reaches the Str\"omgren radius. Here, it attains a value of
  approximately $3.1 \times 10^{49}~\mathrm{ergs}$, which is about half an
  order of magnitude less than for the 60 $\Msun$ case due to the lower
  Lyman continuum luminosity of the 35 $\Msun$ star.
  The subsequent dip in ionization energy is much less pronounced than for
  the 60 $\Msun$ case, since at that time there is less structure formation
  in the \HII region, as is evident in Figures 
  \ref{ion_uchii_apj230.030.001.3.med.mono.eps} through
  \ref{ion_uchii_apj230.048.002.3.med.mono.eps}.
  Afterward, the ionization energy rises smoothly to
  $1.8 \times 10^{50}~\mathrm{ergs}$ before the star enters the RSG stage.
  During the RSG stage the ionization energy drops by an order of
  magnitude since the H-ionizing radiation from the star is completely
  switched off and the photoionized gas recombines. The gas is reionized
  when the star evolves to the W-R phase; the ionization energy reaches a
  global maximum of $2.8 \times 10^{50}~\mathrm{ergs}$. Contrary to
  the 60 $\Msun$ case, the ionization energy remains the dominant form
  of energy in the system for the entire evolution except for a brief period
  at the end of the RSG phase.

  A feature which we already described for the 60 $\Msun$ case can also
  be seen in the 35 $\Msun$ case: The evolution of the thermal energy of
  warm gas follows that of the ionization energy over the lifetime of the
  star with a shift of $0.7-0.9$ dex due to the fact that for both cases,
  photoionization is responsible for the bulk production of ionized gas at
  typically 8000 K. The shift is smaller during the RSG phase of the star
  because cooling is less efficient in the formerly photoionized regimes
  when the hydrogen recombines and the plasma becomes neutral.

  The evolution of the kinetic energy of bulk motion and the thermal energy of
  hot gas are also very similar; they do not deviate from each other by
  more than 0.2 dex until the star enters the RSG phase. Whereas the kinetic
  energy remains basically constant during the RSG phase, the thermal energy
  of hot gas decreases as the supply of hot gas by the reverse shock dies off.
  The cooling of the hot gas and mixing with cooler gas continues. Both values
  rise when the non-radiative reverse shock reestablishes itself in the
  W-R phase. When the calculations are stopped, the kinetic energy is
  $4.9 \times 10^{49}~\mathrm{ergs}$ and the thermal energy of hot gas
  $1.1 \times 10^{50}~\mathrm{ergs}$; the latter value is almost the same
  as at the end of the 60 $\Msun$ calculation at 4.065 Myr.

  The thermal energy of cold quiescent gas in the entire computational domain 
  at the beginning of the calculation is $1.6 \times 10^{49}~\mathrm{ergs}$
  (the same energy density as in the 60 $\Msun$ case, but with a smaller
  computational volume). This value grows smoothly during the lifetime
  of the star as more and more ambient gas becomes swept up and weakly
  shocked by the outer shell. The thermal energy of the cold gas at the
  end of the calculation is $3.5 \times 10^{49}~\mathrm{ergs}$; i.e.,
  $1.9 \times 10^{49}~\mathrm{ergs}$ have been added during the evolution.

\begin{figure}
  \plotone{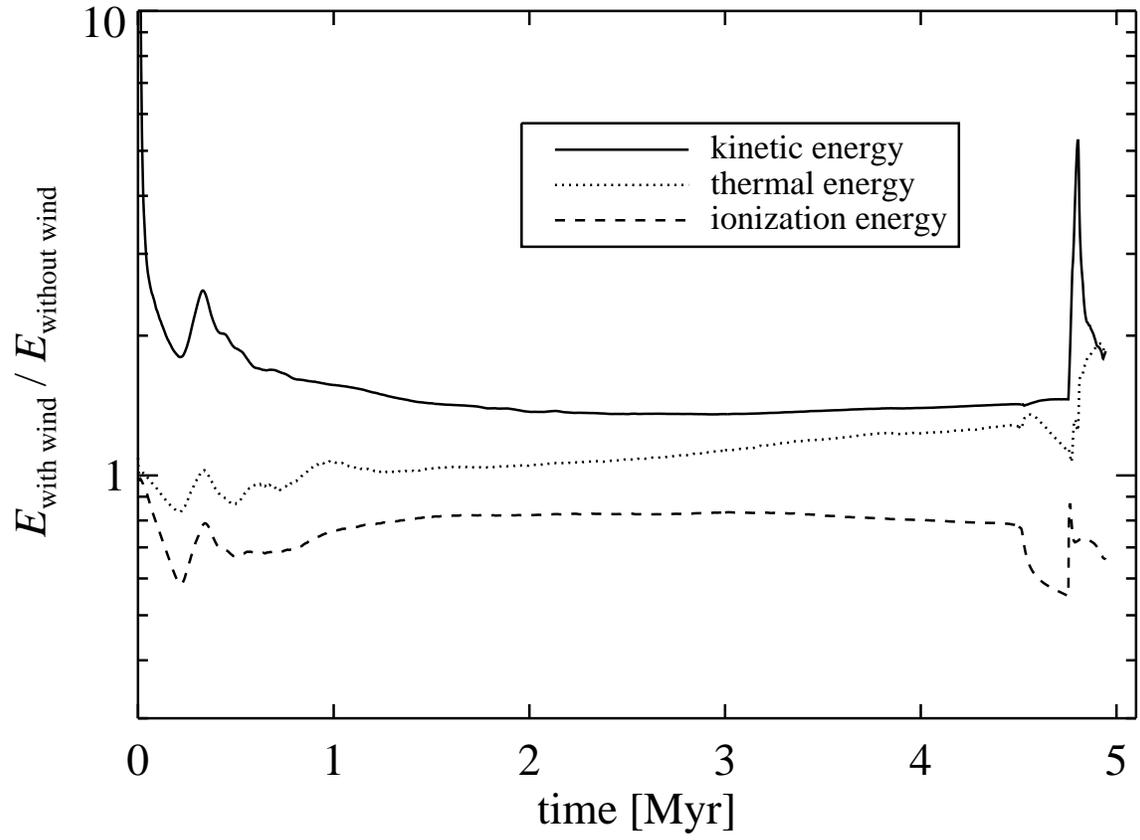}
  \caption{Ratio of energies for the 35 $\Msun$ case with and without
           a stellar wind.
           \label{E_with_without_wind35apj_up.eps}
          }
\end{figure}

  To study the impact of the stellar wind on the energy transfer in the
  circumstellar gas, we compare the energy in the system as a function of
  time for the two cases: 1) the standard case with wind and
  2) the \HII region evolution without wind
  (Figure \ref{E_with_without_wind35apj_up.eps}, see also
  Figure 18 in \citeauthor{freyer03}). The kinetic energy of bulk motion is
  enhanced due to the added kinetic energy of the stellar wind shell.
  On the other hand, the compression of the \HII region into a shell with
  higher density reduces the amount of ionization energy stored in the system
  compared to the windless case. Thus, the ratio of ionization energy for the
  calculation with wind to that of the windless model is below 1
  throughout the lifetime of the star. Both features are well recognized from
  the 60 $\Msun$ case. In general, the deviations of the energies between the
  models with and without stellar wind are smaller in the 35 $\Msun$ case,
  especially during the MS evolution. For most of the MS time the ratios are
  well within the interval $0.5-2.0$. This difference to the 60 $\Msun$ case
  exists because the ratio of mechanical luminosity to Lyman continuum
  luminosity of the 35 $\Msun$ star is smaller than that of the 60 $\Msun$
  star.

  There is a drop in the ratio of the ionization energies with and without
  wind during the RSG phase of the star. At first glance, this is surprising
  since in the calculation with stellar wind there is additional ionization
  energy in the hot bubble. But the density in the hot bubble is very low and
  although the thermal energy of the hot gas in the bubble is important, there
  is not much ionization energy involved. The reason for the drop in the ratio
  of the ionization energies with and without wind is the fact that in the
  calculation with wind the \HII region is compressed into a shell surrounding
  the hot bubble and the density in this shell is higher than in the spherical
  \HII region of the windless simulation. The higher density results in
  shorter recombination times and thus a faster loss of ionization energy when
  the Lyman continuum radiation of the star ceases at the beginning of the
  RSG phase.

  Due to the acceleration of the slow RSG wind by the shocked W-R wind, there
  is a strong peak in the ratio of the kinetic energies with and without wind.
  This additional kinetic energy is partially dissipated when the accelerated
  RSG wind material hits the \HII region. At the end of the simulation the
  kinetic energy in the calculation with stellar wind is enhanced by 85\%,
  whereas the ionization energy is reduced by 34\%, and the thermal energy
  that is added to the system during the evolution is increased by 88\%
  compared to the calculation without stellar wind.

\begin{figure}
  \plotone{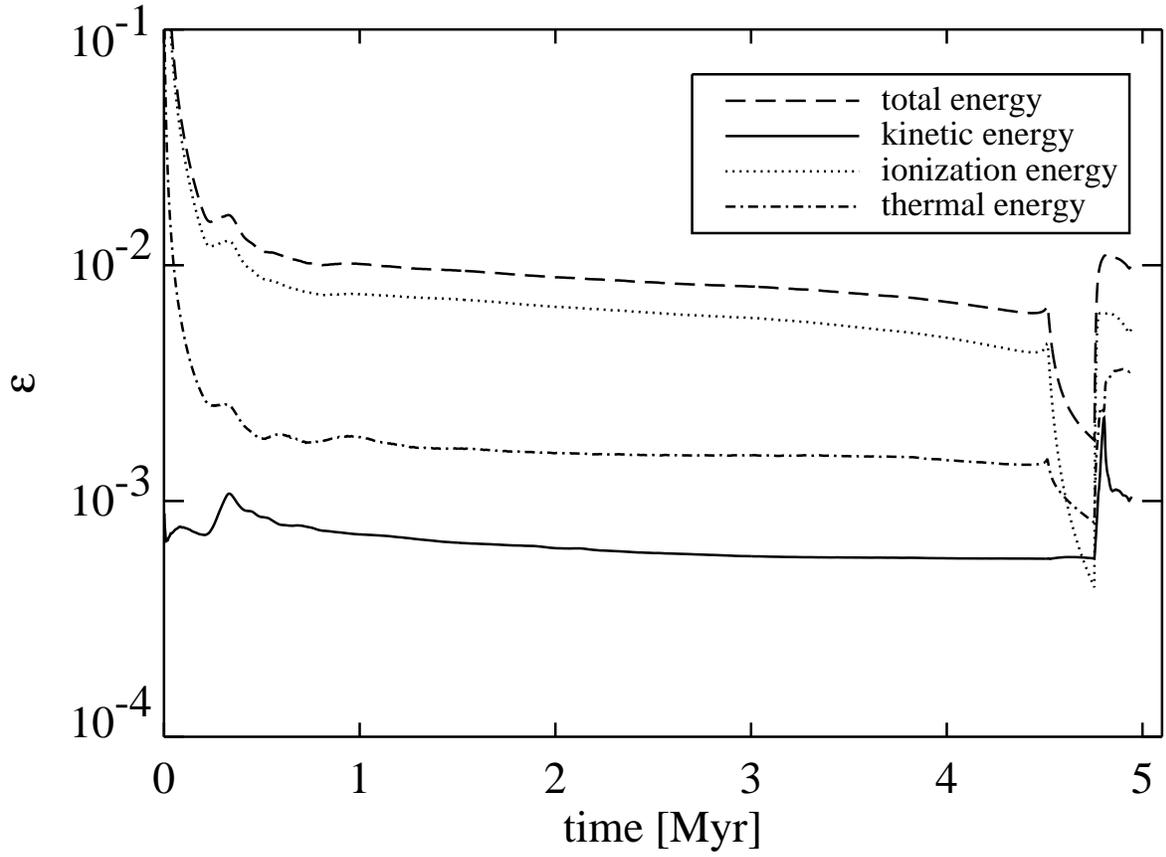}
  \caption{Energy transfer efficiency with respect to the total energy input
           for the 35 $\Msun$ case.
           \label{Ecompare229apj_up.eps}
          }
\end{figure}

  One of the goals of this work is to determine the efficiencies with which
  the stellar input energy is converted into the different forms of energy
  in the circumstellar medium. We show these values for the 35 $\Msun$ case
  in Figure \ref{Ecompare229apj_up.eps} (compare Figure 21 in
  \citeauthor{freyer03}). We recall here that we define the transfer
  efficiency as the cumulative fraction of the stellar input energy that has
  been converted into a particular form of energy in the circumstellar medium
  up to the time $t=\tau$, where $\tau$ is the age of the star. Figure
  \ref{Ecompare229apj_up.eps} shows the transfer efficiencies into kinetic,
  ionization, and thermal energy and their sum for the 35 $\Msun$ case. 

  As we have already seen in Figure \ref{e_distri229apj_up.eps}, the 
  energization of the circumstellar gas occurs fairly smoothly
  until the star reaches the RSG stage. The transfer efficiency
  into kinetic energy reaches $5.6 \times 10^{-4}$ before and during
  the RSG phase and peaks up to $2.2 \times 10^{-3}$ when the shocked
  W-R wind accelerates the RSG wind. At the end of the simulation it is
  $10^{-3}$. The transfer efficiency into thermal energy is higher
  than that into kinetic energy by a factor of 2 to 3 during most of the
  MS time. It reaches $1.4 \times 10^{-3}$ before the RSG phase, drops by
  almost a factor of 2 during the RSG phase, rises again when the star enters
  the W-R stage, and ends up at $3.6 \times 10^{-3}$ at the end of the
  simulation. The transfer efficiency into ionization energy is the highest
  except during the RSG stage of the star. Before the RSG phase it is
  $4.4 \times 10^{-3}$ and at the end of the simulation it is
  $5.5 \times 10^{-3}$.

  The total energy transfer efficiency (into kinetic, thermal, and ionization
  energy) at the end of the simulation is $10^{-2}$; i.e., 1\% of
  the total input energy from the star is transferred to the circumstellar
  gas. This is comparable to the net efficiency of the windless simulation.
  This result shows that, globally speaking, the role of
  the stellar wind is less important than for the 60 $\Msun$ case where its
  presence doubled the total energy transfer efficiency at the end of the
  calculation. However, the total energy transfer efficiency at the end of
  the 35 $\Msun$ calculation is 2.7 times as high as for
  the 60 $\Msun$ case. Since the circumstellar energy in the 35 $\Msun$
  case is dominated by ionization energy, a substantial fraction will be
  lost quickly when the star ultimately turns off. On the other hand, the SN
  will inject a huge amount of additional energy into the circumstellar gas.
  Both processes are beyond the scope of this paper.

  We summarize the values of the individual energy components
  ($E_k$, $E_i$, $E_{t,\mathrm{cold}}$, $E_{t,\mathrm{warm}}$, and
  $E_{t,\mathrm{hot}}$) at the end of the 35 $\Msun$ simulations with and
  without stellar wind in Table \ref{table_num_results}. The values of the
  energy transfer efficiency into kinetic energy ($\varepsilon_k$), ionization
  energy ($\varepsilon_i$), thermal energy ($\varepsilon_t$), and their sum
  ($\varepsilon_{\mathrm{tot}}$) at the end of the 35 $\Msun$ simulations
  with and without stellar wind are given in Table \ref{table_num_eff}.
  For comparison purposes the corresponding 60 $\Msun$ results from
  \citeauthor{freyer03} are also given there. During the lifetime of the
  35 $\Msun$ star, the total energy emitted in the Lyman continuum is
  $E_{\mathrm{LyC}} = 4.61 \times 10^{52}~\mathrm{ergs}$ and the mechanical
  energy of the stellar wind amounts to
  $E_w = 4.77 \times 10^{50}~\mathrm{ergs}$.

\subsubsection{Comparison with Analytical Results}

  Taking mean values for the relevant stellar parameters over the lifetime of
  the 35 $\Msun$ star, we can calculate the kinetic, ionization, and
  thermal energy in the system according to the analytical solutions given
  in {\S} 2 of \citeauthor{freyer03}. 
  With $\langle T_{\mathrm{eff}} \rangle = 3.78 \times 10^4~\mathrm{K}$,
  $\langle L_{\mathrm{LyC}} \rangle =
  2.95 \times 10^{38}~\mathrm{ergs~s^{-1}}$, and using
  $\alpha_B = 3.37 \times 10^{-13}~\mathrm{cm^3~s^{-1}}$
  as hydrogen recombination coefficient and
  $c_{s,\mathrm{\scriptscriptstyle{II}}} =
  1.15 \times 10^6~\mathrm{cm~s^{-1}}$ for the isothermal sound speed in the
  \HII region (corresponding to
  $T_{\mathrm{\scriptscriptstyle{II}}} = 8000~\mathrm{K}$), we obtain for
  $n_0 = 20~\mathrm{cm^{-3}}$ after $\tau = 4.945~\mathrm{Myr}$
  from equations (3), (4), and (5) in \citeauthor{freyer03}
  for the 35 $\Msun$ case without wind:
  \begin{eqnarray}
    E_k &=& 2.7 \times 10^{49}~\mathrm{ergs}\ , \nonumber \\
    E_i &=& 2.9 \times 10^{50}~\mathrm{ergs}\ , \nonumber \\
    E_t &=& 4.5 \times 10^{49}~\mathrm{ergs}\ . \nonumber
  \end{eqnarray}
  With the same definition of the energy transfer efficiency according to
  equation (11) in \citeauthor{freyer03} we get the corresponding
  energy transfer efficiencies in the analytical approach:
  \begin{eqnarray}
    \varepsilon_k &=& 5.8 \times 10^{-4}\ , \nonumber \\
    \varepsilon_i &=& 6.4 \times 10^{-3}\ , \nonumber \\
    \varepsilon_t &=& 9.7 \times 10^{-4}\ . \nonumber
  \end{eqnarray}

  The analytical value for the transfer efficiency into kinetic energy
  deviates from the numerical result by less than 2\%. Bearing in mind all
  the approximations which were made in order to obtain the analytical
  solution, it is clear that the almost perfect correspondence between
  analytical and numerical result is certainly by chance. This is supported
  by the fact that the numerical value for $\varepsilon_k$ is almost constant
  during most of the MS and RSG lifetime, but jumps up by $\approx$40\%
  during the final W-R stage. Analytical and numerical results for the
  transfer efficiency into thermal energy show fairly good correspondence
  immediately before the RSG phase (not shown here), but since the final
  W-R stage boosts the thermal energy and thus also the transfer efficiency
  into thermal energy, the final value deviates from the analytical one by
  approximately a factor of 2. The comparison of analytical and numerical
  results for the transfer efficiency into ionization energy bears similar
  results. The values are fairly close to each other before the star turns
  to the RSG stage, and the deviation at the end of the simulation is also
  only $\approx$30\%.

  As we have already discussed in \citeauthor{freyer03}, the comparison
  of analytical with numerical results is much more difficult in the case
  of the combined SWB/\HII region calculation because no analytical solution
  is yet known for energy transfer efficiencies in the case of \HII regions
  with stellar winds. Again, we construct an analytical approximation by
  simply adding up the energy contributions from the \HII region and the SWB,
  bearing in mind that this is only a rough estimation which actually
  neglects the mutual interactions. In any case the analytical energy
  transfer efficiencies into kinetic and thermal energy are upper limits,
  since cooling in the hot bubble is not considered in the analytical
  approach.

  Equations (9) and (10) from \citeauthor{freyer03} yield the kinetic and
  thermal energy for the SWB only. Inserting the mean value
  of the mechanical wind luminosity, $\langle L_w \rangle =
  3.05 \times 10^{36}~\mathrm{ergs~s^{-1}}$, and adding up the
  results for the pure \HII region, we obtain
  \begin{eqnarray}
    E_k &=& 1.6 \times 10^{50}~\mathrm{ergs}\ , \nonumber \\
    E_i &=& 2.9 \times 10^{50}~\mathrm{ergs}\ , \nonumber \\
    E_t &=& 2.6 \times 10^{50}~\mathrm{ergs}\ . \nonumber
  \end{eqnarray}
  Related to the sum of Lyman continuum radiation energy and mechanical
  wind energy (which is almost negligible for the 35 $\Msun$ star),
  we get the corresponding energy transfer efficiencies according to
  equation (12) in \citeauthor{freyer03}:
  \begin{eqnarray}
    \varepsilon_k &=& 3.4 \times 10^{-3}\ , \nonumber \\
    \varepsilon_i &=& 6.3 \times 10^{-3}\ , \nonumber \\
    \varepsilon_t &=& 5.6 \times 10^{-3}\ . \nonumber
  \end{eqnarray}
  Comparing these values with the results of the analytical solution for the
  windless case and the simulations with and without wind, one can see that
  the increase of the kinetic energy deposit from the windless to the combined
  SWB/\HII region simulation (almost doubled) as well as the increase of the
  thermal energy deposit (also almost doubled) is about a factor of 3 below
  the analytical upper limits.

\subsection{Comparison with Observations}
\label{subsec_comp_obs}

\begin{figure}
  \plotone{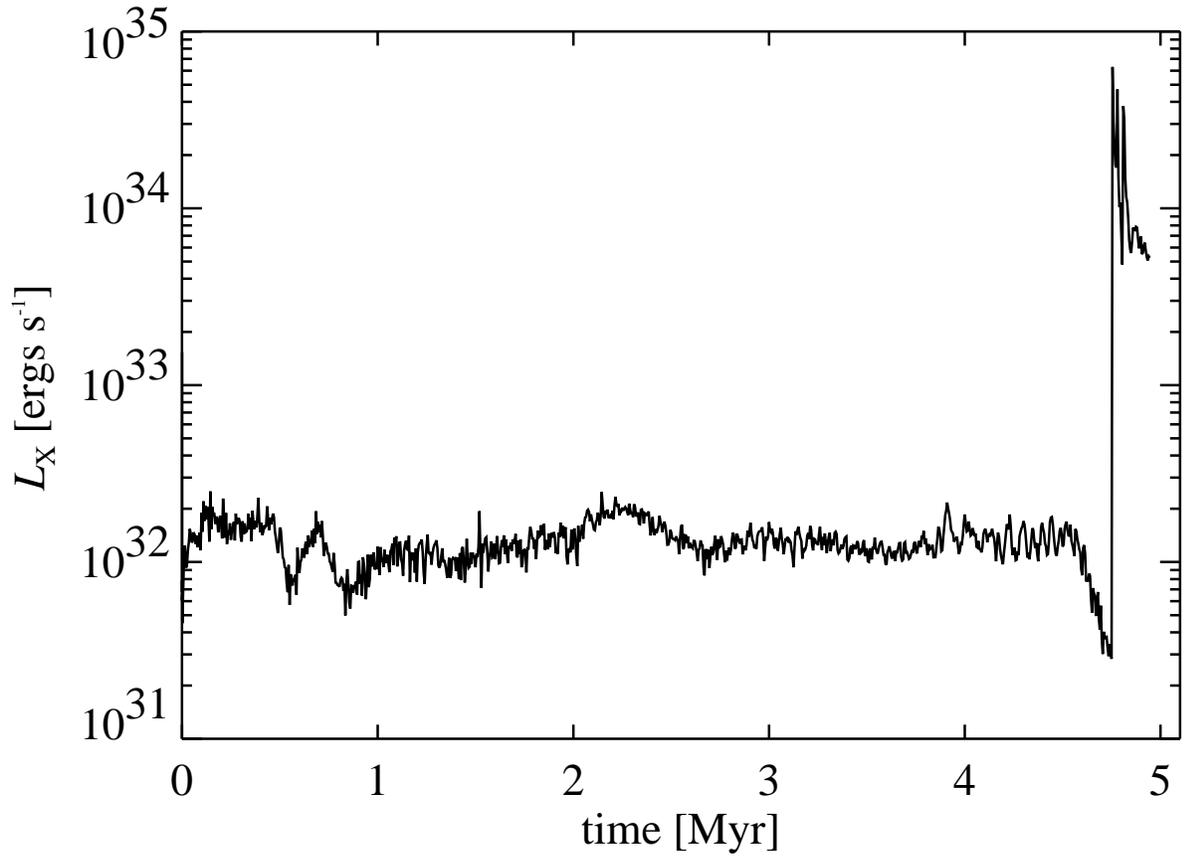}
  \caption{X-ray luminosity in the energy band $0.1-2.4~\mathrm{keV}$
           for the 35 $\Msun$ case with stellar wind.
           \label{L_X_plot229_0.1-2.4kev_S308apj_up.eps}
          }
\end{figure}

  In section \ref{sec_observations} we point out the importance of X-ray
  observations of the hot gas in SWBs to study the physics of SWBs.
  Thus, we have examined the X-ray properties of our numerical model.
  In Figure \ref{L_X_plot229_0.1-2.4kev_S308apj_up.eps}
  we plot the total X-ray luminosity of our model bubble in the energy band
  $0.1-2.4$ keV as a function of time. The emissivity $j_{\nu} (\rho,T,Z)$
  in each grid cell is calculated with the \citet{raymond77} program.
  For temperatures below $10^5~\mathrm{K}$ the emissivity is set to zero.
  The total X-ray luminosity at time $t$ is calculated as 
  \begin{equation}
    L_X(t) = \sum_{\mathrm{cells}}
             \int\limits_{h\nu=0.1~\mathrm{keV}}^{h\nu=2.4~\mathrm{keV}}
             4 \pi j_{\nu}(\rho_\mathrm{cell}(t),T_\mathrm{cell}(t),Z)
             V_\mathrm{cell}~d\nu\ ,
  \end{equation}
  where $\rho_\mathrm{cell}(t)$ and $T_\mathrm{cell}(t)$ are the plasma
  density and temperature in the grid cell, respectively, $V_\mathrm{cell}$
  is the volume of the grid cell, and $h\nu$ the energy of the X-ray photons.
  For the summation over grid cells the finest grid that is available for the
  respective coordinates is always used. As for the calculation of all
  the other global quantities, the grid data are mirrored at the equator.   
  We assume optically thin emission; absorption is neglected.
  For the set of chemical abundances, $Z$, we use the same values which
  \citet{chu03} used for their spectral X-ray fits of S308 and which are
  based on the abundance determination of \citet{esteban92b} for the
  optically visible shell. We assume that the elements not mentioned by
  \citet{chu03} have the same abundance relative to the solar value
  as oxygen (0.13). Thus, for the computation of the X-ray emissivity
  the following elements are considered:
  H, He, C, N, O, Ne, Mg, Si, S, Ar, Ca, Fe, Ni with the following abundances
  relative to the solar values \citep{anders89}:
  1.0, 2.1, 0.1, 1.6, 0.13, 0.22, 0.13, 0.13, 0.13, 0.13, 0.13, 0.13, 0.13.
  This set of elemental abundances is intended to represent the chemistry 
  in the RSG wind. The composition of the hot gas in the MS bubble
  may be different because the radiating material is supposed to originate
  from the MS wind of the star and (if thermal evaporation or ablation is
  important) also from the ambient medium. However, for the sake of simplicity
  we use the same chemical composition for all the emitting gas during the
  whole evolution. We will briefly discuss impacts of this choice below.
  The chemical composition employed for this diagnostic purpose is thus
  inconsistent with the solar chemical composition that is used to calculate
  the cooling of the gas (energy sink term in the radiation-hydrodynamical
  equations).

  Figure \ref{L_X_plot229_0.1-2.4kev_S308apj_up.eps} shows that the X-ray
  luminosity assumes the value $10^{32}~\mathrm{ergs~s^{-1}}$ soon after the
  turn on of the stellar wind. It remains remarkably constant during the
  MS phase and the associated growth of the SWB. The deviation from
  $10^{32}~\mathrm{ergs~s^{-1}}$ is less than a factor of 2.5 during this
  period. Most of the X-ray emission originates at the interface
  between the hot gas in the SWB and the swept-up \HII shell. Although He and
  N are overabundant, the strong underabundance of the other metals reduces
  the X-ray luminosity. Using a solar chemical composition instead of
  the abundances described above would increase the luminosity in
  Figure \ref{L_X_plot229_0.1-2.4kev_S308apj_up.eps} by a factor $3-4$.
  With the onset of the RSG phase the total X-ray luminosity decreases by
  a factor of $3-4$ until the star reaches the W-R stage. This happens because
  the hot gas (or more precisely the gas in the interface region close to the
  shell of swept-up ambient gas) cools and the supply of the bubble with hot
  gas ceases during the RSG phase.

  The subsequent onset of the fast W-R wind drives up the total X-ray
  luminosity by approximately 3 orders of magnitude. It reaches a few times
  $10^{34}~\mathrm{ergs~s^{-1}}$, but varies quite strongly. It is interesting
  to note that in this phase almost all of the X-ray emission comes from the
  W-R shell (instead of the shocked W-R wind). At $t = 4.775~\mathrm{Myr}$ the
  radius of this X-ray emitting shell is approximately 10 pc with 
  a thickness of $1.0-1.3$ pc. The temperature drops from
  $2.5 \times 10^6~\mathrm{K}$ behind the shock to $6 \times 10^5~\mathrm{K}$
  at the inside of the shell, the density rises from
  $2 \times 10^{-25}~\mathrm{g~cm^{-3}}$ to
  $1.2 \times 10^{-24}~\mathrm{g~cm^{-3}}$ and the velocity from 300 to
  500 $\mathrm{km~s^{-1}}$. The shocked W-R wind has (from the reverse shock
  to the contact discontinuity) a density of
  $5 \times 10^{-27}~\mathrm{g~cm^{-3}}$ to
  $1.4 \times 10^{-26}~\mathrm{g~cm^{-3}}$, a temperature of
  $1.0 \times 10^8~\mathrm{K}$ to $5 \times 10^7~\mathrm{K}$ and a velocity
  of 1000 $\mathrm{km~s^{-1}}$ to 500 $\mathrm{km~s^{-1}}$. The $\Ha$ emission
  comes mostly from the RSG shell. The density in this shell is
  $(3-12) \times 10^{-25}~\mathrm{g~cm^{-3}}$, which may be seen as a lower
  limit since this shell is not completely resolved in our calculations.
  The temperature in the shell is $(8-20) \times 10^3~\mathrm{K}$, indicating
  that there is shock heating present in addition to photoionization.
  The shell's velocity has slowed to $25~\mathrm{km~s^{-1}}$ due to its
  interaction with the MS wind material.

  Judging by the geometrical extent, the evolutionary state of the W-R bubble
  at this time is approximately comparable to the currently observable stage
  of S308. The model data show surprisingly good agreement with the
  X-ray observations of S308 described above. The total X-ray luminosity of a
  few times $10^{34}~\mathrm{ergs~s^{-1}}$ is slightly higher but of the same
  order of magnitude as the
  $\le 1.2 \pm 0.5 \times 10^{34}~\mathrm{ergs~s^{-1}}$ observed by
  \citet{chu03}. Since almost all of the X-ray emission in our model comes
  from the W-R shell, the temperature range of
  $(0.6-2.5) \times 10^6~\mathrm{K}$ agrees quite well with the
  $1.1 \times 10^6~\mathrm{K}$ that results from the spectral fit of the
  observed X-rays. Although the observational proof for the existence of a
  substantially hotter gas component is still under debate, in our model it
  could well be identified with the shocked, hot W-R wind which contributes
  only a small fraction to the total soft X-ray emission. The average density
  in the X-ray emitting W-R shell is $n_e \approx 0.4~\mathrm{cm^{-3}}$,
  which is well within the observationally determined range of
  $n_e = 0.28-0.63~\mathrm{cm^{-3}}$ for the assumed range $0.5-0.1$ of the
  hot gas volume filling factor \citep{chu03}. The total mass of the
  W-R shell is approximately 15 $\Msun$; i.e., most of the 18.6 $\Msun$
  RSG wind has already been swept up. This is at the upper end of the
  observationally supported range of $11 \pm 5~\Msun$ for an assumed volume
  filling factor of 0.5. The fact that the W-R shell supplies most of the
  X-ray emission alleviates the necessity to assume that thermal evaporation
  of RSG wind gas raises the mass of the X-ray emitting shocked W-R wind.
  The process of thermal evaporation, which is not implemented in our
  numerical model, would not be very efficient anyway, since the temperature
  gradient between the ${\approx}10^6~\mathrm{K}$ W-R shell and the
  shocked W-R wind at several times $10^7~\mathrm{K}$ is rather modest. 
  
  The conclusion of \citet{wrigge99} that S308 cannot be described by the
  two-wind model seems to be vulnerable because he assumed
  that the X-ray emission originates from the shocked W-R wind and
  that the energy in the forward shock ahead of the W-R shell is completely
  dissipated in a different wavelength range. Consequently, using the formula
  for the X-ray luminosity of the shocked W-R wind under the impact of
  conductive evaporation from \citet{garcia95a} in order to reproduce the
  observed X-ray luminosity, he derived values for the mechanical wind
  luminosity which are much too low compared to the actually observed values
  of the central W-R star. By contrast, the mechanical wind luminosity in our
  model calculation during the first $2 \times 10^4~\mathrm{yr}$ of the
  W-R stage is in the range $(3-10) \times 10^{37}~\mathrm{ergs~s^{-1}}$,
  which is in reasonable agreement with the observed stellar wind
  luminosity (scaled to the distance of $D = 1.5~\mathrm{kpc}$ used here)
  of $4.9 \times 10^{37}~\mathrm{ergs~s^{-1}}$ \citep{hamann93} or
  $1.3 \times 10^{37}~\mathrm{ergs~s^{-1}}$ for a clumping-corrected
  mass-loss rate \citep{nugis98}.

\begin{figure}
  \plotone{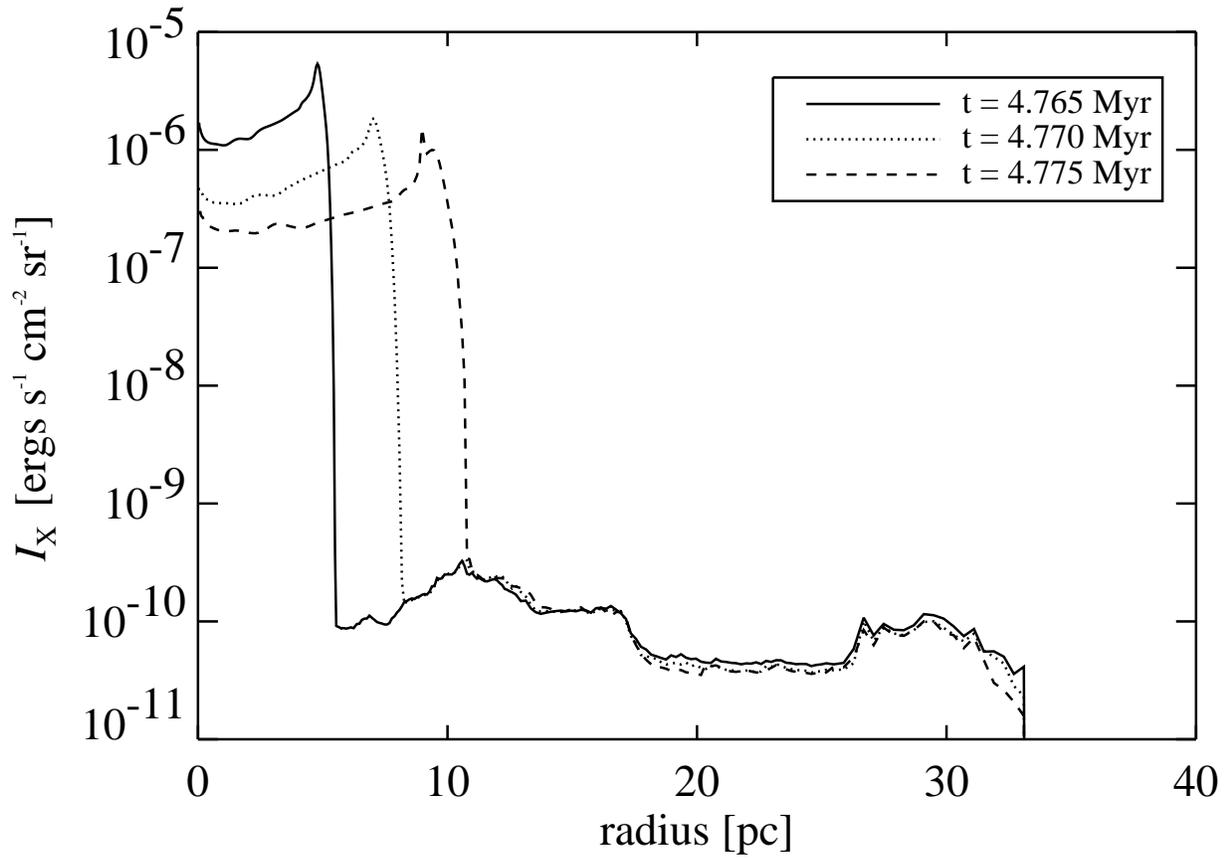}
  \caption{Angle-averaged unabsorbed X-ray intensity in the energy band
           $0.1-2.4~\mathrm{keV}$ for the 35 $\Msun$ case with stellar wind
           at selected evolutionary times.
           \label{IX3_ng_100_2400ev_S308_plot_apj_up.eps}
          }
\end{figure}

  Because the X-ray emitting volume is a thick shell ($1.0-1.3$ pc), another
  observational constraint that is well reproduced in our model is the
  limb brightening of the X-ray emission. In Figure
  \ref{IX3_ng_100_2400ev_S308_plot_apj_up.eps} we display the unabsorbed
  angle-averaged X-ray intensity profile in the energy range $0.1-2.4$ keV at
  three evolutionary times during the early W-R stage, namely
  $t = 4.765~\mathrm{Myr}, 4.770~\mathrm{Myr}$, and $4.775~\mathrm{Myr}$,
  the latter corresponding to the phase discussed above. We see that the 
  ``background'' intensity produced by the hot gas in the MS bubble is of order
  $10^{-10}~\mathrm{ergs~s^{-1}~cm^{-2}~sr^{-1}}$. The intensity for lines
  of sight through the W-R shell is higher by $3-5$ orders of magnitude. The
  limb brightening within the W-R shell can be seen for all three
  profiles as well as the decrease of the peak intensity with time.
  The consideration of absorption would not alter this result because the
  limb brightening is due to the geometry of the emitting plasma rather
  than due to varying absorption.

  Since the shocked W-R wind in our model calculation has not yet swept up the
  entire RSG wind, the X-ray emitting shell is interior to the optical
  shell as has been found in the observations. Differing from \citet{chu03},
  we interpret the gap between the outer rim of the optical emission and the
  outer edge of the X-ray emission as being the RSG ejecta in front of the
  W-R shell rather than being the W-R shell itself. The thickness of the gap
  in our model at this time is approximately 0.5 pc, which is smaller than the
  observed gap ($0.7-1.5$ pc). However, this strongly depends on the exact
  instant of time and on the exact duration of the RSG phase
  as well as the velocity of the RSG wind.

  If our interpretation of the dynamical evolution is true, the age that has
  been attributed to the W-R bubble is probably significantly overestimated
  because of the assumption that the observed velocity $63~\mathrm{km~s^{-1}}$
  is the velocity of the W-R shell rather than the velocity of the RSG shell.
  Based on our model the age of the bubble is
  ${\approx}2 \times 10^4~\mathrm{yr}$ for the stage described above, much
  less than the $1.4 \times 10^5~\mathrm{yr}$ derived from the observed
  velocity of $63~\mathrm{km~s^{-1}}$.

  As the referee pointed out, there is no observational evidence yet of a
  high-velocity (${\approx}400~\mathrm{km~s^{-1}}$) gas component that we
  identify as the W-R shell in our numerical results. The kinematic data
  obtained from the \OIII $\lambda 5007$ emission line observations of S308
  and from the \NV $\lambda\lambda 1239,~1243$ and
  \CIV $\lambda\lambda 1548,~1551$ absorption line studies toward HD 50896
  \citep{boroson97} do not show velocities as high as several
  $100~\mathrm{km~s^{-1}}$. However, if the gas in the W-R shell of our
  model calculation at $T \approx 10^6~\mathrm{K}$ has reached collisional
  ionization equilibrium, oxygen exists mostly as O {\sc{vii}}, nitrogen as
  N {\sc{vi}} or N {\sc{vii}}, and carbon as C {\sc{v}}, C {\sc{vi}}, or
  even completely ionized. Thus, there is almost no O {\sc{iii}}, N {\sc{v}},
  or \CIV in the W-R shell whose emission or, respectively, absorption could
  be observed. Other ionic tracers are needed to find high-velocity gas of
  that temperature.

  The appearance of NGC 6888 is harder to explain within the framework of our
  model. The observed geometrical size implies that the nebula is younger
  than S308, provided that the temporal evolution of the stellar parameters
  during the RSG and W-R stage is about the same. A maximum value of
  $\approx$8000 yr can be derived under the assumptions that the optically
  visible nebula has not yet been swept up by the expanding W-R bubble and
  that our stellar model parameters are appropriate for NGC 6888. The observed
  X-ray luminosity $1.6 \times 10^{34}~\mathrm{ergs~s^{-1}}$ as well as the
  emitting plasma temperature $2 \times 10^6~\mathrm{K}$ \citep{wrigge94} are
  in reasonable agreement with the model data, but the X-ray emission in the
  model originates from a thick shell which is more or less homogeneous rather
  than from small filaments with a volume filling factor of only a few \%.
  Although the hot W-R shell is able to ``hide'' mass from optical observations
  and thus helps to find the yet undetected portion of RSG wind mass, 8000 yr
  after the onset of the W-R wind it contains only $\approx$3 $\Msun$, which is
  not enough to solve the problem completely (assuming that the RSG mass loss
  used in the model is appropriate for NGC 6888).
  The biggest difficulty is probably the appearance of the optical shell.
  In the numerical model the RSG shell already has a radius greater than
  10 pc and the unaffected RSG wind between the W-R shell and the RSG shell
  contributes significantly to the $\Ha$ emission. This conflicts with the
  observation of a very thin, filamentary shell with hydrogen number densities
  as high as $1000~\mathrm{cm^{-3}}$.

  There may be various reasons for this mismatch between our model and the
  observations of NGC 6888. It is possible that physical effects not yet
  covered in our model are more important for NGC 6888 or that the resolution
  applied in our calculations is not yet high enough to allow for the
  formation of high-density filaments. Another possibility is that the actual
  mass-loss history of HD 192163 differs from what is expected and used in our
  calculations.

  Apart from the detailed comparison,
  Figure \ref{L_X_plot229_0.1-2.4kev_S308apj_up.eps} illustrates
  another aspect: The fact that up to now only W-R bubbles have been observed
  in X-rays is not only by chance. At least for the MS $\rightarrow$ RSG
  $\rightarrow$ W-R sequence the X-ray luminosity during the final
  W-R stage exceeds that of the rest of the stellar lifetime by more than an
  order of magnitude. Moreover, during the early W-R phase the emission
  comes from a relatively small volume (compared to most of the MS lifetime),
  resulting in an even higher X-ray surface brightness, which is easier to
  detect.

\section{Summary and Conclusions}
\label{sec_conclusions}

  The basic difference between the simulations presented in this paper and
  those of \citeauthor{freyer03} is the choice of the stellar parameters,
  which here are chosen to model a 35 $\Msun$ star that undergoes the
  evolution from the MS through the RSG and W-R stages.
  The fundamental structures which evolve are basically the same as observed
  in the 60 $\Msun$ case. They are generally smaller because for most of the
  time the stellar wind luminosity and the Lyman continuum luminosity are
  lower than in the 60 $\Msun$ case. At the end of the simulation the entire
  bubble structure has a radius of $\approx$44 pc, which is some 6 pc smaller
  than the final bubble of the 60 $\Msun$ case, although the 35 $\Msun$ star
  lives 0.88 Myr longer than the 60 $\Msun$ star.

  Instability-driven structure formation during the early MS phase of the star
  (the formation of ionized fingers corrugating the ionization front and the
  formation of neutral spokes shadowed by dense clumps), which we found to be
  quite prominent in the case of the 60 $\Msun$ calculation, is also visible
  in the calculations presented here, but it is much less pronounced and only
  short-lived, because of the lower mechanical wind luminosity of the star.
  The lower mechanical wind luminosity of the star reduces the thermal pressure
  of the hot gas in the bubble. The lower pressure in the hot bubble increases
  the geometrical thickness of the shell of swept-up \HII gas, making it less
  sensitive to thin-shell instabilities that could trigger the formation of
  the morphological structures described above. Since this behavior better
  preserves the basic spherical structure, other morphological effects become
  visible, which might be prohibited in the 60 $\Msun$ case by the strong
  corrugation of the bubble shell: When the swept-up \HII shell broadens
  geometrically, the plasma density in the shell decreases. This rarefaction
  reduces the rate of Lyman continuum photons necessary to sustain the
  photoionization of the shell and excess photons become available, which
  drive the ionization front outward and photoevaporate additional material
  from the neutral shell of swept-up ambient medium. The inward flow of
  evaporated material collides with the outward flow of the dissolving
  swept-up \HII shell, temporarily forming a new shell of enhanced density.
  Since all the plasma within the \HII region is almost isothermal, this
  density fluctuation vanishes soon.

  Another consequence of the reduced stellar wind luminosity (compared
  to the 60 $\Msun$ case) is the fact that for most of the time the shape of
  the \HII region is more or less preserved as a broad shell interior to the
  thin shell of swept-up ambient material rather than being so thin as in the
  60 $\Msun$ case where the \HII region is compressed into the illuminated
  inner part of the outer shell. Nevertheless, the geometrical thickness of
  the photoionized shell shrinks at the end of the simulation when the stellar
  wind luminosity reaches its maximum during the final W-R stage.

  The morphological impact of the low-velocity mass-loss phase (the RSG stage)
  is more prominent than that of the 60 $\Msun$ star (the LBV stage), because
  the total mass loss during the RSG stage is higher ($\approx$18.6~$\Msun$)
  than during the LBV stage of the 60 $\Msun$ star ($\approx$7.3~$\Msun$).
  The slowly expanding RSG wind material is subsequently swept up and
  accelerated by the shocked W-R wind so that Rayleigh-Taylor instabilities
  break it into filaments.

  The final, pre-supernova structure that shows up in the 35 $\Msun$ case
  after 4.945 Myr is basically comparable to that of the 60 $\Msun$ case
  at its end. The entire bubble is slightly smaller, the \HII shell at
  the inside of the outer neutral swept-up shell is more extended than in
  the 60 $\Msun$ case. Outer shell and \HII region are less clumpy and show
  less rippling as a consequence of the different evolutionary scenarios.
  
  The reduced substructure formation during the early MS phase is also
  reflected in the circumstellar energetics. The decrease of ionization energy
  and thermal energy of warm gas (compared to the respective case without
  stellar wind) resulting from the formation of dense photoionized structures
  with short recombination times is smaller than for the 60 $\Msun$ case.
  Ionization energy dominates the energy in the circumstellar gas
  for most of the evolution. The kinetic energy of bulk motion as well as
  the warm and the hot component of thermal energy stay fairly close together
  from $t \approx 0.3~\mathrm{Myr}$ until the star enters the RSG phase.

  The energetic variations during the RSG stage are stronger than those
  during the LBV stage of the 60 $\Msun$ star because, due to the duration of
  the RSG phase, the ionizing radiation of the star is switched off for a
  considerable period of time so that the photoionized regions recombine.
  In addition, the mass that is ejected during the RSG stage and accelerated
  during the subsequent W-R stage is about 2.5 times higher than the mass
  ejected during the LBV stage of the 60 $\Msun$ star.

  At the end of the 35 $\Msun$ simulation the total energy transfer efficiency 
  is 1\%. This value is about the same as in the corresponding case without
  stellar wind, but it is 2.7 times higher than the value at the end of the
  60 $\Msun$ simulation. 54\% of the net energy which has been added to the
  system is then in the form of ionization energy, 36\% in thermal energy and
  10\% in kinetic energy of bulk motion. The corresponding values at the end
  of the 60 $\Msun$ calculation are 25\%, 40\%, and 35\%, respectively.
  This is another indication that the stellar wind plays a more prominent role
  in the 60 $\Msun$ case.

  Remarkable agreement of the X-ray properties is found when comparing our
  model calculations during the early W-R phase with observations of S308.
  The order of magnitude of the observed X-ray luminosity as well as the
  temperature of the emitting plasma and the limb brightening of the intensity
  profile are well reproduced. The obvious explanation that our model
  overcomes the ``missing wind problem'' described in
  {\S} \ref{sec_observations} is that almost the entire X-ray emission during
  this phase comes from the W-R shell rather than from the shocked W-R wind.
  Analytical models constructed so far \citep[see e.g.][]{garcia95a} assume
  the W-R shell to be thin and cool so that the energy in the forward shock is
  completely dissipated in the low-energy wavelength range. The source of
  X-rays in these models is the shocked W-R wind and the efficiency of
  heat conduction and thermal evaporation between the hot gas and the cold
  W-R shell strongly influences the luminosity and the spectral shape of the
  X-ray emission. If heat conduction is efficient enough to cool the hot
  shocked W-R wind down to the observed temperature and if the W-R wind
  luminosity in the models is adjusted to reproduce the observed X-ray
  luminosity, the W-R wind luminosity is usually much lower than observed
  \citep{wrigge99}.  

  Another factor that reduces the X-ray luminosity in our model to values
  roughly comparable with S308 is the assumed set of chemical abundances.
  Although He and N are overabundant according to the observations in the
  nebula, the underabundance of the other metals reduces the X-ray luminosity
  by a factor of $3-4$ compared to what can be expected from solar chemical
  composition.

  A further consequence of this interpretation of our results is that
  the $\Ha$ emission originates mostly from the RSG shell.
  This is in agreement with the finding of \citet{chu03} that the X-ray
  emission is completely interior to the optical shell. Since the age of S308
  (and other W-R bubbles) was hitherto derived from the expansion velocity of
  the optical nebula under the assumption that the nebula is part of the
  W-R shell, our results imply an age of S308 which is much younger
  (${\approx}2 \times 10^4~\mathrm{yr}$) than assumed so far.

  However, the match of our model data and the observations is worse for the
  case of NGC 6888. Besides numerical or model restrictions, differences of
  the mass-loss and luminosity history between the central star of NGC 6888
  and our model star might be responsible for the discrepancies.

\acknowledgments

  This work was supported by the Deutsche Forschungsgemeinschaft (DFG) under
  grant number He~1487/17 and by the National Aeronautics and Space
  Administration (NASA) under grant NRA-03-OSS-01-TPF. Part of the research
  described in this paper was conducted at the Jet Propulsion Laboratory
  (JPL), California Institute of Technology. The computations were performed
  at the Rechenzentrum der Universit\"at Kiel, the Konrad-Zuse-Zentrum f\"ur
  Informationstechnik in Berlin, and the John von Neumann-Institut f\"ur
  Computing in J\"ulich. We thank the referee, Guillermo Garc\'{\i}a-Segura,
  for valuable comments and for making available new kinematic data of S308.




\clearpage


\clearpage
\clearpage
\clearpage
%
%






\clearpage

\begin{table}
\begin{center}
\caption{Energy Components at the End of the 35 $\Msun$ Simulations
         \label{table_num_results}}
\begin{tabular}{lccccc}
\tableline\tableline
Model Parameters                 & $E_k$                            &
$E_i$                            & $E_{t,\mathrm{cold}}$            &
$E_{t,\mathrm{warm}}$            & $E_{t,\mathrm{hot}}$             \\
                                 & $(\times 10^{49}~\mathrm{ergs})$ &
$(\times 10^{49}~\mathrm{ergs})$ & $(\times 10^{49}~\mathrm{ergs})$ &
$(\times 10^{49}~\mathrm{ergs})$ & $(\times 10^{49}~\mathrm{ergs})$ \\
\tableline
Windless  & $2.6$  & $39$  &
$3.0$     & $7.6$  & $0$   \\
With wind & $4.9$  & $25$  &
$3.5$     & $4.3$  & $11$  \\
\tableline
\end{tabular}
\tablecomments{The thermal energy of the cold component, $E_{t,\mathrm{cold}}$,
   contains the internal energy of the initially unperturbed ambient medium
   ($1.6 \times 10^{49}~\mathrm{ergs}$) that has to be subtracted whenever
   the input of thermal energy into the system is considered.}
\end{center}
\end{table}

\clearpage

\begin{table}
\begin{center}
\caption{Energy Transfer Efficiencies at the End of the 60 $\Msun$ and 
         35 $\Msun$ Simulations
         \label{table_num_eff}}
\begin{tabular}{lcccc}
\tableline\tableline
Model Parameters             & $\varepsilon_k$ &
$\varepsilon_i$              & $\varepsilon_t$ &
$\varepsilon_{\mathrm{tot}}$ \\
                             & $(\times 10^{-4})$ &
$(\times 10^{-4})$           & $(\times 10^{-4})$ &
$(\times 10^{-4})$           \\
\tableline
60 $\Msun$ windless  & $3.3$ &
$12$                 & $6.4$ &
$22$                 \\
60 $\Msun$ with wind & $13$  &
$9.1$                & $15$  &
$37$                 \\
35 $\Msun$ windless  & $5.7$ &
$84$                 & $19$  &
$109$                \\
35 $\Msun$ with wind & $10$  &
$55$                 & $36$  &
$101$                \\
\tableline
\end{tabular}
\tablecomments{The data of the 60 $\Msun$ case are taken
               from \citeauthor{freyer03}.}
\end{center}
\end{table}






\end{document}